\documentclass[11pt]{preprint}
\usepackage{cite} 
\usepackage{hyperref} 
\usepackage[version=4]{mhchem}
\usepackage{siunitx}
\usepackage{subcaption}
\usepackage{makecell}
\usepackage{booktabs,multirow}
\usepackage{array}
\usepackage{makecell}
\usepackage{xcolor}
\definecolor{markupgreen}{HTML}{228B22}

\usepackage{caption}
\DeclareCaptionFont{customsize}{\fontsize{10}{12}\selectfont}
\hypersetup{
  colorlinks=true,
  linkcolor=black,
  citecolor=blue,   
  urlcolor=blue,
  breaklinks=true
}

\title{\textbf{Uncertainty Aware AFM-IR Hyperspectral Cube Reconstruction via Convolutional-Attentive Conditional Neural Processes}}

\author[1,2]{William F. Kent}
\author[1,2,3]{Roberto dos Reis}
\author[1,2]{Gajendra Shekhawat}
\author[4]{Emma Alexander}
\author[1,2,3]{Vinayak P. Dravid\thanks{Corresponding author. Email: \href{v-dravid@northwestern.edu}{v-dravid@northwestern.edu}}}

\affil[1]{Department of Materials Science and Engineering, Northwestern University, Evanston, IL 60208}
\affil[2]{The NU\textit{ANCE} Center, Northwestern University, Evanston, IL, 60208, USA}
\affil[3]{International Institute for Nanotechnology, Northwestern University, Evanston, IL 60208, USA}
\affil[4]{Department of Computer Science, Northwestern University, Evanston, IL 60208, USA}

\date{\today}

\begin{document}
\maketitle

\begin{abstract}
Atomic force microscopy-infrared spectroscopy (AFM-IR) maps chemical composition at nanoscale spatial resolution, but the time required to acquire individual spectra limits hyperspectral imaging. Here, we introduce a convolutional-attentive conditional neural process (CACNP), a neural process architecture for reconstructing AFM-IR hyperspectral cubes from sparsely sampled spectra and a small set of structure images, each acquired at a selected wavenumber. The model is trained entirely on synthetic cubes generated from experimental infrared spectra and predicts both spectral intensities and their associated uncertainty. Validation on experimental datasets of poly(methyl methacrylate) and polystyrene particles embedded in epoxy demonstrates accurate reconstruction using only 5\% of the available spectra, with competitive performance when compared to other methods. Inference takes less than 0.1 s on $128 \times 128$ hyperspectral cubes, supporting rapid reconstruction updates, while predictive uncertainty provides a basis for adaptive sampling of localized chemical features. This approach advances autonomous, high-throughput AFM-IR workflows and offers an architecture for accelerating other hyperspectral imaging methods.
\end{abstract}

\keywords{Nano-IR, AFM, Neural Process, Adaptive Sampling, Neural Networks}

\section{Introduction}
\label{sec::intro}

Materials science is concerned with elucidating the relationships between the composition, structure, and properties of materials. Establishing these relationships requires characterization methods capable of resolving chemical variation at the length scales over which properties are determined. Infrared (IR) spectroscopy is a widely used method for identifying chemical species through their vibrational absorption bands, but conventional far-field IR microscopy is diffraction limited to spatial resolutions on the order of several micrometers, which is far coarser than the length scales of many functional materials and biological structures \cite{reddyHighDefinitionInfraredSpectroscopic2013,kimberSpectroscopicImagingBiomaterials2017}. Atomic force microscopy-infrared spectroscopy (AFM-IR), also referred to as nano-IR, circumvents this limit by combining atomic force microscopy (AFM) with a pulsed, wavelength-tunable IR source. Absorption of IR radiation produces photothermal expansion of the sample, which is transduced mechanically by the AFM cantilever, resulting in the spatial resolution being set by the tip-sample contact rather than by the wavelength of the probing radiation \cite{dazziAFMIRTechnologyApplications2017,grossChallengesOpportunitiesIR2019,schwartzGuideNanoscaleIR2022,dazziAFMIRCombining2012,mathurinPhotothermalAFMIRSpectroscopy2022}. AFM-IR has been used to characterize a wide variety of materials systems, including a wide range of energy materials \cite{grossChallengesOpportunitiesIR2019}, such as perovskite solar cells \cite{chaeChlorideIncorporationProcess2015}, and biological systems \cite{vddossantosAFMIRNanoscaleChemical2023}, including prostate cancer cells \cite{romanComparisonHighDefinition2020}, bone tissue \cite{imbertDynamicStructureComposition2018}, and subcellular structures in eukaryotic cells \cite{quaroniMidinfraredSpectroscopyMicroscopy2018}. It has also been used to correlate IR spectra with nanoscale mechanical properties \cite{silbernaglBulkChemicalComposition2021} and to characterize chemical distributions within polymer blends and at buried interfaces \cite{morschExploringWhetherBuried2019,morschMolecularOriginsEpoxyAmine2022,morschPersistenceInterphaseRegions2025}.

The principal limitation of AFM-IR is acquisition speed \cite{romanComparisonHighDefinition2020a}. A complete hyperspectral cube requires that an individual spectrum be collected at every position in the image, and the serial nature of this measurement means that acquisition times scale with both the number of spatial positions and the number of wavenumbers sampled. In practice, users commonly avoid acquiring a full cube by imaging at a single wavenumber, or at a small number of wavenumbers selected in advance. While this approach is substantially faster, it presupposes knowledge of which absorption bands are diagnostic of the chemistry of interest, and it leaves open the possibility that spectral features not represented in the selected bands are missed entirely. Long acquisition times also hinder experimental throughput and can result in significant sample drift over the duration of a full hyperspectral scan. These constraints have motivated the broader argument that automated and autonomous methods are required to accelerate scanning probe microscopy (SPM) workflows \cite{kalininAutomatedAutonomousExperiments2021}.

Machine learning has been applied extensively to accelerate SPM acquisition and to improve the information extracted from limited measurements \cite{rahmanlaskarScanningProbeMicroscopy2023}. Convolutional neural networks (CNNs) have been used to accelerate conventional AFM topography by removing fast-scan artifacts \cite{zhengSpeedingTopographyImaging2022}, to reconstruct topography from compressive measurements \cite{zhangAdaptiveBlockImaging2024}, to super-resolve AFM images \cite{xunAFMSuperResolutionReconstruction2024,xuEnhancedReconstructionAtomic2025}, and to denoise low signal-to-noise-ratio fast scans \cite{natinskyReconstructionHighresolutionAtomic2024}. For hyperspectral SPM specifically, Borodinov et al. applied a coupled non-negative matrix factorization (CNMF) algorithm to correlate high-resolution structural channels with sparsely sampled AFM-IR spectra \cite{borodinovApplicationPansharpeningAlgorithm2019}, and K\"astner et al. demonstrated a similar approach of using low-rank matrix reconstruction to recover a full AFM-IR hyperspectral cube from a small fraction of measurements \cite{kastnerCompressedAFMIRHyperspectral2023}, with the associated subsampling schemes assessed by Metzner et al. \cite{metznerAssessmentSubsamplingSchemes2022}. Ziatdinov et al. used Gaussian process (GP) modeling to reconstruct hyperspectral band-excitation piezoresponse force microscopy data of \ce{BiFeO3} films from sparse measurements \cite{ziatdinovImagingMechanismHyperspectral2020a}, and related deep kernel learning (DKL) approaches have been used to correlate inexpensive structural images with expensive spectroscopic outputs \cite{liuLearningRightChannel2023} and to guide automated piezoresponse force microscopy experiments \cite{liuExploringLeakageDielectric2022}. Most directly related to the present work, Gao et al. introduced SS-Net, a self-supervised miniaturized spatial-spectral network that reconstructs nano-IR hyperspectral images without requiring a large training dataset \cite{gaoFastNanoIRHyperspectral2024}.

Analogous strategies have been pursued in electron and X-ray microscopy. Deep learning denoising has been used to enable rapid image acquisition \cite{parkDeepLearningDenoising2025,leeMachineLearningPipeline2026}, and adaptive sampling schemes have been developed for both scanning transmission electron microscopy \cite{hujsakHighSpeedLow2018b} and scanning electron microscopy \cite{dahmenFeatureAdaptiveSampling2016}. Roccapriore et al. applied DKL to automated four-dimensional scanning transmission electron microscopy experiments, in which the model is retrained during the experiment to direct subsequent measurements \cite{roccaprioreAutomatedExperiment4DSTEM2022}. In hyperspectral X-ray imaging, Adrar et al. used a Gaussian mixture variational autoencoder to correlate subsampled scanning transmission X-ray microscopy data with phase-segmented high-resolution data in order to map phases in cycled sodium-ion cathodes \cite{adrarMachineLearningEnabledRobustPhase2026,adrarAIDrivenPhaseIdentification2026}.

Despite these advances, existing hyperspectral reconstruction methods present two limitations that are relevant to autonomous acquisition. The first is inference speed. GP methods scale unfavorably with the number of observations \cite{liuWhenGaussianProcess2020,quinonero-candelaUnifyingViewSparse2005}, and inference times of the order of tens of seconds are not uncommon, which is slow relative to the acquisition time of an individual spectrum and therefore constrains how frequently a reconstruction can be updated during an experiment; Roccapriore et al. reported that retraining a DKL model required one to two seconds on a graphics processing unit and approximately an order of magnitude longer on a central processing unit \cite{roccaprioreAutomatedExperiment4DSTEM2022}. The second limitation is the absence of an uncertainty estimate. Matrix factorization \cite{kastnerCompressedAFMIRHyperspectral2023,borodinovApplicationPansharpeningAlgorithm2019} and SS-Net \cite{gaoFastNanoIRHyperspectral2024} both achieve strong reconstruction accuracy, but neither returns a calibrated estimate of the uncertainty associated with a reconstructed spectrum. Without such an estimate, there is neither a principled quantity on which users can estimate the quality of their reconstruction, nor a quantity on which to determine how much acquisition is required or which next measurement position will be most informative, which is the central requirement of an adaptive acquisition strategy. A method suited to autonomous AFM-IR must therefore reproduce the underlying data faithfully, perform inference quickly enough to be run between measurements, and report a meaningful estimate of its own uncertainty. 

The requirement for uncertainty-aware inference is shared by the broader effort to develop autonomous microscopy. Autonomous SPM workflows have been demonstrated in which the instrument selects its own measurement positions on the basis of a model fitted to the data acquired so far \cite{vasudevanAutonomousExperimentsScanning2021}, and have been extended to hypothesis-driven experimentation in which competing physical models compete to explain the observations \cite{liuAutonomousScanningProbe2023}. These developments form part of a wider movement toward artificial-intelligence-driven materials discovery and characterization \cite{madikaArtificialIntelligenceMaterials2025}, in which the role of the human operator is increasingly one of supervising and constraining an automated decision process rather than selecting each measurement directly \cite{kalininHumanintheLoopFutureMachine2024}. To further accelerate hyperspectral SPM through AI-driven adaptive sampling, we make use of neural processes. 

Neural processes are a family of meta-learning models that learn to map a set of observations directly to an approximation of the corresponding posterior predictive distribution, and are therefore well matched to the requirements outlined above. Conditional neural processes (CNPs) were introduced as an amortized alternative to GP regression in which a permutation-invariant encoder summarizes the observed set and a decoder returns a predictive mean and variance at any query location \cite{pmlr-v80-garnelo18a}, and were subsequently extended to include a latent variable that captures correlations between predictions \cite{garneloNeuralProcesses2018}. Attentive neural processes (ANPs) address the tendency of neural processes to underfit at the observed locations by allowing each query to attend to the observations it most resembles \cite{kimAttentiveNeuralProcesses2019}. Convolutional conditional neural processes (ConvCNPs) embed the observed set onto a grid and process it with a convolutional architecture, which introduces translation equivariance as an inductive bias \cite{gordonConvolutionalConditionalNeural2020}. More recent variants replace the convolutional backbone with a transformer \cite{nguyenTransformerNeuralProcesses2022} or modify the convolutional operations to act in the spectral domain \cite{mohseniSpectralConvolutionalConditional2026}. Because a neural process is trained across many realizations of a task rather than fitted to a single dataset, it offers the possibility of meta-learning a prior appropriate to IR spectra, in place of the hand-specified kernel required by a GP.

In this work, a convolutional-attentive conditional neural process (CACNP) is introduced for the reconstruction of AFM-IR hyperspectral cubes from limited spatial and spectral context. The model returns both a reconstructed cube and a per-voxel estimate of the predictive uncertainty, which is used to direct subsequent measurements. The CACNP is trained entirely on synthetic AFM-IR hyperspectral cubes, generated by assigning spectra drawn from a corpus of 15,829 experimental spectra obtained from the NIST Chemistry Web Book to randomly generated regional morphologies, and is evaluated on two experimental datasets of polystyrene (PS) and poly(methyl methacrylate) (PMMA) embedded in epoxy. CACNP is compared with nearest-neighbor filling, coupled non-negative matrix factorization, GP regression, and SS-Net in terms of reconstruction accuracy, inference time, and uncertainty. Uncertainty-driven acquisition is evaluated for its ability to recover localized spectral features and its effect on overall reconstruction error.

\section{Methods}
\label{sec::methods}

\subsection{Infrared Spectral Corpus}
\label{subsec::corpus}

Synthetic cubes were constructed from 15,829 experimental IR spectra obtained from five collections in the NIST Chemistry WebBook \cite{nist_coblentz_2026,nist_infrared_2026,nist_quantitative_ir_2026,nist_iarpa_pnnl_liquid_2026,nist_iarpa_pnnl_solid_2026}. Spectra were interpolated onto a common grid spanning 761 to \SI{1865}{cm^{-1}} in \SI{4}{cm^{-1}} increments, giving 277 channels, and each spectrum was min-max normalized to $[0,1]$.  28 spectra (0.2\% of the corpus) had no data within the desired wavenumber range and were fully zero-filled, and 68 spectra (0.4\% of the corpus) had spectral ranges that partially overlapped the desired wavenumber range and were partially zero-filled. The grid matches the experimental datasets described in \S~\ref{subsec::exp_samp}. Collection and phase counts are provided in Tables~\ref{tab:nist_collections} and \ref{tab:phase_dist}, and corpus construction is described in \S~\ref{subsec::ir_corpus}. 

The corpus contains conventional IR spectra rather than AFM-IR measurements, as there is no publicly-available large corpus of AFM-IR spectra. Gas- and condensed-phase spectra can differ in peak positions, widths, and relative intensities because of intermolecular interactions and rotational structure \cite{burshtein1994introduction,yuanClustermodelDFTSimulations2018,ozakiAdvancesChallengesPerspectives2021,balaQuantitativeAnalysisInfrared2020}. AFM-IR introduces additional differences from bulk FTIR \cite{romanComparisonHighDefinition2020a}, including sensitivity to the orientation of vibrational transition dipoles relative to the incident polarization \cite{hinrichsPolarizationDependentAtomicForce2018,piergiesPolarizationEffectTipenhanced2018,mathurinPhotothermalAFMIRSpectroscopy2022,waeytensProbingAmyloidFibril2021}. Nevertheless, the corpus supplies a broad distribution of spectral features throughout the target range. There are no solid phase PMMA and PS spectra within the corpus nor any spectra exactly matching the experimental data, avoiding data leakage between the training data and experimental tests. Comparison with the training corpus identified chemically distinct nearest matches that retained substantial spectral differences from the experimental PS, PMMA, and epoxy spectra (see Fig.~\ref{fig:si_corpus_matches}). Together with the exclusion of the experimental cubes from training, this confirms that reconstruction was evaluated on independent experimental data rather than exact copies of training spectra. The nearest peak and functional-group distributions for the whole corpus are shown in Figs.~\ref{fig:peak_dists} and \ref{fig:chem_dists}.

\subsection{Synthetic Hyperspectral Cube Generation}
\label{subsec::synth}

The CACNP was trained entirely on synthetic AFM-IR hyperspectral cubes generated during training from the corpus in \S~\ref{subsec::corpus}. This avoids the need for a large library of experimentally acquired cubes and permits direct control over the morphologies, spectra, and artifacts presented to the model. A comparable strategy has been used for generating traditional AFM topography data \cite{millan-solsonaSyntheticDatadrivenDeep2026}. The generation code is available at \cite{Kent_CACNP_AFM_IR_2026}.

\begin{figure}[htbp!]
    \centering
    \includegraphics[width=\linewidth,height=0.78\textheight,keepaspectratio]{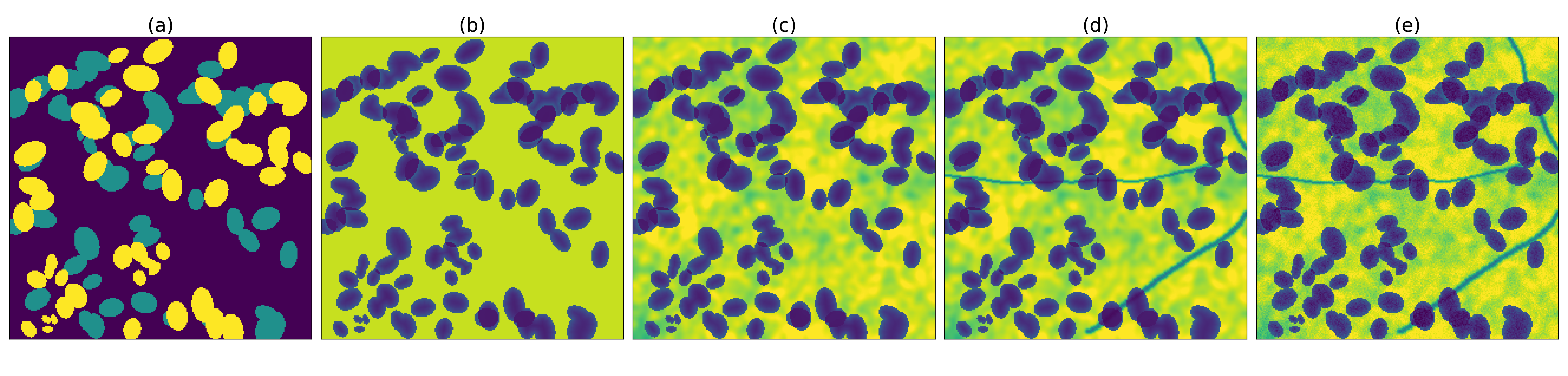}
    \caption{Illustration of synthetic AFM-IR cube generation. (a) Three-phase dispersed morphology within a continuous matrix. (b) Spectral slice after assignment of phase spectra and edge-distance mixing with underlying material. (c) Addition of phase-specific intensity texture. (d) Addition of scratch-like defects. (e) Addition of Gaussian detector noise. Panels (b--e) show the same spectral channel with a common intensity scale. Colors in (a) indicate phase identity.}
    \label{fig:synth_gen}
\end{figure}

Fig.~\ref{fig:synth_gen} illustrates the generation stages. Each cube contained $128\times128$ spatial positions, 277 spectral channels, and two to four phases. Phase maps were generated either as dispersed elliptical particles within a matrix or as Voronoi grains, with equal probability (Fig.~\ref{fig:synth_gen}a). One corpus spectrum was assigned randomly to each phase. Subsurface mixing was approximated by a contribution from the underlying phase that decays exponentially with inward distance from the visible phase boundary (Fig.~\ref{fig:synth_gen}b). This provides a geometric approximation to the depth sensitivity of AFM-IR \cite{dazziPhotothermalAFMIRDepth2024}, without assigning physical thicknesses or probing depths. 

Spatially correlated intensity texture (Fig.~\ref{fig:synth_gen}c) and scratch-like multiplicative defects (Fig.~\ref{fig:synth_gen}d) were applied to the assembled cube, followed by independent Gaussian noise with a standard deviation drawn uniformly from 0.01 to 0.08 in normalized units (Fig.~\ref{fig:synth_gen}e). Component spectra were drawn uniformly from the spectral corpus, inevitably resulting in combinations of spectra that would not be experimentally feasible, but ensuring a highly diverse training set to avoid the CACNP overfitting on a subset of hyperspectral cubes. The complete procedure and context selection are described in \S~\ref{subsec::si_synthetic_generation}; morphology, artifact, and context parameters are listed in Tables~\ref{tab:si_generation_morphology}, \ref{tab:si_generation_artifacts}, and \ref{tab:si_generation_context}.

\subsection{Experimental Samples and Data Acquisition}
\label{subsec::exp_samp}

Two hyperspectral cubes were acquired on a Bruker IconIR system in resonance-enhanced contact mode from a standard sample of PS and PMMA particles embedded in epoxy. The single-particle dataset covers a $1.5\times\SI{1.5}{\micro\meter}$ field of view on a $64\times64$ grid (\SI{23}{nm} pixels) and resolves a PMMA--epoxy interface. The multi-particle dataset covers a $20\times\SI{20}{\micro\meter}$ field of view on a $128\times128$ grid (\SI{156}{nm} pixels) containing both polymers. Both cubes span 761--\SI{1865}{cm^{-1}} in \SI{4}{cm^{-1}} increments. Acquisition took approximately 5 and 20 hours, respectively. While some thermal drift is expected over such long acquisition times, the features in the AFM-IR hyperspectral cubes and the topography images acquired prior to spectral characterization correspond well, as can be seen in Figs.~\ref{fig:si_small_fov} and \ref{fig:si_cubes_basic}. For the analysis performed in this work, the hyperspectral data is treated as the ground truth data, and all structure images are obtained by sampling the appropriate image from the GT hyperspectral cube. Registration between the acquired spectra and physical locations on the sample was performed using Nanoscope (Bruker). Additional sample information is provided in \S~\ref{subsec::si_exp_data}. During deployment of the CACNP in long experimental campaigns, a more rigorous treatment of thermal drift and spectrum-topography registration would be required.

\subsection{CACNP Architecture}
\label{subsec::arch}

\begin{figure}[htbp!]
    \centering
    \includegraphics[width=\linewidth,height=0.78\textheight,keepaspectratio]{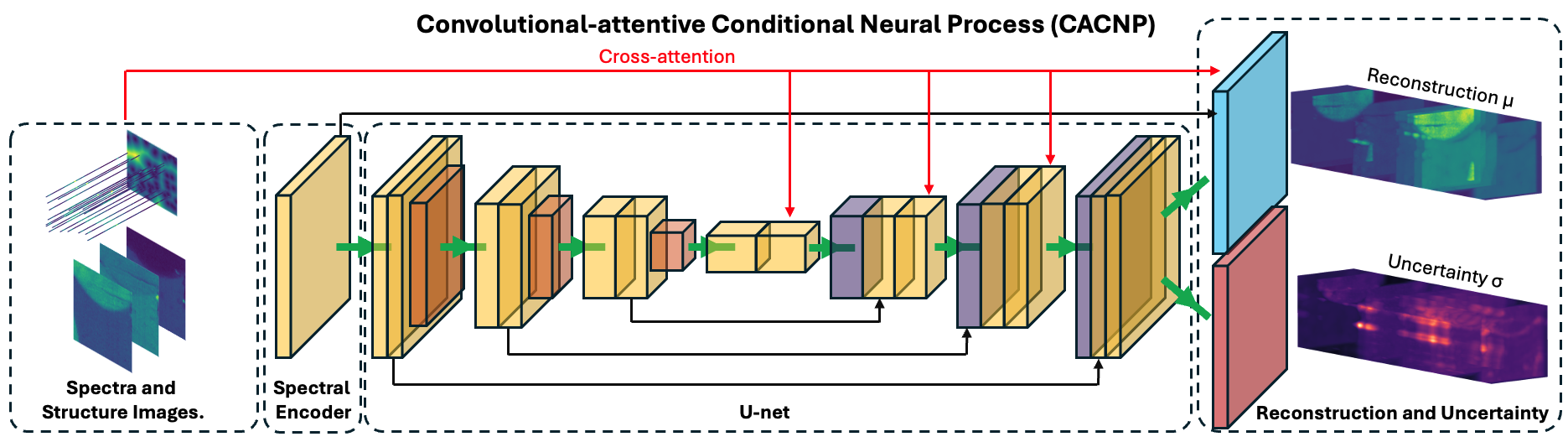}
    \caption{Convolutional-attentive conditional neural process (CACNP) architecture schematic. Context spectra, distance transform, and embedded structure images are fed into a spectral encoder. The encoded input information is passed through a U-net (skip connections shown by black arrows). At the bottleneck, first two decoder stages, and $\mu$ output head, cross-attention is performed with context spectra (shown by red arrows). Final output heads reproduce the reconstruction $\mu$ (blue) and uncertainty $\sigma$ (red). Convolution, pooling, and bilinear interpolation layers are represented by yellow, orange, and purple blocks, respectively. }
    \label{fig:arch}
\end{figure}

The CACNP was implemented in Python using PyTorch and PyTorch Lightning \cite{paszkePyTorchImperativeStyle2019,falcon2019pytorchlightning}, and is depicted in Fig.~\ref{fig:arch}. The model takes as input a set of context spectra acquired at a subset of positions in the field of view, together with one to four full-field structure images acquired at single wavenumbers, and returns a predictive mean and a predictive standard deviation for every voxel of the cube. The architecture was inspired by the original model shown in Gordon et al. \cite{gordonConvolutionalConditionalNeural2020}.

Context spectra are mapped onto the spatial grid by nearest-neighbor filling, with an additional channel recording the distance to the nearest observation. Each structure image is labeled with a learned spectral-ID embedding. These structure images are concatenated with the nearest-neighbor filled cube and distance map, then projected through $1\times1$ convolution spectral encoder before entering a U-Net.

The key architectural feature that yields improved performance is cross-attention, which is implemented at the bottleneck and two decoder stages, allowing spatial features to query the measured context spectra directly. Separate convolutional heads return the predicted mean and standard deviation. A final attention operation adds a correction to the mean using spectral content alone, without positional encoding, providing a direct route from measured spectral features to the reconstructed cube. Attention output projections are initialized to zero. Layer dimensions, parameter counts, and architecture settings are given in \S~\ref{subsec::si_cacnp_architecture} and Tables~\ref{tab:si_cacnp_layers} and \ref{tab:si_cacnp_settings}. Fig.~\ref{fig:si_attn_ablation} shows reconstruction results for an attention ablation, showing worse reconstruction results for an architecture with all hyperparameters and architecture features being held the same except with no cross-attention implemented. The impact of cross-attention is discussed further in \S~\ref{subsec::si_attn_ablation}. 

\subsection{Training Procedure}
\label{subsec::train}

New synthetic cubes were generated for each training task. Context positions were sampled uniformly without replacement, with their number drawn from a truncated log-normal distribution. Between one and four structure images were selected at peaks in the spectral variance of the context observations, ensuring images with meaningful phase contrast were selected as structure images. Each cube was normalized using a single mean and standard deviation computed over all context spectra. The sampling distributions are specified in \S~\ref{subsubsec::si_context_select}.

The model was trained on an NVIDIA A100 GPU for 621 epochs (approximately 40 hours) using AdamW with a learning rate of $3\times10^{-4}$, linear warm up, and cosine decay. The objective was

\begin{align*}
\mathcal{L} ={}&
\mathcal{L}_{\mathrm{NLL}}
+\lambda_{\mathrm{ctx}}\mathcal{L}_{\mathrm{ctx}}
+\lambda_{\mathrm{mse}}\mathcal{L}_{\mathrm{mse}} 
+\lambda_{\mathrm{grad,img}}\mathcal{L}_{\mathrm{grad,img}}
+\lambda_{\mathrm{grad,full}}\mathcal{L}_{\mathrm{grad,full}}
+\lambda_{\mathrm{TV}}\mathcal{L}_{\mathrm{TV}}.
\end{align*}

The Gaussian negative log-likelihood $\mathcal{L}_{\mathrm{NLL}}$ couples the predicted mean and uncertainty. The context loss $\mathcal{L}_{\mathrm{ctx}}$ penalizes high uncertainty values at the provided context. Early training runs yielded models that predicted equal confidence at context and unobserved voxels, so $\mathcal{L}_{\mathrm{ctx}}$ was added to drive the model to be more confident reconstructing spectra at context locations. The mean-squared error $\mathcal{L}_{\mathrm{mse}}$ penalizes reconstruction error at unobserved positions, providing additional incentive on top of $\mathcal{L}_{\mathrm{NLL}}$ for the model to minimize predicted error at unobserved voxels. Separate spatial-gradient terms match the reconstructed gradients to those of the supplied structure images $\mathcal{L}_{\mathrm{grad,img}}$ at their spectral channels and to those of the complete ground-truth cube across all channels $\mathcal{L}_{\mathrm{grad,full}}$. Separate gradient losses were used to drive the model to use the region information within the structure images, while also prioritizing recreating the gradients of the underlying hyperspectral cube. The total-variation term promotes spatially smooth reconstructions while retaining sharp boundaries. Loss definitions and weights are provided in \S~\ref{subsec::loss_func} and Table~\ref{tab:hparams}; the training schedule is listed in Table~\ref{tab:si_cacnp_training}.

\subsection{Uncertainty Calibration}
\label{subsec::calibration}

Predictive standard deviations were calibrated using a held-out subset of the measured context. The model was conditioned on 70\% of the context positions, and one scale factor was computed as the root-mean-square standardized residual over the remaining 30\% and all spectral channels. The model was then evaluated using the full context, and its standard deviations were multiplied by this factor. The predicted mean and model parameters were unchanged by the scaling. Held-out calibration of a pretrained reconstruction network has also been explored in conformal prediction \cite{angelopoulos2023conformal}, with a growing body of applications in image reconstruction \cite{yeQUTCCQuantileUncertainty2026,angelopoulos2022image,teneggi2023trust,adame2026image,ye2025learned}; 
the present procedure rescales Gaussian standard deviations rather than calibrating quantile predictions. The implementation and its limitations are described in \S~\ref{subsec::si_calibration}. Coverage denotes the fraction of voxels at spatial positions without measured context spectra satisfying $|y-\mu|\leq2\sigma$, evaluated across all spectral channels, including the supplied structure-image channels. The nominal Gaussian coverage is 95.45\%. This calibration assumes that the standardized residuals at calibration locations are representative of those at unobserved locations under full-context conditioning. While this may not necessarily guarantee spatially uniform calibration, the results in the following section will show the calibration brings the coverage close to the 95.45\% target at unobserved locations (see Table~\ref{tab:si_multi_struct_metrics} and Table~\ref{tab:si_single_struct_metrics}).

\subsection{Acquisition Strategies}
\label{subsec::acq}

Three acquisition strategies were compared. Random acquisition selects uniformly from unmeasured positions. Combined random-uncertainty acquisition selects positions randomly except for every tenth measurement, which is assigned to the position with the largest spectral-mean predictive standard deviation. Combined distance-uncertainty acquisition uses a weighted sum of spectral-mean standard deviation and distance to the nearest measured position, with both quantities normalized over unmeasured positions. The distance term encourages spatial coverage. Reconstructions are updated after each acquisition round using all context measured to that point.

Acquisition can be terminated at a prescribed measurement budget or when an uncertainty-based stopping criterion is met. The latter requires uncertainty estimates that remain reliable as the context distribution changes; its practical limitations are examined in \S~\ref{subsec:acq_strat}.

\section{Results}
\label{sec:results}

\subsection{Performance on Experimental Data}
\label{subsec:exp_per}

Reconstruction performance on the experimental datasets was evaluated using RMSE, spectral angle, and segmentation misclassification rate. RMSE quantifies errors in absorption amplitude. Spectral angle treats IR spectra as vectors and quantifies their relative angles, quantifying the agreement of peak positions and relative intensities. Misclassification rate, which quantifies the disagreement between a segmentation of the ground truth hyperspectral cubes performed with principal component analysis and $K$-means clustering, assesses whether the reconstructed spectra preserve the spatial regions of similar chemistry identified by clustering the ground-truth data. Together, these metrics evaluate the fidelity of the reconstructed signal and its utility for distinguishing chemical domains. Evaluation included measured context positions to assess whether the reconstructed cube reproduces the supplied observations as well as the unmeasured data. Apart from nearest-neighbor filling, the compared implementations do not enforce exact reproduction of context spectra, so agreement at measured positions is itself an aspect of reconstruction fidelity. Metric definitions and the evaluation conventions are detailed in \S~\ref{subsec::error_metrics}.

\begin{figure}[htpb!]
    \centering
    \includegraphics[width=\linewidth,height=0.85\textheight,keepaspectratio]{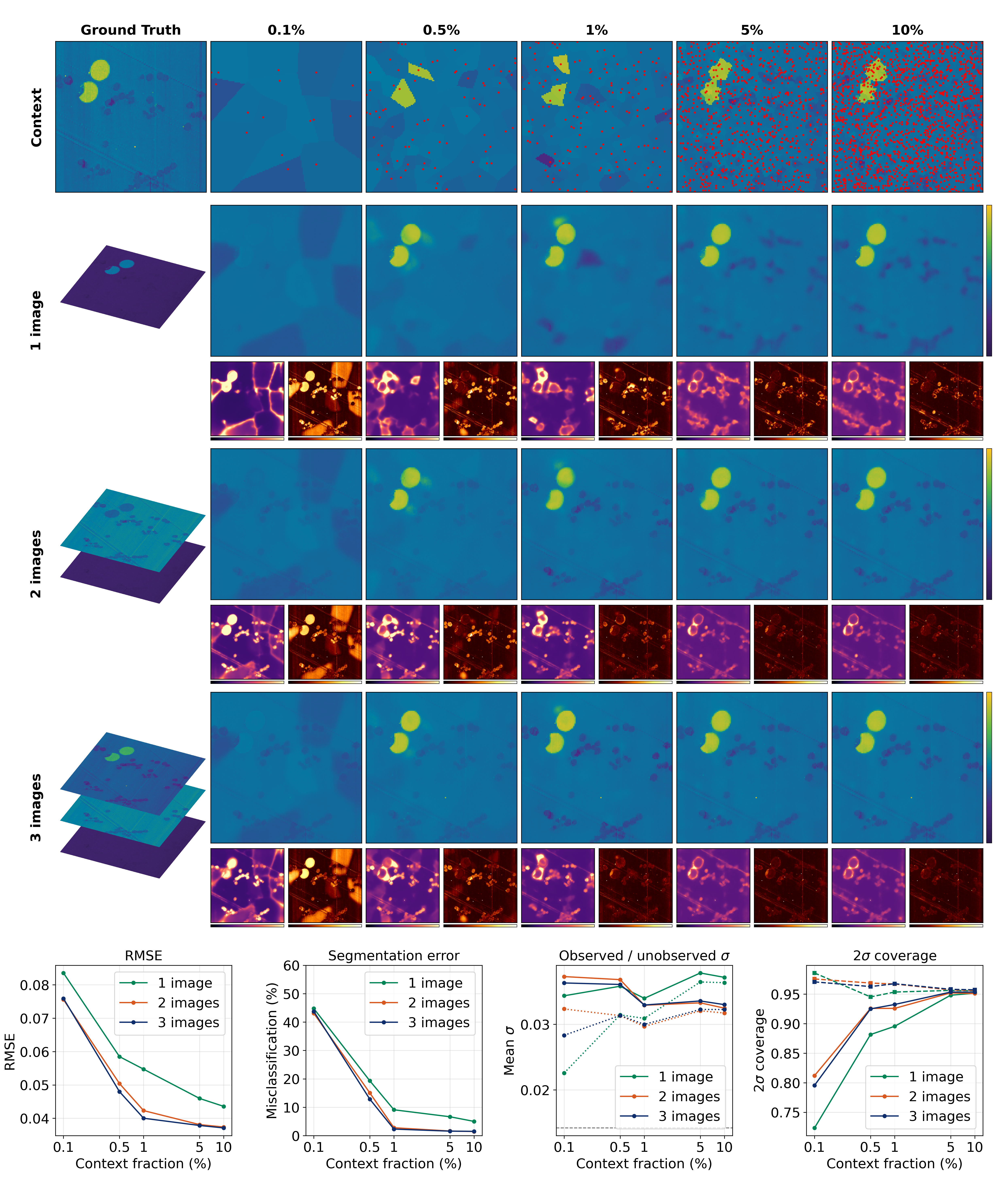}
    \caption{CACNP reconstructions of the multi-particle dataset using one to three structure images at context fractions from 0.1\% to 10.0\%, as well as corresponding plots of RMSE, segmentation error, $\sigma$, and $2\sigma$ coverage vs context fraction. Reconstructed slices $\mu$ at the \SI{1149}{cm^{-1}} energy are shown, with the pixel-average uncertainty $2\sigma$ and mean error $|y-\mu|$ below. The ground truth \SI{1149}{cm^{-1}} slice is shown in the top left, with nearest neighbor-filled reconstructions at the context points (shown in red) in the top row. The structure images are shown on the left of each reconstruction row. The one-image case uses \SI{1729}{cm^{-1}}, the two-image case uses \SI{1729}{cm^{-1}} and \SI{1049}{cm^{-1}}, and the three-image case additionally uses \SI{1189}{cm^{-1}}. Color bars are shown to the right of the reconstruction plots, and below the $2\sigma$ and $|y-\mu|$ plots, with ranges of (0.0,2.5), (0.0,0.2), and (0.0,0.2), respectively. Dashed/solid lines correspond to observed/unobserved $\sigma$, and raw/calibrated $2\sigma$ coverage.}
    \label{fig:recon_vs_img_multi}
\end{figure}

Fig.~\ref{fig:recon_vs_img_multi} shows slices of multi-particle reconstructions using one to three structure images and context fractions from 0.1\% to 10.0\%, together with the corresponding uncertainty and error maps shown below each reconstruction, and graphs of RMSE, segmentation error, observed and unobserved $\sigma$, and $2\sigma$ coverage.  As context fraction increases and structure images are added, reconstruction accuracy improves. Reconstruction error decreases rapidly as context is added, with improvements in RMSE and misclassification rapidly decreasing as context is increased from 0.1\% to 5\%. Improvements beyond 5\% are more modest, with the three image, 5\% context reconstruction having an RMSE of \SI{0.0375}{mV} and a misclassification rate of 1.57\%, compared with \SI{0.0369}{mV} and 1.37\% at 10\% context (Table~\ref{tab:si_multi_struct_metrics}). The variation is similar for the single-particle dataset, where the RMSE decreases from \SI{0.0200}{mV} at 5\% context to \SI{0.0198}{mV} at 10\% context in the three structure image case. These reconstruction errors are close to the experimental noise levels in each sample, which have estimated noise standard deviations, calculated using the \texttt{estimate\_sigma} function from \texttt{skimage.restoration} Python module, of \SI{0.018}{mV} and \SI{0.014}{mV} for the multi- and single-particle datasets, respectively. These values provide a reference noise scale for interpreting the reconstruction RMSE.

Structure images provide information about phase boundaries and regions of similar chemistry. To assess the impact of providing the CACNP with varying levels of spatial information, multiple structure images with different phase contrast were provided. The first image, at \SI{1729}{cm^{-1}}, separates PMMA from epoxy but provides little contrast between PS and epoxy. Consequently, the resulting reconstruction exhibits sharp boundaries between the PMMA (bright) and epoxy (light blue), but blurry boundaries between PS (dark blue) and the epoxy (light blue). Adding the \SI{1049}{cm^{-1}} image, which contrasts both PMMA and PS from the epoxy substrate, clarifies these boundaries, while the third image at \SI{1189}{cm^{-1}} containing contrast between all three phases provides a smaller improvement. For an experiment using only one structure image, \SI{1189}{cm^{-1}} would be a more informative choice for this sample. While the CACNP is able to approximate the morphology of the PS particles in the single image case, the model performance is significantly improved by having all interphase boundaries represented in the structure images, evidenced by the qualitatively sharper particles and the improved RMSE and misclassification metrics in Fig.~\ref{fig:recon_vs_img_multi}. Reconstructions do not improve significantly by adding more structure images in the single-particle dataset, whose two main phases, PMMA and epoxy, are accurately captured by a single image, as can be seen in Fig.~\ref{fig:si_recon_vs_img_single}.

The uncertainty $\sigma$ maps broadly follow the error $|y-\mu|$ maps in Fig.~\ref{fig:recon_vs_img_multi} and Fig.~\ref{fig:si_recon_vs_img_single}, indicating that the model has learned a spatially meaningful uncertainty metric, although uncertainty is underestimated at low context fractions. At lower context fractions, the CACNP has to ascertain the signal noise and spectral diversity from a limited subset of data, so the model's overconfidence is unsurprising. As more context is provided, the uncertainty initially increases. For the multi-particle dataset, the CACNP reaches a maximum $\sigma$ at 1\%, and then decreases as more context is added. Additional structure information improves the models uncertainty estimates: for the multi-particle dataset with 0.5\% context, raw $2\sigma$ coverage increases from 86.3\% with one structure image to 91.9\% with three. With three images, coverage reaches 95.0\% at 1\% context and remains close to the nominal 95.4\% at higher fractions.

Calibration brings low-context coverage closer to the nominal value in the multi-particle dataset, with $2\sigma$ coverage fractions ranging from 95.3\% to 97.1\% for 0.5\% to 10.0\% context, close to the target 95.4\% of a normal distribution. Raw and calibrated coverages are tabulated in Table~\ref{tab:si_multi_struct_metrics} and Table~\ref{tab:si_single_struct_metrics} , and visualized against absolute error in Fig.~\ref{fig:si_multi_sigma_cov} and  Fig.~\ref{fig:si_single_sigma_cov} for the multi- and single-particle datasets, respectively. 

\begin{figure}[htpb!]
    \centering
    \includegraphics[width=\linewidth,height=0.78\textheight,keepaspectratio]{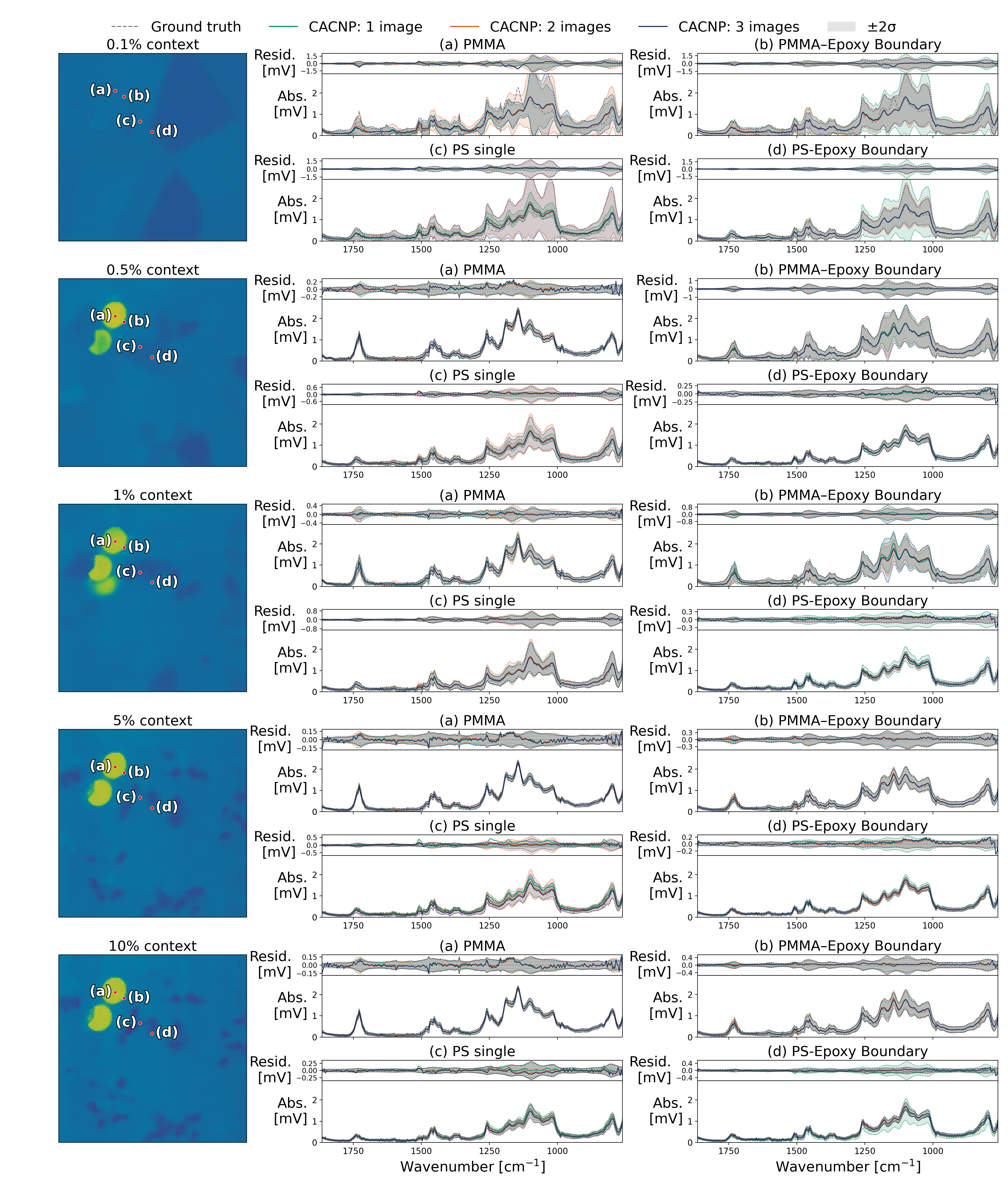}
    \caption{Spectral comparison of CACNP reconstructions of the multi-particle dataset using one to three structure images at context fractions of 0.1\%, 0.5\%, 1.0\%, 5.0\%, and 10.0\% (top to bottom). The left column shows reconstructed slices at \SI{1149}{cm^{-1}}, with four locations marked: (a) PMMA, (b) a PMMA--epoxy boundary, (c) an isolated PS particle, and (d) PS-epoxy boundary. The corresponding spectral panels compare ground truth (gray dashed lines) with CACNP predictions using increasing numbers of structure images. Colored solid lines indicate the predicted mean $\mu$, while shaded regions and dashed bounds indicate $\mu \pm 2\sigma$ using calibrated standard deviations. Residual strips above each spectrum show the signed error $\mu-y$, with uncertainty bounds of $\pm 2\sigma$ about zero.}
    \label{fig:spectra_vs_img}
\end{figure}

Fig.~\ref{fig:spectra_vs_img} compares reconstructed spectra at particle interiors (positions a,c) and boundaries (positions b,d) for the multi-particle dataset. The principal spectral features are recovered, while uncertainty is concentrated around peaks with variable intensity, including the PS bands near 1500 and \SI{1600}{cm^{-1}}. Boundary spectra have wider uncertainty intervals than spectra from particle interiors, as can be seen in the PMMA-epoxy and PS-epoxy boundary spectra (positions b,d) in Fig.~\ref{fig:spectra_vs_img}, and these intervals generally narrow as context is added. Information from the structure images has a clear impact on uncertainty, as can be seen when analyzing the PS-epoxy boundary (position d) in Fig.~\ref{fig:spectra_vs_img}, which has significantly higher predicted uncertainties for the single image reconstruction (green) than the two (orange) and three (blue) image reconstructions, highlighting the benefit of interphase contrast within structure images. 

Similar results are observed for the single-particle dataset in Fig.~\ref{fig:si_spectra_vs_img_single}. A standout result in Fig.~\ref{fig:si_spectra_vs_img_single} is seen in the model's ability to successfully reconstruct an outlier region of approximately $2 \times 2$ pixels, which contains a spectrum that is like epoxy in most of its peak locations except for two additional peaks between \SI{1550}{cm^{-1}} and \SI{1400}{cm^{-1}}. The model also correctly identifies with multiple structure images at low context fraction that the outlier region is of high uncertainty. Notably, it takes reaching context fractions of 5\% before the model starts to correctly reconstruct the peak amplitudes. This is likely due to the model requiring more sample statistics to determine that this region is a substantive outlier and not a result of acquisition noise.

\subsection{Comparison with Other Reconstruction Methods}
\label{subsec:method_comp}

\begin{figure}[htpb!]
    \centering
    \includegraphics[width=\linewidth]{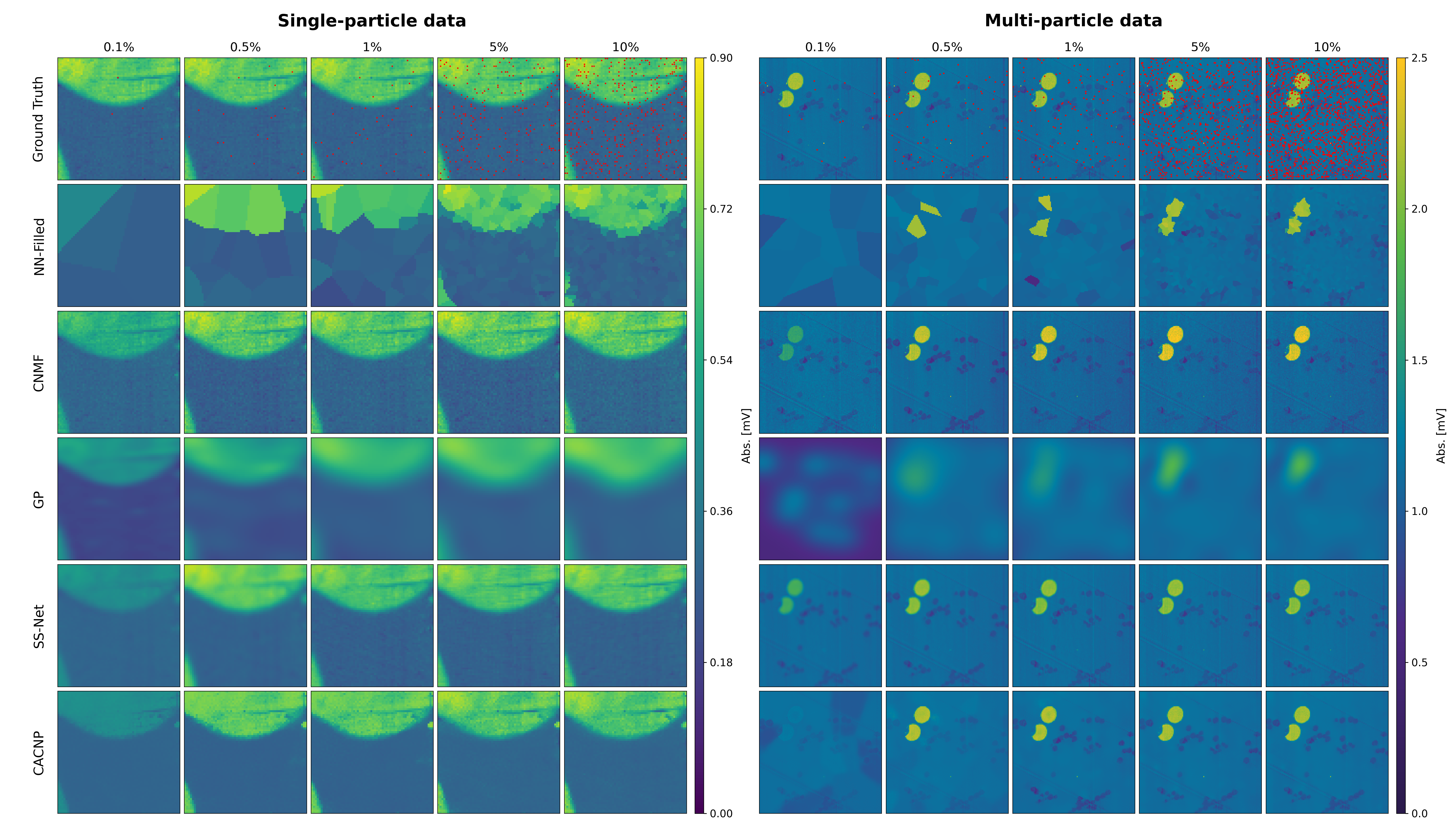}
    \caption{Comparisons of different reconstruction methods for the single- (left) and multi- (right) particle datasets. The \SI{1149}{cm^{-1}} slice for each dataset is shown, with nearest-neighbor filling, CNMF, GP, SS-Net, and CACNP reconstructions shown below. The context pixels, drawn randomly, are shown in the top as red dots. All single- (left) and multi- (right) particle images  are $1.5 \times \SI{1.5}{\micro\meter}$ ($64 \times 64$ pixels) and $20 \times \SI{20}{\micro\meter}$ ($128 \times 128$ pixels), respectively. The single- and multi-particle datasets were reconstructed using the \SI{1505}{cm^{-1}}, \SI{1141}{cm^{-1}}, \SI{1077}{cm^{-1}}, and the \SI{1729}{cm^{-1}}, \SI{1189}{cm^{-1}}, \SI{1049}{cm^{-1}} slices as structure images, respectively.}
    \label{fig:method_comp_imgs}
\end{figure}

In order to assess the performance of the CACNP relative to other hyperspectral reconstruction methods, the two experimental cubes were reconstructed using nearest-neighbor (NN) filling, coupled non-negative matrix factorization (CNMF) reconstruction \cite{borodinovApplicationPansharpeningAlgorithm2019}, GP regression, Spatial-Spectral Network (SS-Net) reconstruction \cite{gaoFastNanoIRHyperspectral2024}, and the CACNP using the same context observations. CNMF used a rank-three factorization, representing each reconstructed spectrum as a nonnegative linear combination of three learned spectral factors. The GP was fitted with a Mat\'ern 5/2 kernel and optimized with an Adam optimizer. Additional information on these methods, such as CNMF rank choice, SS-Net implementation, and GP optimization, is provided in \S~\ref{subsec::si_other_methods}. Fig.~\ref{fig:method_comp_imgs} provides a visual comparison of reconstructions using 3 structure images. The methods that accept structure images were supplied with the same images.

At 10\% context, NN filling reproduces measured spectra exactly but does not use structure image contrast to reconstruct other channels, producing jagged interphase boundaries (second row, Fig.~\ref{fig:method_comp_imgs}). The GP reconstruction (fourth row), recovers broad spectral variation but smooths sharp interfaces. A stationary kernel applies the same covariance rule throughout the field of view, which limits its ability to accommodate both sharp phase boundaries and smooth interiors. The sharper GP boundaries at low context do not persist as additional spectra constrain the fit across channels with weaker spatial variation. CNMF offers strong reconstruction performance, recovering the principal morphology and spectral variation, but producing noisy reconstructions and overshooting the PMMA (bright particle) and PS (dark particle) intensities. SS-Net closely reproduces the ground-truth morphology and intensity in these examples. The CACNP, like SS-Net and CNMF, performs well at reconstructing the boundaries and spatial intensity variations of the ground truth dataset. 

CNMF, SS-Net, and CACNP all recover sharp phase boundaries, but CNMF exhibits more fine-scale intensity variation, consistent with fitting spatial coefficients without an explicit spatial-smoothness penalty, whereas SS-Net and CACNP use spatial neighborhoods through their convolutional architectures. Fig.~\ref{fig:cnmf_ssnet_cacnp_img_comp} shows comparisons of \SI{1149}{cm^{-1}} slices of CNMF, SS-Net, and CACNP reconstructions of the single- and multi-particle datasets using 1 to 3 structure images. Reconstruction quality of the single-particle dataset does not change significantly with more or fewer structure images, as the phase contrast in the \SI{1149}{cm^{-1}} slice is well described by a single structure image. The reconstructions of the multi-particle dataset depend much more strongly on the supplied structure images, with all three methods improving in their boundary sharpness and intensity assignment as additional structure images are added. The contrast between the PS (dark) particles and background epoxy is weak in the first supplied structure image, resulting in poor reconstructions of the PS particles in all three methods. However, the CACNP method is able to approximately reconstruct the dark intensity of the PS particles with only a single structure image, albeit with blurry boundaries.

Fig.~\ref{fig:si_meth_spec_single} also reveals a limitation that is less apparent from the aggregate error metrics: CNMF reproduces the spectra of the principal phases more closely than the localized outlier spectrum at position (c). Even at 5--10\% context, the CNMF reconstruction does not reproduce the distinctive outlier spectral shape and overestimates other peak intensities, whereas CACNP captures more of the outlier features. This behavior is consistent with the restriction of all reconstructed spectra to combinations of three shared nonnegative spectral factors: a rare spectral component may be poorly represented by factors dominated by the principal phases. The present comparison does not isolate this explanation from the effects of context selection and factorization optimization.

Fig.~\ref{fig:si_line_recon} compares line profiles of CNMF, SS-Net, and CACNP reconstructions using 1, 2, and 3 structure images and 10\% context. The SS-Net (yellow) and CACNP (blue) reconstructions closely follow the ground truth (black) profiles, while the CNMF profiles exhibit noisier profiles and overpredict the intensity of the bright PMMA particles in the \SI{1149}{cm^{-1}} slice. The CNMF overprediction moderates as more structure images are used for reconstruction, but persists.

\begin{figure}[htpb!]
    \centering
    \includegraphics[width=\linewidth,height=0.78\textheight,keepaspectratio]{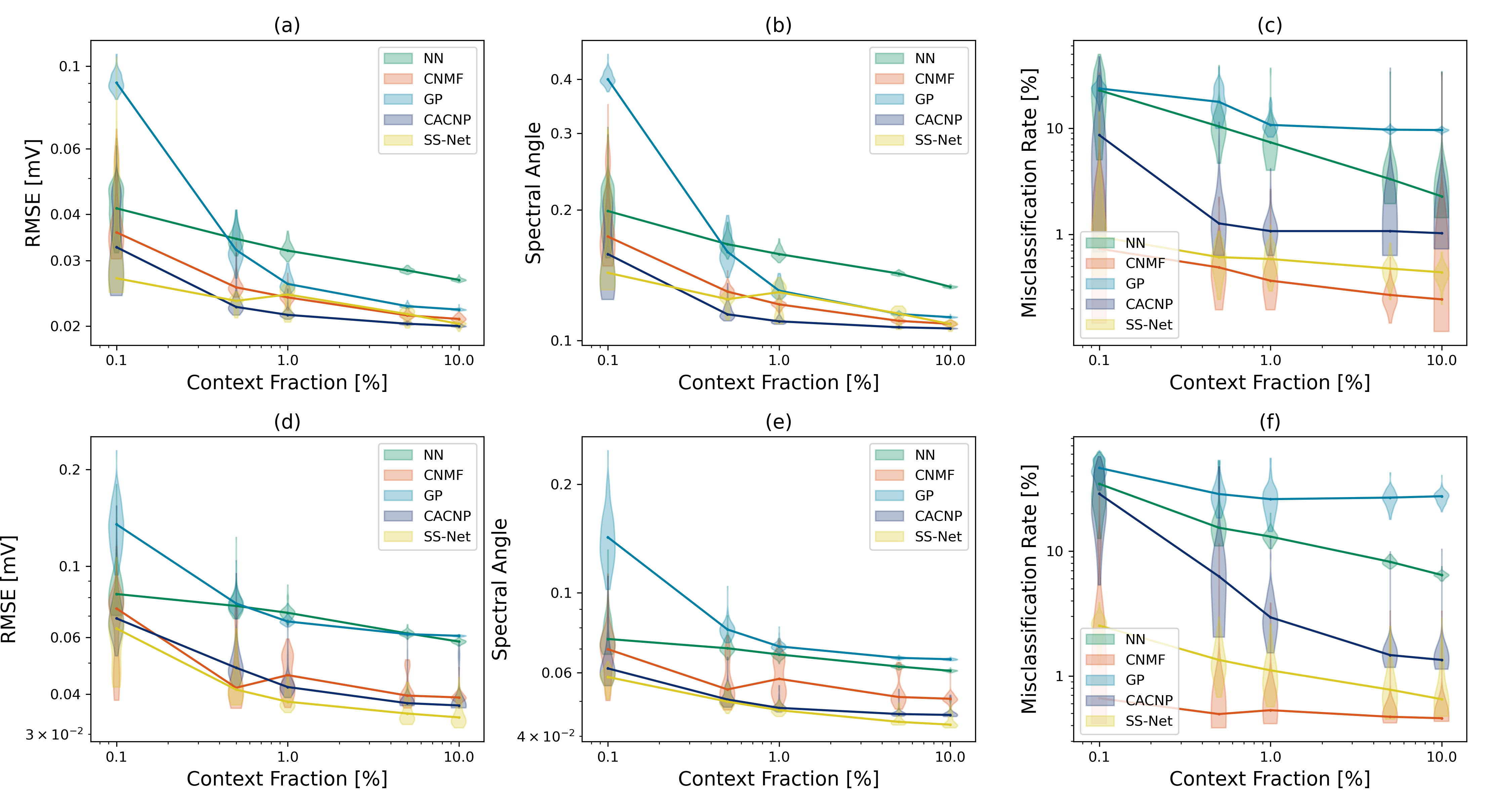}
    \caption{RMSE (a,d), spectral angle (b,e), and misclassification rate (c,f) for the  NN-filled, CNMF, GP, SS-Net, and CACNP reconstructions of the single- (a,b,c) and multi- (d,e,f) particle dataset. Each line plot shows the median value for the given method and context. Each context fraction contains 100 random context sets, with the same draws being used for each method. The distributions of the context draws are displayed as violin plots. The single- and multi-particle datasets were reconstructed using the \SI{1505}{cm^{-1}}, \SI{1141}{cm^{-1}}, \SI{1077}{cm^{-1}}, and the \SI{1729}{cm^{-1}}, \SI{1189}{cm^{-1}}, \SI{1049}{cm^{-1}} slices as structure images, respectively.}
    \label{fig:method_comp_metrics}
\end{figure}

Fig.~\ref{fig:method_comp_metrics} compares RMSE, spectral angle, and misclassification rate over 100 shared random context draws at each sampling fraction. For the single-particle dataset, SS-Net provides the lowest median RMSE and spectral angle at 0.1\% context, achieving \SI{0.0269}{mV} and 0.1433, compared with \SI{0.0326}{mV} and 0.1582 for CACNP and \SI{0.0358}{mV} and 0.1737 for CNMF. However, CACNP achieves lower median RMSE and spectral angle than SS-Net and CNMF from 0.5--10\% context. At 10\%, the three methods are closely matched: CACNP, SS-Net, and CNMF achieve RMSEs of \SI{0.0200}{mV}, \SI{0.0203}{mV}, and \SI{0.0209}{mV}, respectively, with corresponding spectral angles of 0.1067, 0.1092, and 0.1093. CNMF and SS-Net produce lower median misclassification rates than CACNP at every evaluated fraction. This advantage is substantial at 0.1\%, where their errors are 0.73\% and 0.95\%, compared with 8.63\% for CACNP, but becomes modest in absolute terms from 0.5\% context onward, with differences of less than one percentage point. At 10\%, misclassification rates are 0.24\% for CNMF, 0.44\% for SS-Net, and 1.03\% for CACNP. These segmentation results depend on the relatively simple PCA--$K$-means procedure; more sophisticated segmentation methods could improve outcomes for all reconstructions, although their effect on the relative rankings remains to be evaluated. Overall, the CACNP offers competitive reconstruction performance when compared with SS-Net and CNMF.

For the multi-particle dataset, SS-Net achieves the lowest median RMSE and spectral angle across all evaluated context fractions, while CNMF yields the lowest misclassification rates. At 10\% context, SS-Net, CACNP, and CNMF achieve RMSEs of \SI{0.0338}{mV}, \SI{0.0369}{mV}, and \SI{0.0391}{mV}, respectively, with corresponding spectral angles of 0.0431, 0.0458, and 0.0508. Their misclassification rates are 0.65\%, 1.34\%, and 0.46\%, respectively. Thus, CACNP provides competitive spectral reconstruction, although its segmentation disadvantage is more pronounced at low context. At higher context, the modest accuracy benefits of SS-Net should be weighed against its substantially greater computational cost and lack of predictive uncertainty estimates.

Tables~\ref{tab:si_time_errors_small} and \ref{tab:si_time_errors_big} report estimated cumulative acquisition and reconstruction times. These estimates include acquisition of three structure images at one line per second, requiring 64 or 128 seconds per image (assuming \SI{1}{Hz} line scan rate) for the single- and multi-particle datasets, respectively, and four seconds per acquired spectrum. Reconstruction times were averaged across repeated timing trials and interpolated between measured context sizes, then accumulated assuming a complete reconstruction after each newly acquired spectrum. For CNMF, GP, and SS-Net, each update includes model fitting and reconstruction; CACNP requires no retraining, but its timing includes uncertainty calibration. Acquisition and computation are assumed to occur sequentially. More information on timing estimation is provided in \S~\ref{subsec::si_timing}.

CNMF and GP refit their models at each update, while SS-Net trains a separate network; CACNP requires only forward inference and uncertainty calibration. For the single-particle dataset at 1\% context, cumulative times are approximately 6 minutes for both CACNP and CNMF, 46 minutes for GP, and 53 minutes for SS-Net. At 10\% context, these increase to approximately 31 minutes, 32 minutes, 7 hours 45 minutes, and 11 hours 25 minutes, respectively. For the multi-particle dataset at 1\% context, CACNP and CNMF both require approximately 18 minutes, compared with 3 hours 29 minutes for GP and 4 hours 45 minutes for SS-Net. At 10\%, the corresponding times are approximately 2 hours 1 minute for CACNP, 2 hours 2 minutes for CNMF, 33 hours 51 minutes for GP, and 51 hours 20 minutes for SS-Net. CACNP and CNMF therefore have similar total acquisition times under the assumed workflow, whereas repeated GP fitting and SS-Net training impose substantially larger cumulative computational costs. These totals describe the complete simulated acquisition sequence rather than the duration of a single reconstruction.

\begin{figure}
    \centering
    \includegraphics[width=1.0\linewidth]{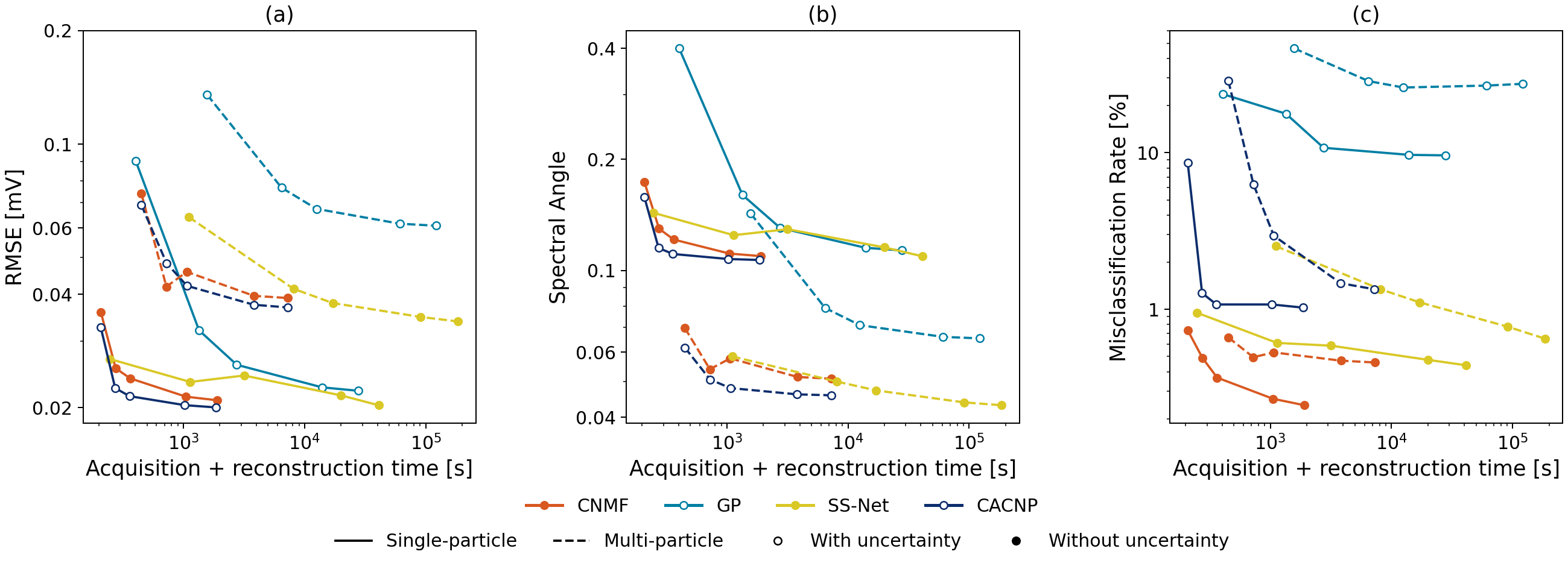}
    \caption{Median reconstruction errors versus estimated acquisition and reconstruction time for CNMF, GP, SS-Net, and CACNP on the single-particle (solid lines) and multi-particle (dashed lines) datasets. NN filling is not shown. Panels show (a) RMSE, (b) spectral angle, and (c) misclassification rate over 100 shared random context draws at fractions of 0.1\%, 0.5\%, 1\%, 5\%, and 10\%. Both axes use logarithmic scales.}
    \label{fig:err_vs_time}
\end{figure}

Fig.~\ref{fig:err_vs_time} plots median RMSE, spectral angle, and misclassification rate against estimated cumulative acquisition and reconstruction time, with lower errors toward the bottom and shorter times toward the left. The CACNP curves occupy the bottom-left region of the RMSE and spectral-angle panels, indicating a particularly favorable balance between reconstruction accuracy and acquisition-to-reconstruction time. CACNP achieves competitive spectral fidelity with substantially shorter cumulative times than GP and SS-Net and times similar to CNMF. CNMF and SS-Net provide lower misclassification rates in several comparisons, so CACNP does not dominate every metric. Nevertheless, its combination of low spectral errors, rapid updates, and calibrated voxel-wise uncertainty makes it well suited to acquisition workflows in which both reconstruction quality and measurement time matter.

\subsection{Acquisition Stopping Conditions}
\label{subsec:acq_stop}

\begin{figure}[htbp!]
    \centering
    \includegraphics[width=1.0\linewidth]{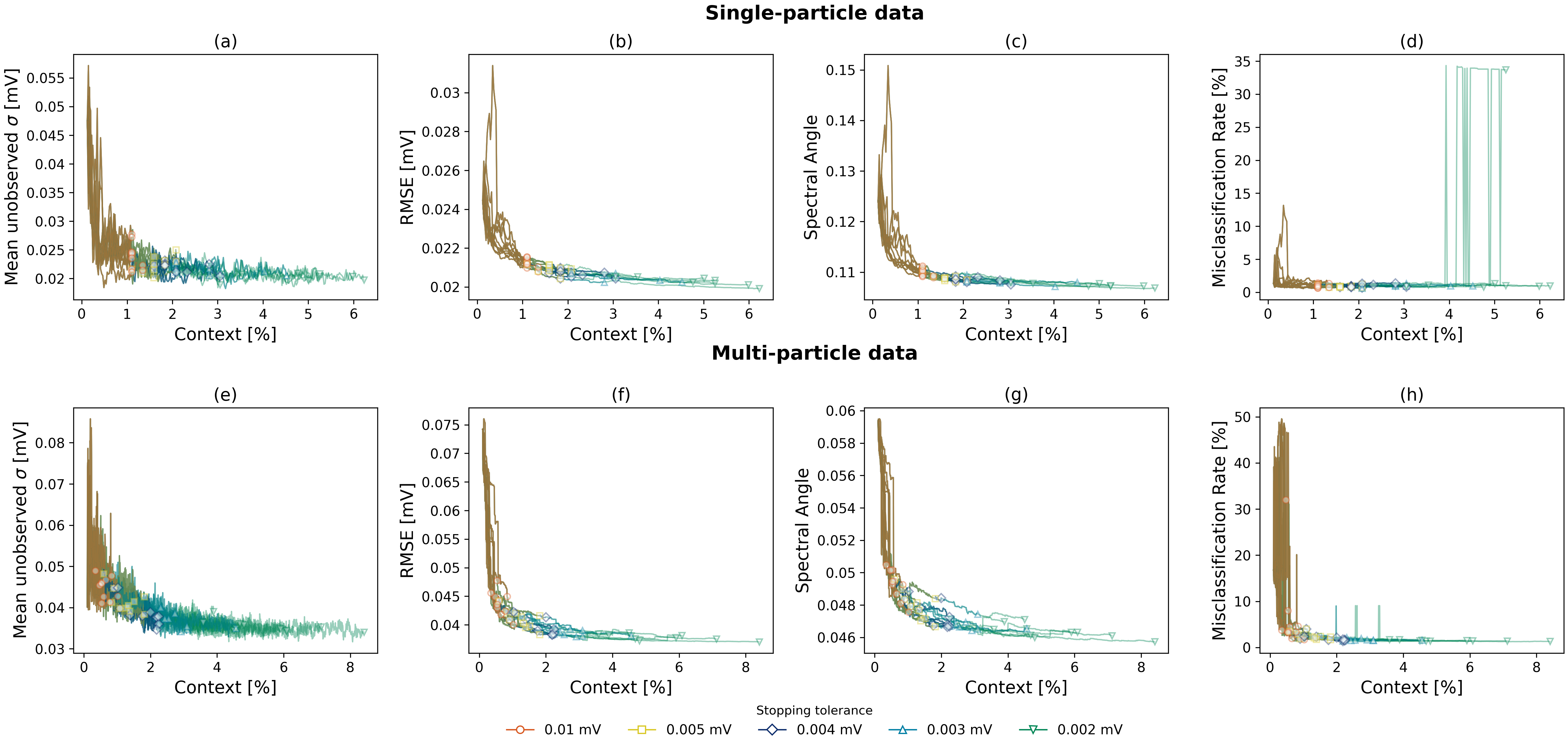}
    \caption{Mean unobserved $\sigma$ (a,e), RMSE (b,f), spectral angle (c,g), and misclassification rate (d,h) for 10 random acquisitions of the single- (a-d) and multi- (e-h) particle datasets reconstructed using the CACNP. Markers identify the first point at which the maximum-to-minimum variation in mean unobserved $\sigma$ remained below the specified tolerance for three consecutive windows of ten new acquisitions.}
    \label{fig:stopping}
\end{figure}

The predictive uncertainty provided by CACNP offers a means of terminating acquisition when additional measurements no longer substantially change the model's confidence, potentially reducing the number of spectra and the total acquisition-to-reconstruction time. To evaluate this approach, ten random acquisition trajectories were simulated for each experimental dataset, and stopping points were identified using the calibrated mean uncertainty at unobserved positions. Hypothetical stopping points were recorded on a common trajectory when the maximum-to-minimum variation in uncertainty remained below a specified threshold for three consecutive, non-overlapping windows of ten newly acquired spectra. Stability was used rather than an absolute uncertainty target because the uncertainty magnitude depends on the instrument's signal scale and the noise present in the measurements. Although the tolerances used here are also expressed in signal units and therefore require appropriate selection, they assess stabilization rather than requiring a universally applicable uncertainty value. Fig.~\ref{fig:stopping} compares the acquisition trajectories and their stopping points, while Fig.~\ref{fig:si_stopping} shows representative reconstructions for both datasets.

Stricter stopping tolerances generally required more measurements and yielded lower RMSE and spectral angle, with diminishing improvements as acquisition progressed. The strictest tolerance, \SI{0.002}{mV}, stopped at approximately 2.8--6.2\% context for the single-particle dataset and 3.6--8.4\% for the multi-particle dataset, whereas \SI{0.003}{mV} typically stopped earlier, at approximately 1.8--4.6\% across the two datasets. A tolerance of \SI{0.004}{mV} provided a less demanding compromise, producing useful reconstructions at approximately 0.9--3.1\% context. Occasional large jumps in misclassification rate, particularly for the single-particle dataset, were not accompanied by comparable changes in RMSE or spectral angle. This discrepancy suggests sensitivity of the relatively simple PCA--$k$-means segmentation procedure, rather than a corresponding abrupt deterioration in spectral reconstruction; more robust segmentation methods could help assess and mitigate this sensitivity. These results support uncertainty stabilization as a practical stopping criterion for balancing measurement effort and reconstruction quality. Uncertainty stabilization remains a heuristic: it does not guarantee a specified reconstruction error or detection of every localized spectral feature.

\subsection{Impact of Acquisition Strategies}
\label{subsec:acq_strat}

\begin{figure}[htpb!]
    \centering
    \includegraphics[width=\linewidth,height=0.78\textheight,keepaspectratio]{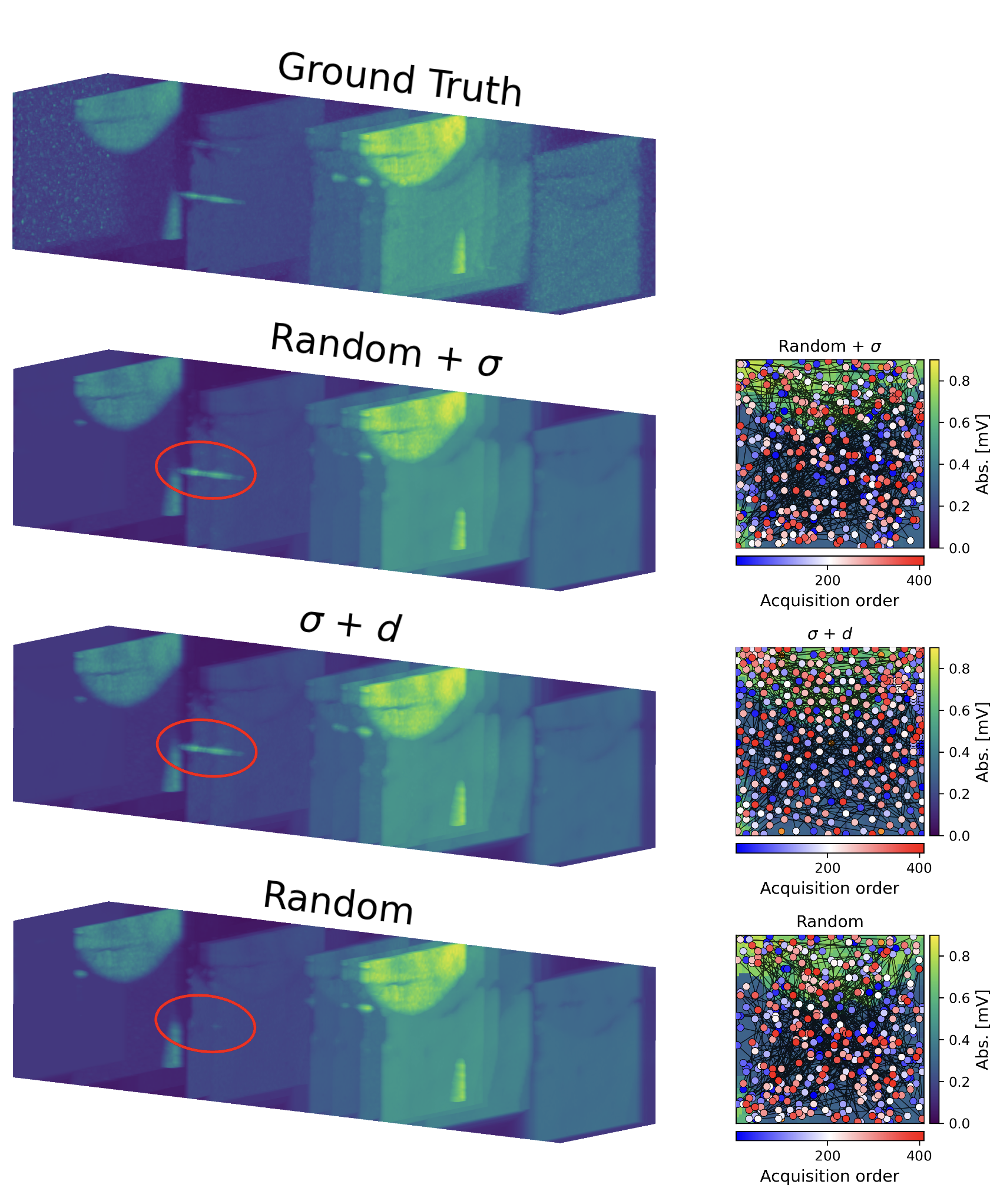}
    \caption{3D renderings of CACNP reconstructions and corresponding sample points for simulated acquisitions using combined random-uncertainty (Random + $\sigma$), combined distance-uncertainty ($d$ + $\sigma$), and random acquisition function of the single-particle dataset. The sampled points are displayed over their NN-filled images, color-coded according to their acquisition order.}
    \label{fig:acq_rend}
\end{figure}

The CACNP's predictive uncertainty presents the option to prioritize sampling where the model is least confident, improving reconstruction quality and revealing subtle or unexpected spectral features. To assess this potential, the three acquisition strategies described in \S~\ref{subsec::acq} were compared: uniformly random selection, random selection with every tenth measurement directed to the highest-uncertainty position (random+$\sigma$), and selection using a weighted combination of uncertainty and distance from existing measurements ($\sigma$+$d$). Fig.~\ref{fig:acq_rend} shows three-dimensional renderings of the ground-truth and reconstructed single-particle hyperspectral cubes, together with the measured positions colored by acquisition order. Both uncertainty-guided strategies concentrate measurements around the anomalous region in the upper left of the context maps and recover its distinctive spectral feature (shown in red). Random acquisition samples this region insufficiently, and the corresponding feature is not accurately reconstructed. These results demonstrate that CACNP uncertainty can guide the detection and characterization of localized spectral features that sparse random sampling may miss. Fig.~\ref{fig:si_strat_comp_single} compares RMSE, spectral angle, misclassification rate, and uncertainty coverage throughout acquisition. Although the strategies converge at different rates, their RMSEs approach similar values by 10\% context, and calibrated $2\sigma$ coverage approaches the nominal 95.4\% target.

However, directing measurements toward uncertain regions does not consistently improve reconstruction quality. In the multi-particle dataset, the uncertainty-guided strategies erroneously propagate the intensity of an outlier near the bottom center of the image into surrounding regions (see Fig.~\ref{fig:si_acq_img_comp}). This artifact is more pronounced for $\sigma$+$d$ acquisition (Fig.~\ref{fig:si_acq_img_comp}c) than for random+$\sigma$ acquisition (Fig.~\ref{fig:si_acq_img_comp}b), likely due to the $\sigma$+$d$ method concentrating more samples on high $\sigma$ pixels, and is absent from the random reconstruction (Fig.~\ref{fig:si_acq_img_comp}d). The corresponding acquisition metrics in Fig.~\ref{fig:si_strat_comp_multi} show that RMSE and spectral angle do not converge to common values over the evaluated context range. Raw $2\sigma$ coverage for $\sigma$+$d$ acquisition remains approximately 90--93\%, while calibrated coverage fluctuates below the nominal 95.4\% target. High predictive uncertainty may therefore identify noisy or anomalous observations rather than measurements that most improve the reconstruction. Concentrating measurements in these regions can amplify their influence while leaving other parts of the sample insufficiently characterized. An alternative explanation is the CACNP training procedure: each training task, the model was shown uniformly random context positions, but in the case of adaptive acquisition, the model selects points concentrated mostly on interphase boundaries. These boundary-concentrated points sharpen the boundary reconstruction, but also provide the model with less context in regions of uniform chemistry, and thus less information on the true noise and pure phase spectra of the sample. These results illustrate that the utility of predictive uncertainty for acquisition depends on both the sample and the selection strategy, consistent with active-learning benchmarks in which random sampling can outperform uncertainty-based selection \cite{scheinActiveLearningLogistic2007,yangBenchmarkComparisonActive2018,holzmullerFrameworkBenchmarkDeep2023}.

Adaptive acquisition therefore offers a tradeoff: localized spectral features can be recovered more effectively, but sampling concentrated around these features can increase errors elsewhere. More sophisticated acquisition strategies may address this limitation, but their development is beyond the scope of this work.

\section{Discussion}
\label{sec:discussion}

The CACNP combines accurate hyperspectral reconstruction, rapid inference, and predictive uncertainty in a single model. For the experimental datasets examined here, reconstructions at 5\% context closely reproduce the measured cubes using 20$\times$ fewer point spectra than a full reconstruction. Including structure-image acquisition, point-spectrum acquisition, and repeated reconstruction and uncertainty calibration, the estimated acquisition times decrease from approximately 5 hours to 17 minutes for the single-particle dataset and from 20 hours to 64 minutes for the multi-particle dataset, excluding additional instrument overhead. Once trained, the model predicts the reconstructed mean and uncertainty without fitting a new model for each context set, and its uncertainty estimates can be calibrated using the held-out procedure described above. This computational efficiency enables reconstruction to be updated during the experiment and allows convergence of the calibrated uncertainty to inform when acquisition can stop.

The model also reduces the need to prescribe the spatial and spectral correlations governing reconstruction. Conventional GP regression expresses these assumptions through a selected covariance function, whereas the CACNP learns relationships between sparse observations and complete cubes across synthetic training tasks constructed from experimental spectra. Its prior therefore reflects the spectral corpus, synthetic morphologies, and measurement artifacts represented during training. Conditioning on up to four structure images further allows the model to combine complementary sources of chemical contrast. Boundaries that are poorly resolved at one wavenumber can become apparent at another, providing the U-Net with spatial information that supports reconstruction across the spectral range. The structure-image comparisons demonstrate that the relevance of this contrast matters more than simply increasing the number of images.

A central architectural feature is cross-attention between the reconstruction features and the measured context spectra. In the Attention U-Net of Oktay et al., attention gates use internal feature maps to suppress irrelevant spatial features passed through skip connections \cite{oktayAttentionUNetLearning2018}. In the CACNP, the measured spectra instead provide an external source of keys and values that can be queried by features within the U-Net and by the preliminary reconstructed spectra. This supplies a direct route for measured spectral information to influence reconstruction, complementing the spatial relationships captured by convolution. External conditioning of U-Net features through cross-attention has a precedent in latent diffusion models \cite{rombachHighResolutionImageSynthesis2022a}. Here, the conditioning information consists of experimentally acquired spectra, and the output is obtained through direct reconstruction rather than iterative generation.

The combination of sparse spectral observations, structure images, and predictive uncertainty also suggests a flexible acquisition workflow. An initial set of spectra could be used to select informative imaging wavenumbers through significant wavenumber analysis (SWA), as demonstrated by Gao et al. \cite{gaoFastNanoIRHyperspectral2024}, or through criteria tailored to the expected chemistry. Subsequent uncertainty maps could guide additional point measurements and help identify when the existing structure images provide insufficient contrast. A localized increase in uncertainty following the observation of an unexpected spectral feature could motivate acquisition of another structure image at a diagnostic wavenumber, within the model's four-image capacity. Such increases would require interpretation alongside the measured spectra, since uncertainty can also reflect noise or reconstruction failure and does not uniquely identify a new chemical feature.

Several limitations remain. Rapid inference is enabled by substantial pretraining, and the current architecture is tied to a fixed spectral grid. Applying it to a different wavenumber grid requires a model trained or adapted for that grid, rather than simply changing the channel labels. The CACNP performs well at reconstructing the PS-PMMA-epoxy experimental datasets in this work, despite the experimental spectra not being present in the training corpus. Nonetheless, the CACNP is a pretrained deep learning method, and thus poor out-of-distribution performance is a potential concern. For applications with a narrower chemical scope, a broadly pretrained model could be fine-tuned using a targeted corpus containing representative experimental spectra. This would allow the learned prior to be adapted to the relevant chemistry, while requiring renewed evaluation of reconstruction accuracy and uncertainty calibration.

The propagation of outlier intensities presents a further challenge for adaptive acquisition. High uncertainty can direct measurements toward anomalous regions whose influence then extends incorrectly into the surrounding reconstruction. More careful handling of measurement artifacts and training with contexts resembling those encountered during acquisition could help reduce this behavior. One approach would use a pretrained model to simulate adaptive acquisition on synthetic cubes and then include the resulting context sets during further model training. Reinforcement learning could additionally be used to optimize acquisition policies against reconstruction improvement and measurement cost. These approaches address related but distinct limitations: the distribution of context seen by the CACNP and the criterion used to select subsequent measurements.

More broadly, meta-learning from related spectral tasks offers a route to materials-science applications in which useful priors are difficult to specify analytically. Neural processes have already been used to learn spectral representations for autonomous phase mapping of gold-nanoparticle synthesis \cite{vaddiAutonomousPhaseMapping2025}. This provides a precedent for extending learned spectral priors and predictive uncertainty to exploration of composition space, including the construction of phase maps. With suitable training data and modality-specific adaptations, the present reconstruction framework could also be extended to electron energy-loss spectroscopy and spectroscopic SPM measurements, including bias-dependent Kelvin probe force microscopy. Fast updates from sparse measurements could then support larger fields of view and autonomous experiments connecting spatial structure, chemistry, and functional properties.

\section{Conclusions}
\label{sec:conclusions}

A CACNP trained entirely on synthetic hyperspectral cubes reconstructs experimental AFM-IR data from sparse context spectra and a small number of structure images. The model performs well at low context, achieves competitive accuracy when compared to CNMF and SS-Net, and returns a reconstructed cube and voxelwise uncertainty in less than \SI{0.1}{s} for the datasets examined. Held-out context calibration improves uncertainty coverage, particularly when raw predictions are overconfident. Uncertainty-driven acquisition resolves localized spectral features but can increase global error by concentrating measurements near outliers.

These results support faster AFM-IR characterization and uncertainty-aware acquisition in SPM. Improving acquisition policies and adapting the learned spectral prior to specific experiments could extend the approach to larger fields of view, other hyperspectral techniques, and autonomous exploration of materials composition--structure--property relationships.

\section*{Data and Code Availability}

The training codes, experimental hyperspectral PS-PMMA-Epoxy datasets, and trained models are available on GitHub: \href{https://github.com/wkent2/CACNP-AFM-IR-Reconstruction}{https://github.com/wkent2/CACNP-AFM-IR-Reconstruction}.

\section*{Acknowledgments}

This work made use of the SPID facility of Northwestern University’s NU\textit{ANCE} Center, which has received support from the IIN and Northwestern’s MRSEC program (NSF DMR-2308691). This research was supported in part through the computational resources by the Quest high performance computing facility at Northwestern University, jointly supported by the Office of the Provost, Office for Research, and Northwestern University Information Technology. During the preparation of this study and manuscript, the authors used ChatGPT-6 Astra for the purposes of coding assistance and manuscript editing. The authors have reviewed and edited the manuscript and codes, and take full responsibility for the content of this publication.

\section*{Author Contributions}

W.F.K. proposed the concept, led the paper writing, acquired experimental data, coded and trained models, and performed formal analysis of results. E.A. advised on model architecture development, problem formulation, and assisted with writing. G.S. assisted with acquiring experimental data, proposed the concept, assisted with paper writing, and supervised work. R.dR. assisted with paper writing and supervised work.  V.P.D. acquired funding, proposed concept, and assisted with paper writing.

\newpage 

\clearpage
\setcounter{section}{0}
\setcounter{figure}{0}
\setcounter{table}{0}
\setcounter{equation}{0}
\renewcommand{\thesection}{S\arabic{section}}
\renewcommand{\thesubsection}{\thesection.\arabic{subsection}}
\renewcommand{\thefigure}{S\arabic{figure}}
\renewcommand{\thetable}{S\arabic{table}}
\renewcommand{\theequation}{S\arabic{equation}}
\renewcommand{\theHsection}{S\arabic{section}}
\renewcommand{\theHfigure}{S\arabic{figure}}
\renewcommand{\theHtable}{S\arabic{table}}
\renewcommand{\theHequation}{S\arabic{equation}}

\begin{center}
{\LARGE \textbf{Supplementary Information: Uncertainty Aware AFM-IR Hyperspectral Cube Reconstruction via Convolutional-Attentive Conditional Neural Processes}}
\end{center}

\section{Introduction}
\label{sec::si_intro}

The main paper introduced a CACNP trained on synthetic hyperspectral cubes for reconstruction and adaptive acquisition of experimental AFM-IR data. It compared the effects of structure images, reconstruction methods, and acquisition strategies on accuracy and predictive uncertainty. This Supplementary Information provides the methodological details and additional experimental results supporting those comparisons.

\S~\ref{sec::si_methods} provides additional details on the spectral corpus (\S~\ref{subsec::ir_corpus}), loss function (\S~\ref{subsec::loss_func}), synthetic cube generation and context selection (\S~\ref{subsec::si_synthetic_generation}), experimental datasets (\S~\ref{subsec::si_exp_data}), CACNP architecture and training configuration (\S~\ref{subsec::si_cacnp_architecture}), and uncertainty calibration (\S~\ref{subsec::si_calibration}). \S~\ref{si_sec::exp_res} defines the reconstruction error metrics (\S~\ref{subsec::error_metrics}), evaluates the cross-attention ablation (\S~\ref{subsec::si_attn_ablation}) and the influence of structure images (\S~\ref{subsec::si_struct_img}), and describes the comparison methods (\S~\ref{subsec::si_other_methods}) and their reconstruction results (\S~\ref{subsec::si_method_comp}). It also explains the estimation of cumulative acquisition and reconstruction times (\S~\ref{subsec::si_timing}), the uncertainty-based stopping criterion (\S~\ref{subsec::si_stopping}), and the effects of acquisition strategy (\S~\ref{subsec::si_acq_strat}).

\section{Methods}
\label{sec::si_methods}

The methods introduced in \S~\ref{sec::methods} are expanded here. \S~\ref{subsec::ir_corpus} describes corpus compilation and spectral diversity; \S~\ref{subsec::loss_func} defines the training objective; and \S~\ref{subsec::si_synthetic_generation} details synthetic generation and context selection. \S~\ref{subsec::si_exp_data} describes the experimental fields of view, \S~\ref{subsec::si_cacnp_architecture} specifies the network and training configuration, and \S~\ref{subsec::si_calibration} defines the held-out uncertainty scaling procedure.

\subsection{Experimental IR Corpus}
\label{subsec::ir_corpus}

The corpus of experimental IR spectra used for generating synthetic hyperspectral cubes was introduced and discussed in \S~\ref{subsec::corpus}. Here we expand more on the origin of the spectra, how they were compiled, and the distributions of chemical information. 

As mentioned in the main document, the spectral corpus is composed of spectra downloaded from the NIST Chemistry Web Book. Spectra were obtained from the five collections shown in Table~\ref{tab:nist_collections}, with the majority of spectra coming from the Evaluated Infrared Reference Spectra \cite{nist_coblentz_2026} and NIST/EPA Gas-Phase Infrared Database \cite{nist_infrared_2026}. Infrared spectra were downloaded from the NIST Chemistry WebBook using the \texttt{nistchempy} Python package \cite{nistchempy}, with \texttt{requests} used as a fallback for compounds whose spectra \texttt{nistchempy}'s HTML-based discovery failed to locate. A list of candidate compounds was obtained from an IR-flagged compound index bundled within \texttt{nistchempy}, and each compound's spectra were then retrieved from NIST's JCAMP-DX data endpoint. When more than one IR spectrum was available for a given compound, the spectrum whose native wavenumber range covered the largest fraction of the target grid was selected. If multiple phases were available, condensed phases were selected. 

Each spectrum was linearly interpolated onto a common wavenumber grid spanning 761 to \SI{1865}{cm^{-1}} in \SI{4}{cm^{-1}} increments. Grid points falling outside the native wavenumber range of the source spectrum were assigned an absorbance of zero. Absorbance values were min-max normalized to span $[0,1]$. Downloaded spectra that were in units of transmittance were converted to absorbance spectra using the Beer-Lambert conversion \[A = -\log_{10}(T)\] where $A$ is absorbance and $T$ is transmittance.

\begin{table}[htbp]
    \centering
    \caption{Number of infrared spectra used from each NIST Chemistry WebBook collection.}
    \label{tab:nist_collections}
    \begin{tabular}{lr}
        \hline
        \textbf{NIST IR Collection} & \textbf{Number of Spectra} \\
        \hline
        Evaluated Infrared Reference Spectra (Coblentz) \cite{nist_coblentz_2026} & 7,886 \\
        NIST/EPA Gas-Phase Infrared Database \cite{nist_infrared_2026} & 7,928 \\
        Quantitative Infrared Database \cite{nist_quantitative_ir_2026} & 11 \\
        IARPA/PNNL Liquid Phase IR Spectra \cite{nist_iarpa_pnnl_liquid_2026} & 3 \\
        IARPA/PNNL Solid Phase IR Spectra \cite{nist_iarpa_pnnl_solid_2026} & 1 \\
        \hline
        \textbf{Total} & \textbf{15,829} \\
        \hline
    \end{tabular}
\end{table}

As discussed in \S~\ref{subsec::corpus}, the corpus spectra were obtained using many sample preparation methods, yielding gas, liquid, and solid phase spectra. The number of spectra originating from each phase is shown in Table~\ref{tab:phase_dist}. The corpus is split approximately evenly between gas (51.1\%) and condensed phases (46.7\%), with 2.3\% of spectra having unknown phase. Manual searches of unknown phase spectra indicated that these data originate from digitized hard copies of spectra acquired during or prior to the 1970s. 

\begin{table}[htbp]
    \centering
    \caption{Number of infrared spectra from each phase. }
    \label{tab:phase_dist}
    \begin{tabular}{llr}
        \hline
        \textbf{Phase} & Count &\textbf{\% of Corpus} \\
        \hline
        Gas & 8,081 (gas: 8,045; vapor: 36) & 51.1\% \\
        Solid  & 4,573 (kbr: 2083; solid: 1,418; nujol: 1009; pellet: 57;  & 28.9\% \\
        & film: 6; powder: 4; mull: 1; melt: 1) & \\
        Liquid & 2,815 (solution: 1,738; liquid: 1,063; neat: 13; visc.  paste: 1) & 17.8\% \\
        Unknown & 360 & 2.3\% \\
        \hline
        \textbf{Total} & \textbf{15,829} & 100.0\%\\
        \hline
    \end{tabular}
\end{table}

To ascertain the distribution of IR absorption peaks within the training data, the training corpus was analyzed using \texttt{find\_peaks} function from the \texttt{scipy.signal} Python library \cite{2020SciPy-NMeth}. A candidate point was counted as a peak only if it satisfied three criteria simultaneously: a minimum prominence of 0.03 (when normalized to a $[0,1]$ scale), a minimum absolute height of 0.05, and a minimum spacing of 2 band indices from any other detected peak in the same spectrum. These parameters were chosen to avoid counting spurious noise as peaks. Every peak's wavenumber was then pooled across the entire corpus and binned into \SI{8}{cm^{-1}}-wide histogram bins spanning the full wavenumber axis, giving both a raw peak count per bin and a coverage fraction. 

\begin{figure}[htpb!]
    \centering
    \includegraphics[width=\linewidth,height=0.78\textheight,keepaspectratio]{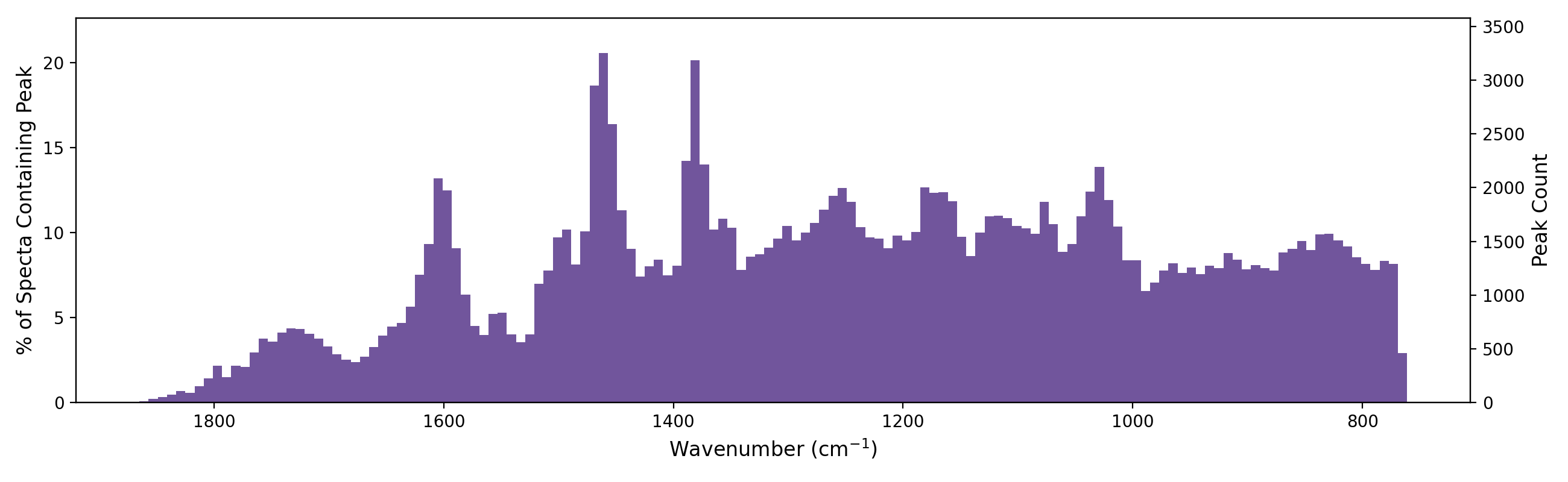}
    \caption{Distribution of peak locations within experimental IR corpus. Distribution was computed from all 15,829 experimental spectra post interpolation. Histogram bin widths are \SI{8}{cm^{-1}} wide.}
    \label{fig:peak_dists}
\end{figure}

The distribution of peak locations is shown in Fig.~\ref{fig:peak_dists}. This distribution clearly shows peak locations spanning the entire spectral range of 761 to \SI{1865}{cm^{-1}}, ensuring that the model saw spectra with peaks across the entire spectral range. The distribution is also clearly non-uniform, reflecting the distributions of common functional groups observed in organic compounds. When deploying the CACNP to reconstruct AFM-IR samples, users should first obtain several sample spectra before acquisition to ensure that their experimental peaks are represented in the training data. 

\begin{figure}[htpb!]
    \centering
    \includegraphics[width=\linewidth,height=0.78\textheight,keepaspectratio]{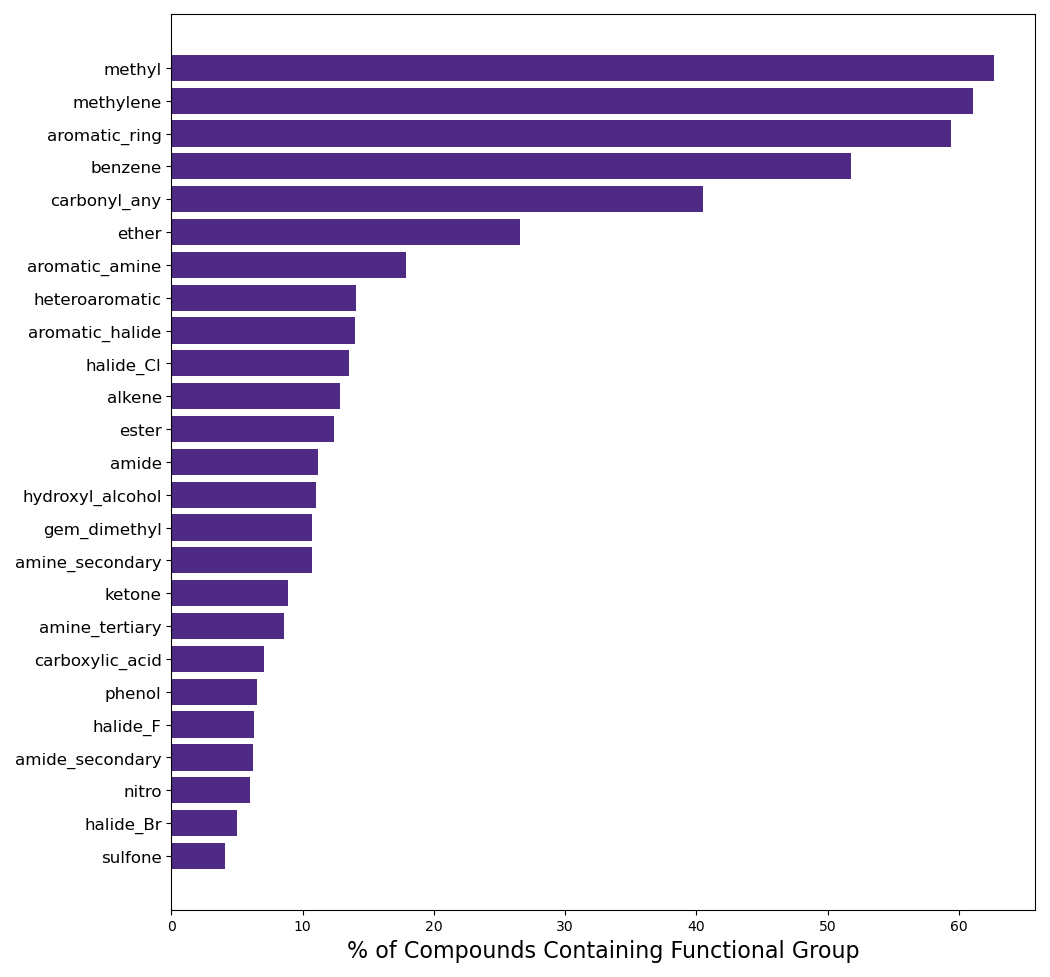}
    \caption{Percent of compounds in IR spectral corpus that contain each of the functional groups for the top 25 functional groups. Functional groups were identified by using the \texttt{rdkit.Chem} library \cite{rdkit}. }
    \label{fig:chem_dists}
\end{figure}

To obtain an understanding of the different functional groups present in the compounds composing the spectral corpus, a Python script was written to identify unique functional groups. A library of 94 functional groups was first constructed, each specified as a SMARTS pattern and compiled once into \texttt{rdkit} query objects. For every molecule in the input corpus, its SMILES string was parsed into an \texttt{rdkit} molecule, and each compiled SMARTS pattern was tested against it with a substructure match. Every pattern that matched incremented that group's presence count. The percentage of compounds containing the top 25 most common functional groups are shown in Fig.~\ref{fig:chem_dists}, which show the most common groups within the spectral corpus are methyl, methylene, and aromatic carbon rings. Functional groups other than those exclusively containing hydrogen, oxygen, and carbon are also represented in the corpus, such as sulfone \ce{R-SO2-R}, nitro \ce{R-NO-2}, and halides \ce{R-F^-}. The corpus also includes inorganic compounds, like metal oxalates. A spreadsheet containing metadata for all compounds in the spectral corpus is available along with the published code \cite{Kent_CACNP_AFM_IR_2026}.

\begin{figure}[htbp!]
    \centering
    \includegraphics[width=1.0\linewidth]{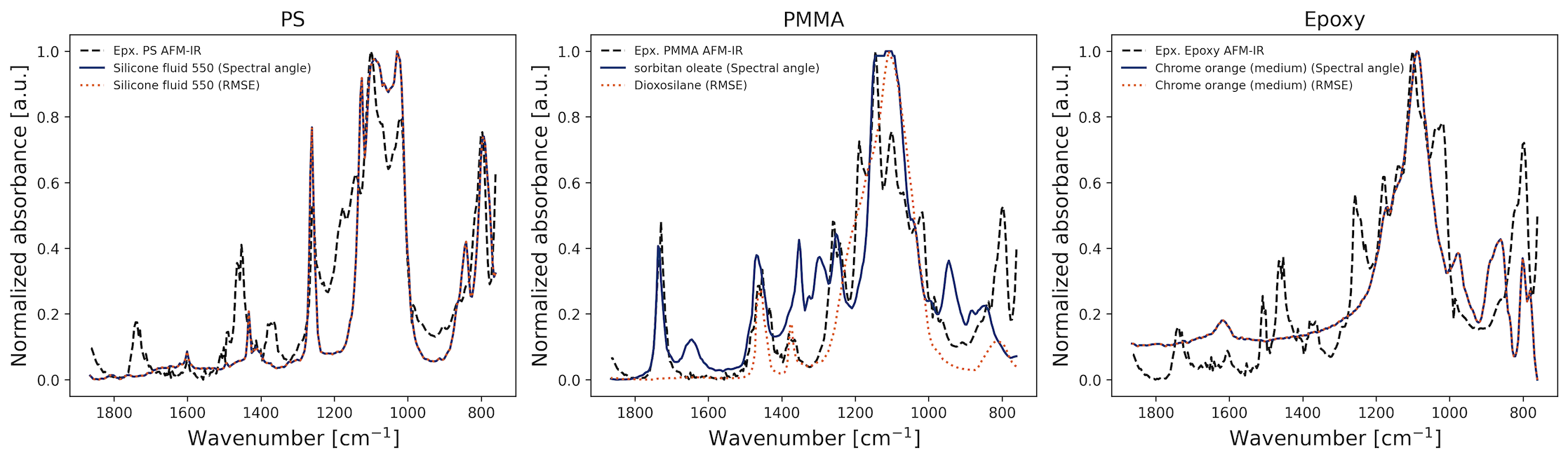}
    \caption{Comparison of experimental AFM-IR spectra of (a) PS, (b) PMMA, and (c) epoxy with their nearest spectra in the training corpus. Experimental spectra (black dashed lines) were min--max normalized to $[0,1]$ and compared with the normalized corpus spectra over the full 761--\SI{1865}{cm^{-1}} range. Blue solid and orange dotted lines identify the corpus spectra minimizing spectral angle and RMSE, respectively. Both criteria select Silicone fluid 550 for PS and Chrome orange (medium) for epoxy. For PMMA, spectral angle selects sorbitan oleate, whereas RMSE selects Dioxosilane (silicon dioxide). Curves overlap where both criteria select the same spectrum.}
\label{fig:si_corpus_matches}
\end{figure}

To assess the similarity between the experimental test spectra and the training corpus, representative PS, PMMA, and epoxy spectra were compared with every finite, nonzero corpus spectrum using spectral angle and RMSE. Experimental spectra were min--max normalized to $[0,1]$, matching the corpus normalization, so RMSE is dimensionless. Fig.~\ref{fig:si_corpus_matches} shows the nearest matches under each criterion. Both criteria selected Silicone fluid 550 for PS, with a spectral angle of 0.3977~rad and RMSE of 0.1389, and Chrome orange (medium) for epoxy, with a spectral angle of 0.4109~rad and RMSE of 0.1458. For PMMA, sorbitan oleate minimized spectral angle at 0.4391~rad (RMSE 0.1502), whereas Dioxosilane (silicon dioxide) minimized RMSE at 0.1400 (spectral angle 0.4500~rad). The nearest corpus matches are chemically distinct from the experimental polymers and retain substantial spectral differences. Combined with the exclusion of the experimental cubes from training, these results demonstrate that the reported reconstruction performance was obtained on independent experimental data, without leakage of the test cubes into training, and that the three examined experimental spectra were not exact duplicates of corpus spectra.

\subsection{Loss Function and Hyperparameters}
\label{subsec::loss_func}

The loss function and training procedure were discussed in \S~\ref{subsec::train}. Here we define the individual loss components and their relative weights.

The CACNP model was trained on an NVIDIA A100 GPU using the AdamW optimizer with a learning rate of $3\times10^{-4}$, linear warmup, and a cosine decay schedule. The training objective combines six loss terms:

\begin{align*}
    \mathcal{L}
    &=\mathcal{L}_{\mathrm{NLL}}
    +\lambda_{\mathrm{ctx}}\mathcal{L}_{\mathrm{ctx}}
    +\lambda_{\mathrm{mse}}\mathcal{L}_{\mathrm{mse}}\\
    &\quad+\lambda_{\mathrm{grad,img}}\mathcal{L}_{\mathrm{grad,img}}
    +\lambda_{\mathrm{grad,full}}\mathcal{L}_{\mathrm{grad,full}}
    +\lambda_{\mathrm{TV}}\mathcal{L}_{\mathrm{TV}}.
\end{align*}

The negative log-likelihood, context uncertainty, and reconstruction error terms are

\begin{align*}
    \mathcal{L}_{\mathrm{NLL}}
    &=\frac{1}{|\mathcal{V}|}
    \sum_{v\in\mathcal{V}}
    \left[
    \frac{1}{2}\log(2\pi)+\log\sigma_v
    +\frac{(y_v-\mu_v)^2}{2\sigma_v^2}
    \right],\\
    \mathcal{L}_{\mathrm{ctx}}
    &=\frac{1}{|\mathcal{C}|}
    \sum_{v\in\mathcal{C}}\sigma_v,\\
    \mathcal{L}_{\mathrm{mse}}
    &=\frac{1}{|\mathcal{T}|}
    \sum_{v\in\mathcal{T}}(y_v-\mu_v)^2,
\end{align*}

where $\mu_v$ and $\sigma_v$ are the predicted mean and standard deviation at voxel $v$, and $y_v$ is the ground-truth value. The sets $\mathcal{V}$, $\mathcal{C}$, and $\mathcal{T}=\mathcal{V}\setminus\mathcal{C}$ contain all voxels, context voxels, and unobserved target voxels, respectively.

Spatial-gradient errors are penalized separately at the structure-image channels and across the full cube. Let $\mathcal{I}_b$ denote the set of structure-image channels supplied for sample $b$, and let $N_{\mathrm{img}}=\sum_{b=1}^{B}|\mathcal{I}_b|$ be the total number of supplied images in the batch. The structure-image gradient loss is

\begin{align*}
    \mathcal{L}_{\mathrm{grad,img}}
    &=\frac{1}{N_{\mathrm{img}}(H-1)W}
    \sum_{b=1}^{B}\sum_{\omega\in\mathcal{I}_b}
    \sum_{r=1}^{H-1}\sum_{c=1}^{W}
    \left[
    (\Delta_h\mu_{b\omega})_{rc}
    -(\Delta_h I_{b\omega})_{rc}
    \right]^2\\
    &\quad+\frac{1}{N_{\mathrm{img}}H(W-1)}
    \sum_{b=1}^{B}\sum_{\omega\in\mathcal{I}_b}
    \sum_{r=1}^{H}\sum_{c=1}^{W-1}
    \left[
    (\Delta_w\mu_{b\omega})_{rc}
    -(\Delta_w I_{b\omega})_{rc}
    \right]^2,
\end{align*}

where $I_{b\omega}$ is the supplied structure image at channel $\omega$. During training, these images are selected slices of the normalized ground-truth cube. Only supplied images contribute to this term; padded image slots are excluded.

The full-cube gradient loss is
\begin{align*}
    \mathcal{L}_{\mathrm{grad,full}}
    &=\frac{1}{BC(H-1)W}
    \sum_{b=1}^{B}\sum_{\omega=1}^{C}
    \sum_{r=1}^{H-1}\sum_{c=1}^{W}
    \left[
    (\Delta_h\mu_{b\omega})_{rc}
    -(\Delta_h y_{b\omega})_{rc}
    \right]^2\\
    &\quad+\frac{1}{BCH(W-1)}
    \sum_{b=1}^{B}\sum_{\omega=1}^{C}
    \sum_{r=1}^{H}\sum_{c=1}^{W-1}
    \left[
    (\Delta_w\mu_{b\omega})_{rc}
    -(\Delta_w y_{b\omega})_{rc}
    \right]^2.
\end{align*}

The isotropic total-variation penalty is
\begin{align*}
    \mathcal{L}_{\mathrm{TV}}
    &=\frac{1}{BC(H-1)(W-1)}
    \sum_{b=1}^{B}\sum_{\omega=1}^{C}
    \sum_{r=1}^{H-1}\sum_{c=1}^{W-1}
    \sqrt{
    (\Delta_h\mu_{b\omega})_{rc}^{2}
    +(\Delta_w\mu_{b\omega})_{rc}^{2}
    +\epsilon
    }
\end{align*}
where $\Delta_h$ and $\Delta_w$ denote forward differences along the two spatial axes, $B$ is the batch size, $C$ is the number of spectral channels, $H$ and $W$ are the spatial dimensions, and $\epsilon=10^{-6}$.

The Gaussian negative log-likelihood couples the predicted mean and uncertainty. The context uncertainty penalty encourages confidence at measured positions, while the mean-squared error on unobserved voxels penalizes reconstruction errors directly. The two gradient terms encourage agreement with the spatial boundaries in the supplied structure images and throughout the ground-truth cube, respectively. Structure-image channels contribute to both terms, giving additional weight to agreement at these channels. The total-variation penalty promotes spatially smooth reconstructions while allowing sharp boundaries. The loss weights and essential training hyperparameters are listed in Table~\ref{tab:hparams}.

\begin{table}[ht]
\centering
    \caption{CACNP training hyperparameters.}
    \label{tab:hparams}
    \begin{tabular}{cc}
        \hline
        Param. & Value \\
        \hline \hline
        $\lambda_{\mathrm{ctx}}$ & 0.1 \\
        $\lambda_{\mathrm{mse}}$ & 1.0 \\
        $\lambda_{\mathrm{grad,img}}$ & 0.5 \\
        $\lambda_{\mathrm{grad,full}}$ & 0.2 \\
        $\lambda_{\mathrm{TV}}$ & 0.02 \\
        \hline
        \end{tabular}
        \hspace{1cm}
        \begin{tabular}{cc}
        \hline
        Param. & Value \\
        \hline \hline
        Learning rate & $3\times10^{-4}$ \\
        Maximum epochs & 800 \\
        Tasks/epoch & 2000 \\
        Warmup epochs & 5 \\
        Batch size $B$ & 4 \\
        \hline
    \end{tabular}
\end{table}

\subsection{Synthetic Hyperspectral Cube Generation and Context Selection}
\label{subsec::si_synthetic_generation}

The synthetic generation procedure was introduced in \S~\ref{subsec::synth}. \S~\ref{subsubsec::si_synth_gen} describes phase maps, spectral assignment and mixing, and measurement artifacts; \S~\ref{subsubsec::si_context_select} describes selection of context spectra and structure images.

\subsubsection{Synthetic Cube Generation}
\label{subsubsec::si_synth_gen}

This expands the generation procedure in \S~\ref{subsec::synth}.

The training code calls an on-the-fly dataset, which samples the generation parameters and passes them to the cube generator. The cube generator constructs the morphology, assigns phase spectra, and applies depth mixing. The dataset then applies texture, scratches, and detector noise before selecting the context positions and structure images. This separation permits a new combination of morphology, spectra, and artifacts to be constructed for each training task. The implementation is provided with the published code \cite{Kent_CACNP_AFM_IR_2026}.

Each cube contains $128\times128$ spatial positions and 277 spectral channels on the grid described in \S~\ref{subsec::corpus}, spanning 761 to 1865 cm$^{-1}$ in 4 cm$^{-1}$ increments. The on-the-fly dataset loads the experimental spectral corpus and wavenumber array directly. The requested number of phases $K$ is drawn uniformly from $\{2,3,4\}$, including a background phase labeled zero. One of two morphologies is selected with equal probability: dispersed particles within a matrix, or a Voronoi grain tessellation. These morphologies approximate curved particle and straight grain boundaries, respectively.

For the dispersed morphology, the generator draws a cube-level feature radius $r_0$ between 1.5 and 40 pixels and an ellipse shape parameter $c_0$ between 0.4 and 1.0. For each non-background phase $k$, these values are multiplied by independent factors drawn from $[0.70,1.30]$ and $[0.85,1.15]$, respectively, and clipped to their original ranges. A target area fraction $f_k$ is assigned sequentially from $[0.002,0.40]$, with the upper bound reduced to the remaining allocation budget divided by the number of phases still to be assigned. The initial budget is 0.85. The number of ellipses is
\begin{equation}
 n_k=\max\left[1,\operatorname{round}\left(\frac{f_kHW}{\max(\pi r_k^2c_k,1)}\right)\right].
\end{equation}
Centers are drawn uniformly within a border margin of $\max(2,0.25r_k)$ pixels. For phases containing at least four ellipses, a Voronoi diagram of the centers and four exterior points supplies a spacing radius, defined by the nearest finite ridge vertex. The first semiaxis is drawn as $r_kU(0.70,1.30)$ and clipped between 1.5 pixels and 0.90 times this spacing radius. The second is $\max[1,a c_kU(0.85,1.15)]$, and the orientation is drawn uniformly between $0$ and $180^\circ$. With fewer than four ellipses, the spacing restriction and additional second-semiaxis variation are omitted. Particle sizes, overlaps, and clipping at the field boundary cause the visible area fractions to differ from their requested values.

Phases are painted onto the background in increasing label order. Where a different phase is covered, its previous label is stored in a second map specifying the material immediately beneath the surface. This records one underlying phase at each position. For the Voronoi morphology, the number of grain centers is $\max[4,\lfloor HW/\max(1,\pi\overline{r}^{\,2})\rfloor]$, where $\overline{r}$ is the mean sampled phase radius. Centers and their phase labels are drawn uniformly, and each pixel receives the label of its nearest center. The underlying-phase map is set to zero for this morphology.

One spectrum is drawn independently with replacement for each phase from the experimental library described in \S~\ref{subsec::ir_corpus}. With spectral perturbations disabled, the assigned spectrum is
\begin{equation}
 S_k(\omega)=\operatorname{clip}_{[0,1]}\left[L_{j_k}(\omega)\right],
\end{equation}
where $L_{j_k}$ is the selected library spectrum. The per-spectrum normalization of the library is retained, and no additional phase-intensity scaling is applied. Each pixel initially receives its visible phase spectrum. The lateral boundary-mixing width is zero, so this initial assignment has sharp boundaries.

A spectral mixing effect due to thermal expansion of the substrate and subsurface phases is approximated as a subsurface contribution that is strongest near particle edges and decreases toward their interiors. An edge-mixing coefficient $\alpha_0\sim U(0,1)$ and decay length $\ell\sim U(0,10)$ pixels are drawn once per cube. For a visible non-background phase $k$, let $d_k(\mathbf{x})$ denote the inward Euclidean distance to the nearest pixel outside that phase, and let $b(\mathbf{x})$ be the stored underlying label. The mixing fraction and resulting spectrum are
\begin{align*}
 \alpha(\mathbf{x})&=\alpha_0\exp[-d_k(\mathbf{x})/\ell],\\
 Y^{(d)}(\mathbf{x},\omega)&=[1-\alpha(\mathbf{x})]S_k(\omega)
                         +\alpha(\mathbf{x})S_{b(\mathbf{x})}(\omega).
\end{align*}
Mixing is disabled if $\alpha_0$ or $\ell$ is zero, and $\alpha$ is set to zero at background positions or where the underlying label matches the visible phase. At overlaps, the subsurface contribution is the base spectrum of the earlier phase; elsewhere it is the background spectrum. For Voronoi grains, non-background regions mix toward the background spectrum. Because the outermost included pixels have positive inward distance, their mixing fraction is smaller than $\alpha_0$. 

To mimic intra-phase variations in intensity, spatial intensity texture is applied to every cube after depth mixing. For each visible phase, two independent Gaussian white-noise fields are smoothed with Gaussian-filter widths drawn from 3 to 15 pixels. Each field is divided by its spatial standard deviation, multiplied by a coefficient drawn from $[0.02,0.06]$, and added to that phase's texture field. Values within each phase mask form the combined field $F(\mathbf{x})$, which is clipped to $[-0.10,0.10]$ and applied to all spectral channels:
\begin{equation*}
 Y^{(t)}(\mathbf{x},\omega)=\operatorname{clip}_{[0,1]}
 \left\{Y^{(d)}(\mathbf{x},\omega)[1+F(\mathbf{x})]\right\}.
\end{equation*}
The modulation scales the complete local mixture according to its visible phase.

Scratches are applied to mimic sample damage with probability 0.3 per cube. An affected cube receives one to four candidate scratches, each starting at a uniformly sampled position and orientation. A persistent random walk grows in two directions with one-pixel steps and normally distributed angular increments of standard deviation 0.0375 radians. Its total length budget is 0.4--1.2 times the image diagonal, divided between the two arms by a fraction drawn from $[0.2,0.8]$. Each arm ends at the field boundary, and paths with fewer than three points are discarded. The scratch profile is $P_j(\mathbf{x})=\exp[-(d_j/w_j)^2]$, where $d_j$ is the distance to the rasterized path and $w_j\sim U(0.8,2.5)$ pixels. The attenuation coefficient $a_j$ is drawn from $[0.20,0.60]$; with probability 0.15 it is multiplied by $-0.6$ to produce a bright defect. The combined field $A=\sum_j a_jP_j$ is clipped to $[-0.95,0.95]$, and the cube is multiplied by $1-A$ and clipped to $[0,1]$.

Random noise is added after texture and scratches. A standard deviation $\sigma_n\sim U(0.01,0.08)$ is drawn for each cube, and independent zero-mean Gaussian noise is added to every voxel. The resulting cube is clipped to $[0,1]$ before normalization. Noise is disabled inside the underlying cube generator and added once by the on-the-fly dataset. 

\subsubsection{Context Selection}
\label{subsubsec::si_context_select}

This expands the context and structure-image selection described in \S~\ref{subsec::train}.

Context positions are selected after all artifacts have been applied. The number of complete context spectra is sampled from a rounded log-normal distribution, with draws outside $[4,4000]$ rejected. This spans approximately 0.024\% to 24.4\% of the spatial positions. For the underlying untruncated distribution, the mode is $0.05HW$ and the 0.99 quantile is $0.20HW$. If $\log N\sim\mathcal{N}(\mu_N,\sigma_N^2)$ and $z_{0.99}$ is the corresponding standard-normal quantile,
\begin{align*}
 \sigma_N&=\frac{-z_{0.99}+\sqrt{z_{0.99}^2+4\log(0.20/0.05)}}{2},\\
 \mu_N&=\log(0.05HW)+\sigma_N^2.
\end{align*}
For the $128\times128$ grid, $(\mu_N,\sigma_N)\approx(6.9503,0.4919)$. The sampled number of context positions is selected uniformly without replacement from the full grid.

Structure-image bands are selected from the population variance among the sampled context spectra at each spectral channel. The variance curve is min-max normalized, and local maxima with normalized height at least 0.2 are identified using \texttt{scipy.signal.find\_peaks} \cite{2020SciPy-NMeth}. Candidate bands are ranked by decreasing variance. A requested image count of one, two, three, or four is drawn with probabilities 0.40, 0.30, 0.20, and 0.10, respectively. The count is reduced if fewer peaks are available, and the highest-ranked bands are selected. If no peaks are found, the maximum-variance band is used.

The complete artifact-containing cube is normalized using one mean and standard deviation pooled over all context pixels and spectral channels, as described above. The normalized cube supplies both the training target and the full-field images at the selected bands. The dataset returns the cube, context mask, four padded band indices, and number of valid images. The training code extracts the image planes and masks unused guidance slots.

Two thousand synthetic tasks are generated per training epoch, with seeds derived from the base seed, split, epoch, worker offset, and item index. Validation uses 200 tasks with a separate base seed and a fixed epoch value of zero, so the same validation tasks are reconstructed at each evaluation. Both splits draw from the same spectral library. Generation and context settings are summarized in Tables~\ref{tab:si_generation_morphology}--\ref{tab:si_generation_context}.

\begin{table}[htbp!]
\centering
\small
\caption{Morphology, spectral assignment, and depth-mixing parameters for the published training configuration. Ranges denote uniform sampling except when otherwise specified.}
\label{tab:si_generation_morphology}
\begin{tabular}{p{0.46\linewidth}p{0.47\linewidth}}
\hline
Parameter & Value \\
\hline
Spatial grid / spectral channels & $128\times128$ / 277 \\
Spectral grid & 761--1865 cm$^{-1}$; 4 cm$^{-1}$ increments \\
Morphology probabilities & Dispersed 0.5; Voronoi grains 0.5 \\
Base feature radius / shape parameter & 1.5--40 pixels / 0.4--1.0 \\
Per-phase radius / shape multipliers & 0.70--1.30 / 0.85--1.15; clipped to base ranges \\
Requested non-background fraction & 0.002--0.40 per phase; sequential budget 0.85 \\
Particle first-semiaxis multiplier & 0.70--1.30 relative to phase radius \\
Voronoi size control & At least 4 particles per phase; 4 exterior points \\
First-semiaxis bounds with size control & 1.5 pixels and $0.90r_{\mathrm{spacing}}$ \\
Second-semiaxis minimum / multiplier & 1 pixel / 0.85--1.15 with size control \\
Particle orientation / center margin & $0$--$180^\circ$ / $\max(2,0.25r_k)$ pixels \\
Minimum grain-center count & 4 \\
Depth edge coefficient $\alpha_0$ & 0--1, one draw per cube \\
Depth decay length $\ell$ & 0--10 pixels, one draw per cube \\
\hline
\end{tabular}
\end{table}

\begin{table}[htbp!]
\centering
\small
\caption{Texture, scratch, and detector-noise parameters.}
\label{tab:si_generation_artifacts}
\begin{tabular}{p{0.46\linewidth}p{0.47\linewidth}}
\hline
Parameter & Value \\
\hline
Texture probability / fields per phase & 1.0 / 2 \\
Texture Gaussian-filter width & 3--15 pixels per field \\
Texture amplitude per field & 0.02--0.06 after standard-deviation scaling \\
Combined texture bounds & $-0.10$ to $+0.10$ \\
Gaussian-filter boundaries / truncation & Reflection / 4 standard deviations \\
Scratch probability / candidate count & 0.30 / 1--4 per affected cube \\
Scratch width $w_j$ / attenuation $a_j$ & 0.8--2.5 pixels / 0.20--0.60 \\
Bright-defect probability / multiplier & 0.15 / $-0.6$ on attenuation coefficient \\
Scratch length budget / arm split & 0.4--1.2 image diagonals / 0.2--0.8 \\
Step length / turn standard deviation & 1 pixel / 0.0375 radians \\
Minimum retained path / termination & 3 points / field boundary \\
Combined scratch-field bounds & $-0.95$ to $+0.95$ \\
Detector-noise mean / standard deviation & 0 / 0.01--0.08; one scale per cube \\
\hline
\end{tabular}
\end{table}

\begin{table}[htbp!]
\centering
\small
\caption{Context, structure-image selection, and dataset parameters used with on-the-fly generation.}
\label{tab:si_generation_context}
\begin{tabular}{p{0.46\linewidth}p{0.47\linewidth}}
\hline
Parameter & Value \\
\hline
Context count distribution & Rounded log-normal \\
Minimum / maximum complete spectra & 4 / 4000 \\
Untruncated mode / 0.99 quantile & 0.05 / 0.20 of spatial positions \\
Underlying log-normal $(\mu_N,\sigma_N)$ & $(6.950293,0.491899)$ for $128\times128$ \\
Spatial context selection & Uniform without replacement \\
Image selection statistic & Per-band population variance over context spectra \\
Variance normalization / peak threshold & Min-max to $[0,1]$ / 0.2 \\
Requested image counts / probabilities & 1, 2, 3, 4 / 0.40, 0.30, 0.20, 0.10 \\
Maximum images / ordering & 4 / descending variance of candidate peaks \\
Flat-variance threshold / no-peak fallback & Range below $10^{-12}$ / maximum-variance band \\
Normalization & One context mean and standard deviation per cube \\
Training tasks per epoch / validation tasks & 2000 / 200 \\
\hline
\end{tabular}
\end{table}

\subsection{Experimental Hyperspectral AFM-IR Data}
\label{subsec::si_exp_data}

The two experimental hyperspectral datasets used in this work were introduced in \S~\ref{subsec::exp_samp}. Both hyperspectral cubes were acquired from the same standard PS and PMMA embedded in an epoxy matrix prepared by Bruker. 

Both hyperspectral datasets were acquired on a Bruker Dimension IconIR system. The single- and multi-particle datasets took $\approx5$ and $\approx20$ hours to acquire, respectively. 

\begin{figure}[htpb!]
    \centering
    \includegraphics[width=\linewidth,height=0.78\textheight,keepaspectratio]{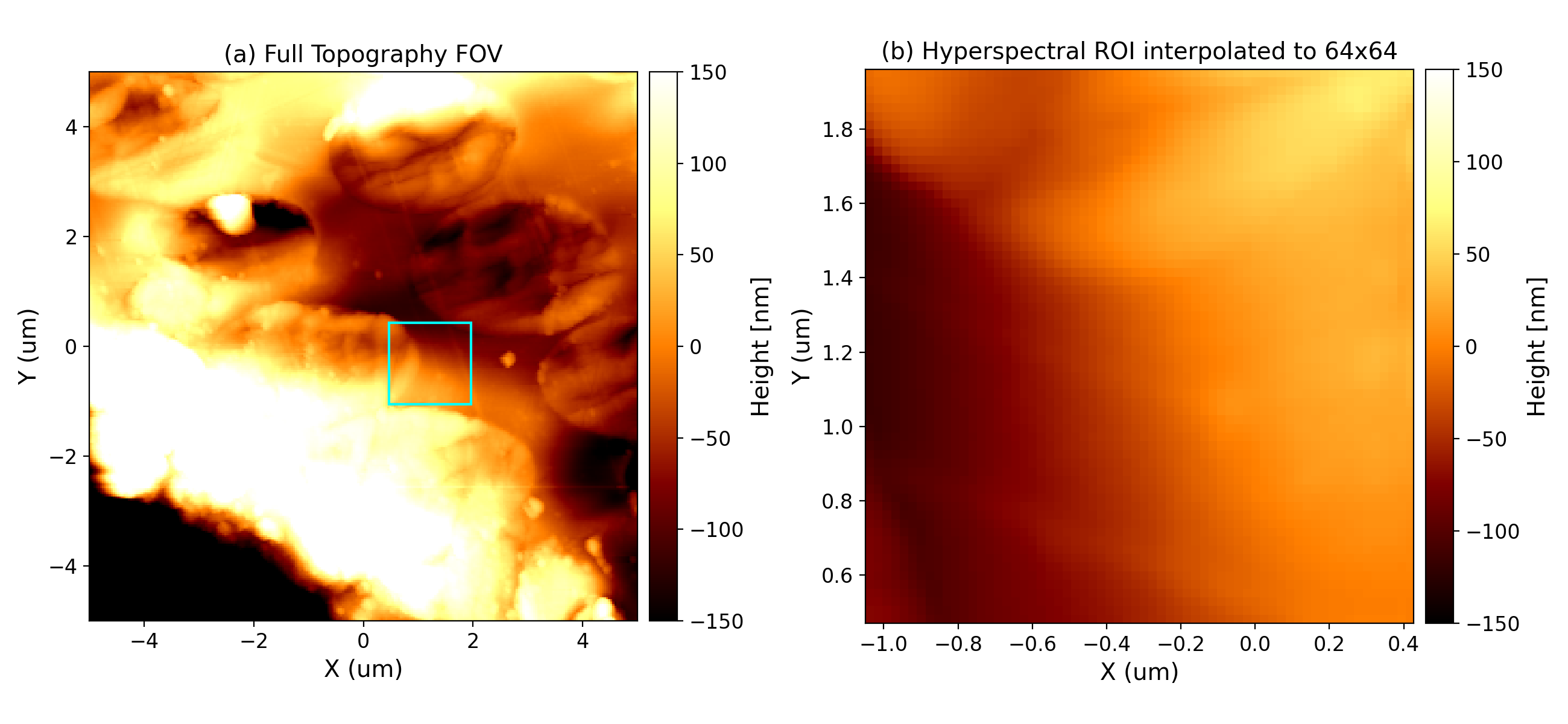}
    \caption{AFM topography images of single-particle ROI. (a) a $ 10 \times \SI{10}{\micro\meter}$ ($128 \times 128$ pixel) large field of view with multiple PMMA particles embedded in epoxy and (b) teal blue ($1.5 \times \SI{1.5}{\micro\meter}$) ROI interpolated to the $64 \times 64$ pixels of the corresponding hyperspectral cube. The PMMA particles can be seen at the top and bottom left of the ROI in (b).}
    \label{fig:si_small_fov}
\end{figure}

Fig.~\ref{fig:si_small_fov} shows a large FOV image showing the area surrounding the single-particle acquisition FOV (Fig.~\ref{fig:si_small_fov}a), as well as an interpolated topography image corresponding to the same area (and pixel resolution) as the hyperspectral cube (Fig.~\ref{fig:si_small_fov}b). Subtle topography variations can be seen in Fig.~\ref{fig:si_small_fov}b, delineating the PMMA particles in the top and bottom-left of the image from the epoxy matrix. This area was chosen for acquisition to evaluate the CACNP's ability to reconstruct spectral variations along particle boundaries. 

\begin{figure}[htbp]
    \centering
    
    \begin{subfigure}[b]{1.0\textwidth}
        \centering
        \includegraphics[width=\textwidth]{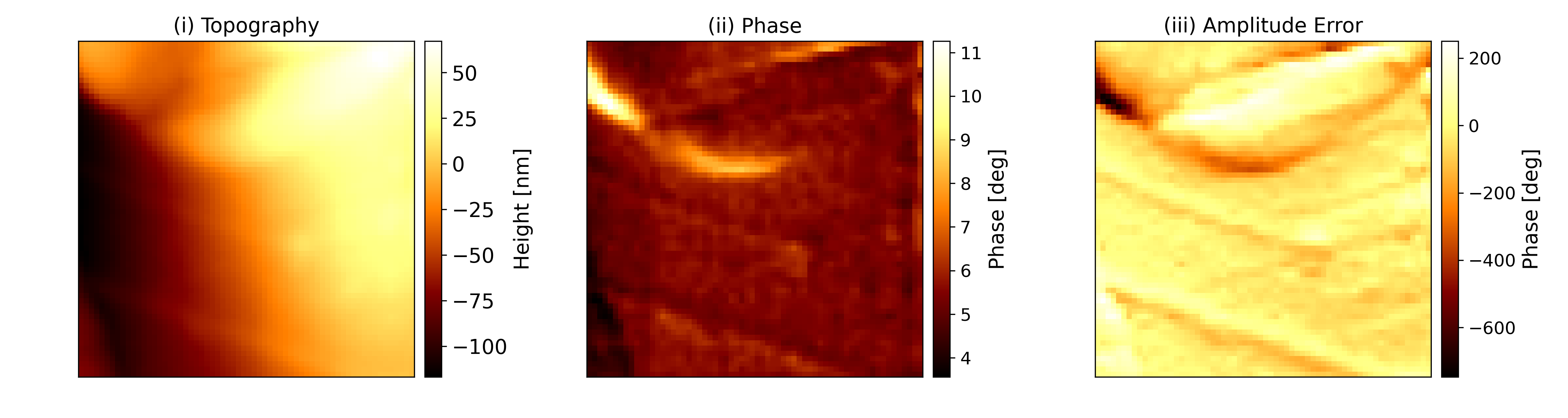}
        \caption{Single-particle dataset}
        \label{fig:top_sub}
    \end{subfigure}
    
    \vspace{0.5cm} 
    
    \begin{subfigure}[b]{1.0\textwidth}
        \centering
        \includegraphics[width=\textwidth]{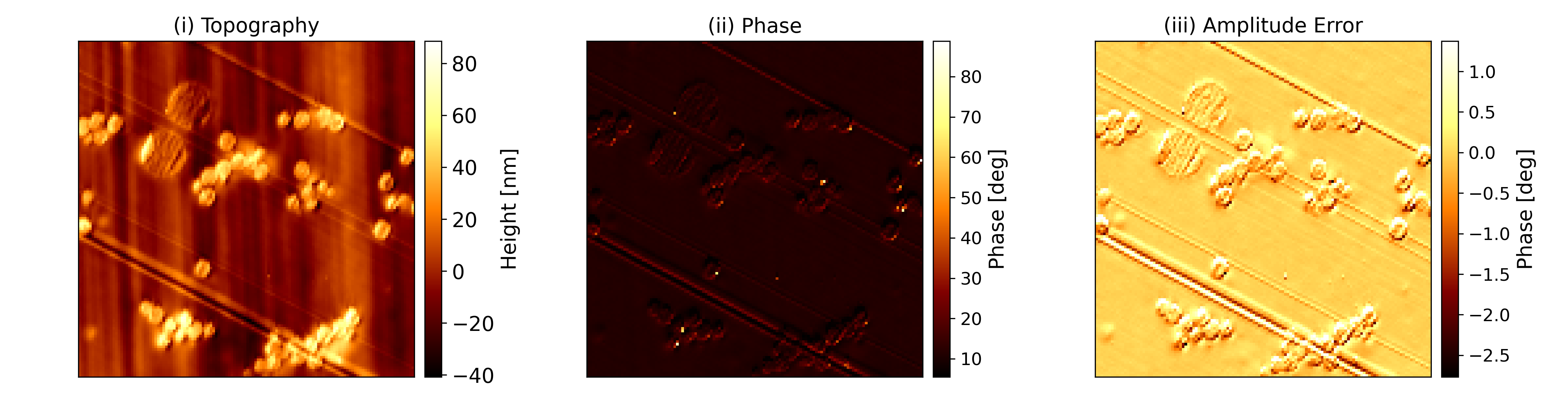}
        \caption{Multi-particle dataset}
        \label{fig:bottom_sub}
    \end{subfigure}
    
    \caption{(i) Topography, (ii) phase, and (iii) amplitude error images of the (a) single-particle and (b) multi-particle datasets. All images in (a) are $1.5 \times \SI{1.5}{\micro\meter}$ ($64 \times 64$ pixels), and all images in (b) are $20 \times \SI{20}{\micro\meter}$ ($128 \times 128$ pixels). Images in (a) are interpolated to reflect the same pixel grid as the corresponding hyperspectral cube. The original FOV is shown in Fig.~\ref{fig:si_small_fov}}
    \label{fig:si_cubes_basic}
\end{figure}

Fig.~\ref{fig:si_cubes_basic} shows topography (i), phase (ii), and amplitude error (iii) images for the single- (a) and multi- (b) particle datasets. As can be clearly seen in Fig.~\ref{fig:si_cubes_basic}bi, the multi-particle sample contains both PMMA, the two large particles in the top left, and PS, the small particles dispersed throughout the sample, particles embedded in epoxy. Vertically-aligned topological variations are observed in the epoxy matrix, as well as several diagonal scratches. 

\subsection{CACNP Model Architecture}
\label{subsec::si_cacnp_architecture}

The CACNP architecture was introduced in \S~\ref{subsec::arch}. Here we expand on how the context spectra and structure images are represented, how these data move through the network, and how the reconstructed mean and predictive uncertainty are obtained. Layer specifications and architecture settings are provided in Tables~\ref{tab:si_cacnp_layers} and \ref{tab:si_cacnp_settings}; the training schedule is summarized in Table~\ref{tab:si_cacnp_training}. 

The model combines a convolutional encoder and decoder with attention to the measured spectra, drawing on convolutional conditional neural processes and attentive neural processes \cite{gordonConvolutionalConditionalNeural2020,kimAttentiveNeuralProcesses2019}. The implementation uses PyTorch and PyTorch Lightning \cite{paszkePyTorchImperativeStyle2019,falcon2019pytorchlightning}. The layer dimensions and hyperparameters described below correspond to the configuration specified in the published training configuration.

Let $Y\in\mathbb{R}^{C\times H\times W}$ denote a hyperspectral cube, where $H$ and $W$ are the two spatial dimensions and $C$ is the number of spectral channels. Training cubes contained $H=W=128$ pixels and $C=277$ channels. The observations consist of context spectra at a set of spatial positions $\mathcal{C}$, together with $N_I$ full-field structure images, where $1\leq N_I\leq4$. Each structure image is accompanied by the integer index of the wavenumber at which it was acquired. The context mask identifying the positions of the context spectra has shape $B\times H\times W$.

Before a cube is passed to the network, it is normalized using the intensities at the context positions. A single mean $m_{\mathcal{C}}$ and standard deviation $s_{\mathcal{C}}$ are computed over all spectral channels and all context positions in that cube. The same transformation is then applied to the context spectra, the structure images, and the training target:
\begin{equation*}
\widetilde{Y}_{xy\omega}=\frac{Y_{xy\omega}-m_{\mathcal{C}}}{s_{\mathcal{C}}}.
\label{eq:si_context_normalization}
\end{equation*}
This is a shared normalization of the observed intensities, rather than a separate normalization at each wavenumber or at each position. Consequently, relative peak heights within and between the context spectra are retained. If the context standard deviation is no greater than $10^{-8}$, the denominator is set to one. Predictions can be returned to the original intensity units using $\mu_{\mathrm{raw}}=m_{\mathcal{C}}+s_{\mathcal{C}}\mu$ and $\sigma_{\mathrm{raw}}=s_{\mathcal{C}}\sigma$, with the same denominator convention.

Before entering the network, nearest neighbor filling is used to assign spectra to unobserved pixels in the hyperspectral cube. This operation copies spectral values without averaging adjacent context spectra. A second input records the Euclidean distance to the assigned observation, divided by $\sqrt{H^2+W^2}$. The distance is zero at the context positions and increases with separation from the measured spectra. Together, the filled spectra and distance map produce $C+1=278$ channels. Nearest-neighbor filling replaced the Gaussian smearing used in the original ConvCNP \cite{gordonConvolutionalConditionalNeural2020}, which gave poor predictions far from observations at low context, including with learned or multiple smearing radii. Filling supplies a complete measured spectrum at every position.

Structure images are encoded separately before being concatenated with the filled spectra. Each of the 277 band indices has a learned 16-dimensional embedding. For a given structure image, this embedding is repeated at every spatial position and concatenated with the normalized image intensity, giving 17 input channels. Two shared $1\times1$ convolutions transform these channels as $17\rightarrow16\rightarrow16$, with a GELU activation between the convolutions. The same embedding network is used for every structure image. This gives the model both the measured intensity and the identity of the spectral channel to which that intensity belongs.

The embedded images are concatenated into four ordered slots, each containing 16 channels. The slots follow descending variance ranking. Slots for images that were not supplied are set to zero after embedding. The resulting guidance tensor therefore has 64 channels regardless of the number of images. This concatenation preserves the separate contribution of each image, although its representation depends on the slot order. 

The 64 guidance channels are concatenated with the 278 channels from the context representation, producing a $342\times128\times128$ tensor for each training example. A spectral input projection, implemented as two $1\times1$ convolutions, transforms the channel dimension as $342\rightarrow512\rightarrow128$. Each convolution is followed by group normalization and GELU. The convolutions mix information across the spectral, distance, and image-guidance channels at each spatial position. Group normalization additionally normalizes groups of channels over their spatial extent. The resulting 128-channel tensor is passed to the U-Net.

The U-Net contains four encoder blocks and three decoder blocks. The encoder channel widths are 128, 256, 512, and 1024, at spatial dimensions of $128\times128$, $64\times64$, $32\times32$, and $16\times16$, respectively. Successive encoder blocks are separated by $2\times2$ max pooling. Each block contains two $3\times3$ convolutions with one-pixel zero padding, each followed by group normalization and GELU. Spatial dropout with probability 0.05 is applied after the first activation. A residual shortcut adds the block input to the transformed features; a bias-free $1\times1$ convolution adjusts the shortcut channel width when necessary. All group-normalized layers in this configuration use 32 groups, and the convolutional layers followed by group normalization do not use bias terms.

The decoder increases the spatial dimensions by bilinear interpolation, with corner alignment disabled. At each level, the upsampled features are concatenated with the encoder features at the corresponding spatial resolution. The three decoder blocks transform the concatenated channel widths as $1536\rightarrow512$, $768\rightarrow256$, and $384\rightarrow256$. Thus, the last decoder block returns 256 features at every position of the original grid. For inputs whose spatial dimensions are not multiples of eight, the U-Net reflect-pads its input on the bottom and right to the next multiple of eight and crops its output back to the original dimensions. No such padding is needed for the $128\times128$ training cubes.

Cross-attention is applied after the bottleneck block and after the first two decoder blocks. The final decoder block does not use cross-attention due to memory constraints during training. At these stages, each spatial feature vector is a query, while the keys and values are constructed from the measured context spectra. The measured spectra are gathered once per forward pass and shared between the three attention stages, although each stage has its own learned projections. Within a batch, context sets are padded to the largest context count and the padded entries are excluded from attention. These context spectra are the actual normalized observations, including simulated measurement noise, rather than denoised spectra or values copied from the filled grid. 

Spatial information is incorporated into the queries and keys using Fourier features. The two coordinates are normalized to the interval $[0,1]$ using pixel-center positions. For either coordinate $u$, the encoding contains $\sin(2^j\pi u)$ and $\cos(2^j\pi u)$ for $j=0,\ldots,7$. Encoding both spatial coordinates gives a 32-dimensional position vector $\phi(\mathbf{x})$. At a given attention stage, a feature vector $\mathbf{f}_i$ and a context spectrum $\mathbf{s}_j$ are projected according to
\begin{align*}
\mathbf{q}_i&=A_Q[\mathbf{f}_i;\phi(\mathbf{x}_i)]+\mathbf{b}_Q,\\
\mathbf{k}_j&=A_K[\mathbf{s}_j;\phi(\mathbf{x}_j)]+\mathbf{b}_K,\\
\mathbf{v}_j&=A_V\mathbf{s}_j+\mathbf{b}_V.
\end{align*}
Each projection returns 256 features, which are divided into four attention heads of width 64. Within each head, attention is computed as
\begin{equation*}
\operatorname{Attention}(Q,K,V)=\operatorname{softmax}\left(\frac{QK^{\mathsf T}}{\sqrt{64}}\right)V,
\end{equation*}
where the softmax is taken over valid context positions. The four outputs are concatenated, projected back to the feature width of that U-Net stage, and added to its input features. The output projection weights and biases are initialized to zero, so the attention branches initially make no change to the convolutional features. This arrangement allows the network to retrieve spectral information from observations throughout the field of view while retaining the spatial information learned by the convolutional backbone. Structure images enter through the guidance tensor and are not additional attention keys or values in this configuration.

After the last U-Net decoder block, the original 64-channel guidance tensor is concatenated directly with the 256-channel decoded features. This additional skip connection makes the full-resolution embedded images available to both output heads. Each head therefore receives 320 channels. The mean and scale heads have separate parameters but the same architecture: a $3\times3$ convolution from 320 to 512 channels, group normalization, GELU, and a second $3\times3$ convolution from 512 to 277 channels. Both convolutions use one-pixel reflection padding. The final convolution includes a bias and is not followed by normalization.

A fourth attention operation is applied directly to the output of the mean head at full spatial resolution. Here, the preliminary reconstructed spectrum at each position supplies the query and the measured context spectra supply the keys and values. Each is projected from 277 to 320 dimensions and divided into four heads of width 80. Attention follows the same equation as the decoder cross-attention, with $\sqrt{80}$ replacing $\sqrt{64}$. No positional encoding is supplied to this operation. Its attention weights therefore depend on the spectral content of the preliminary reconstruction and context spectra, although the preliminary reconstruction itself already contains spatial information from the U-Net. The attended values are projected back to 277 channels and added to the preliminary mean spectrum. This output projection is also initialized to zero. The purpose of this operation is to provide a direct route from measured spectral features to the final reconstruction after the convolutional decoding stages, helping preserve and propagate subtle spectral features such as short peaks. 

The corrected mean is returned without a bounding activation, which permits predictions over the unbounded range of the context-normalized data. The scale-head output $a_{xy\omega}$ is converted to a positive predictive standard deviation using
\begin{equation*}
\sigma_{xy\omega}=\log\left(1+\exp(a_{xy\omega})\right)+10^{-3}.
\end{equation*}
The mean-head attention correction is not applied to the scale head. Together, the two outputs specify a Gaussian predictive distribution at every voxel,
\begin{equation}
p(\widetilde{Y}_{xy\omega}\mid\mathcal{C},\{I_i\})=
\mathcal{N}\left(\mu_{xy\omega},\sigma_{xy\omega}^{2}\right).
\end{equation}
The implementation returns these marginal distributions and does not explicitly predict covariance between voxels. Neither the observed spectra nor the structure-image intensities are forcibly copied into the final mean. 

\begin{table}[htbp]
\centering
\small
\caption{Layer specification of the CACNP. Spatial dimensions are those of the training cubes. Parameter counts include learned weights, biases, and normalization parameters and were calculated from the supplied architecture for 277 spectral channels.}
\label{tab:si_cacnp_layers}
\begin{tabular}{p{0.29\linewidth}p{0.39\linewidth}r}
\hline
Component & Transformation and output grid & Parameters \\
\hline
Image guidance & Band embedding $277\times16$; shared $17\to16\to16$; four slots, $128^2$ & 4,992 \\
Spectral input projection & $342\to512\to128$; $128^2$ & 241,920 \\
Encoder 0 & $128\to128$; $128^2$ & 295,424 \\
Encoder 1 & $128\to256$; $64^2$ & 918,528 \\
Encoder 2 & $256\to512$; $32^2$ & 3,672,064 \\
Encoder 3 / bottleneck & $512\to1024$; $16^2$ & 14,684,160 \\
Bottleneck attention & Feature width 1024; $16^2$ & 684,288 \\
Decoder 2 & $1536\to512$; $32^2$ & 10,225,664 \\
Decoder 2 attention & Feature width 512; $32^2$ & 421,632 \\
Decoder 1 & $768\to256$; $64^2$ & 2,556,928 \\
Decoder 1 attention & Feature width 256; $64^2$ & 290,304 \\
Decoder 0 & $384\to256$; $128^2$ & 1,573,888 \\
Mean head & $320\to512\to277$; $128^2$ & 2,752,277 \\
Scale head & $320\to512\to277$; $128^2$ & 2,752,277 \\
Mean-head attention & $277\to320\to277$; $128^2$ & 355,797 \\
\hline
Total & All parameters trainable & 41,430,143 \\
\hline
\end{tabular}
\end{table}

\begin{table}[htbp]
\centering
\small
\caption{Architecture and normalization settings used for the final training configuration. }
\label{tab:si_cacnp_settings}
\begin{tabular}{p{0.45\linewidth}p{0.48\linewidth}}
\hline
Parameter & Value \\
\hline
Input representation & Nearest-observation spectra and normalized distance \\
Input normalization / denominator threshold & Context z-score / $10^{-8}$ \\
Structure-image slots / embedding width & 4 / 16 per image and per band embedding \\
Guidance aggregation / decoder skip & Concatenation / 64 guidance channels \\
Input and output hidden width & 512 \\
Encoder widths / final decoder width & $(128,256,512,1024)$ / 256 \\
Residual-block kernel / padding & $3\times3$ / one pixel, zeros \\
Pooling / upsampling & $2\times2$ max / bilinear, corners not aligned \\
Group normalization & 32 groups; affine; $\epsilon=10^{-5}$ \\
Activation / spatial dropout & GELU / 0.05 within U-Net blocks \\
Feature-attention stages & Bottleneck, decoder 2, decoder 1 \\
Feature-attention width / heads / head width & 256 / 4 / 64 \\
Fourier frequencies per coordinate & 8, $2^j\pi$, $j=0,\ldots,7$ \\
Position encoding width & 32 \\
Output-head kernels / padding & $3\times3$ / one pixel, reflection \\
Mean-head attention width / heads / head width & 320 / 4 / 80 \\
Mean activation / scale activation & Identity / softplus plus $10^{-3}$ \\
\hline
\end{tabular}
\end{table}

\begin{table}[htbp]
\centering
\small
\caption{Optimization and data-loading settings specified for the final training configuration. The epoch count denotes the maximum training duration.}
\label{tab:si_cacnp_training}
\begin{tabular}{p{0.46\linewidth}p{0.47\linewidth}}
\hline
Parameter & Value \\
\hline
Training mode & Joint Gaussian likelihood; both heads trained \\
Optimizer / learning rate / weight decay & AdamW / $3\times10^{-4}$ / $10^{-5}$ \\
AdamW defaults & $\beta_1=0.9$, $\beta_2=0.999$, $\epsilon=10^{-8}$ \\
Learning-rate schedule & Linear warmup, then cosine decay to zero \\
Warmup / maximum epochs & 5 / 800 \\
Training tasks per epoch / batch size & 2000 / 4 \\
Validation tasks / validation batch size & 200 fixed synthetic tasks / 4 \\
Optimizer steps per full epoch & 500 \\
Planned warmup / maximum optimizer steps & 2500 / 400,000 before early stopping \\
Early-stopping patience & 40 validation checks without improvement \\
Checkpoint and early-stopping monitor & Minimum validation full-cube NLL \\
Training / validation dataset seed & 42 / 43 \\
Data-loader workers / prefetch factor & 7 / 4 \\
Training shuffle / persistent workers & Enabled / disabled \\
Precision / mixed precision & 32-bit / disabled \\
Float32 matrix multiplication setting & \texttt{high} \\
Hardware requested & One NVIDIA A100 GPU; 8 CPU cores; 64 GB RAM \\
Initialization / checkpoint resumption & Fresh initialization / none specified \\
Adaptive-context training & Disabled; uniformly sampled spatial positions \\
\hline
\end{tabular}
\end{table}

\subsection{Uncertainty Calibration}
\label{subsec::si_calibration}

This subsection expands the uncertainty calibration procedure in \S~\ref{subsec::calibration}. Inspired by reconstruction-uncertainty calibration approaches such as QUTCC \cite{yeQUTCCQuantileUncertainty2026}, we rescale the predicted standard deviations using held-out context spectra, without training an additional network. The resulting intervals $\mu \pm 2\sigma$ have a nominal Gaussian predictive coverage of approximately 95.45\%; however, the scaling procedure does not enforce this coverage, which is evaluated empirically at unobserved voxels. Let $\mathcal{C}_{\mathrm{keep}}$ and $\mathcal{C}_{\mathrm{hold}}$ denote a random split containing approximately 70\% and 30\% of the measured spatial positions, respectively. The model is first evaluated using $\mathcal{C}_{\mathrm{keep}}$, retaining the supplied structure images. For the held-out voxels $\mathcal{H}$, comprising every spectral channel at positions in $\mathcal{C}_{\mathrm{hold}}$, the global scale is
\begin{equation}
 s=\left[\frac{1}{|\mathcal{H}|}\sum_{v\in\mathcal{H}}
 \left(\frac{\mu^{\mathrm{keep}}_v-y_v}{\max(\sigma^{\mathrm{keep}}_v,10^{-8})}\right)^2\right]^{1/2}.
\label{eq:si_sigma_scale}
\end{equation}
The model is then evaluated with all measured context, and the calibrated standard deviation is $\sigma^{\mathrm{cal}}_v=s\sigma^{\mathrm{full}}_v$. Neither the model weights nor the full-context mean are altered. The number of held-out positions is the rounded value of $0.3|\mathcal{C}|$. A scale of one is used when fewer than four context positions are available, when the split is empty, or when the estimated scale is nonfinite or nonpositive.

The scale is the root-mean-square standardized residual, rather than an empirical quantile chosen to enforce 95.4\% coverage. The present variance-scaling procedure does not provide that conformal coverage guarantee. It also assumes that a scale estimated with reduced context remains suitable when all context is restored. Coverage is therefore evaluated empirically on unobserved voxels as
\begin{equation}
 \widehat{C}_{2\sigma}=\frac{1}{|\mathcal{U}|}\sum_{v\in\mathcal{U}}
 \mathbf{1}\!\left\{|y_v-\mu_v^{\mathrm{full}}|\leq2\sigma_v\right\},
\label{eq:si_sigma_coverage}
\end{equation}
where $\mathcal{U}$ contains all spectral channels at spatial positions without measured context spectra, and $\sigma$ is either the raw or calibrated standard deviation. The supplied structure-image channels remain included in $\mathcal{U}$, so this is coverage at unobserved spatial positions rather than coverage restricted to entirely unmeasured voxels. Restoring the held-out context changes the predictive distribution, so calibration can produce coverage above or below the nominal Gaussian value.

\section{Results on Experimental Data}
\label{si_sec::exp_res}

These results supplement the reconstruction results on experimental data presented in \S~\ref{sec:results} of the main document. \S~\ref{subsec::si_struct_img} expands the structure-image comparisons, single-particle reconstructions, and uncertainty coverage discussed in \S~\ref{subsec:exp_per}. \S~\ref{subsec::si_method_comp} and \S~\ref{subsec::si_timing} provide the spectral comparisons and timings supporting \S~\ref{subsec:method_comp}, respectively, and \S~\ref{subsec::si_acq_strat} expands the acquisition comparisons in \S~\ref{subsec:acq_strat}.

\subsection{Reconstruction Error Metrics}
\label{subsec::error_metrics}

Reconstruction quality was evaluated using root-mean-squared error (RMSE), spectral angle, and segmentation misclassification rate. These metrics assess complementary aspects of chemical imaging: agreement in absorption amplitude, preservation of spectral shape, and identification of spatial regions with similar chemistry.

The RMSE is

\begin{align*}
\mathrm{RMSE}
=\sqrt{\frac{1}{|\mathcal{E}|}
\sum_{v\in\mathcal{E}}(y_v-\mu_v)^2},
\end{align*}

where $\mathcal{E}$ is the set of evaluated voxels, $y_v$ is the ground-truth absorption, and $\mu_v$ is the reconstructed value. RMSE quantifies the overall magnitude of reconstruction errors in the same units as the absorption signal, with greater weight given to large deviations. It therefore measures how accurately the reconstruction preserves absorption amplitudes and chemical contrast across the dataset, although it does not distinguish errors in spectral shape from errors in overall intensity.

Measured context positions were included when evaluating RMSE, spectral angle, and segmentation misclassification rate to assess consistency with the supplied observations as well as reconstruction of unmeasured information. Apart from nearest-neighbor filling, the compared implementations do not impose a hard constraint that forces reconstructed spectra to equal the measured context spectra. Conditioning on these observations or penalizing their reconstruction error during fitting therefore does not guarantee their exact reproduction. Including context positions evaluates this fidelity directly through RMSE and spectral angle, and assesses whether the resulting spectra preserve their ground-truth cluster assignments through the misclassification rate. All spectral channels, including the supplied structure-image channels, were retained. These metrics consequently characterize complete-cube reconstruction fidelity.

Spectral angle measures the angular separation between the ground-truth and reconstructed spectra at each spatial position:

\begin{align*}
    \theta_i
    =\arccos\left(
    \frac{\mathbf{y}_i\cdot\boldsymbol{\mu}_i}
    {\|\mathbf{y}_i\|_2\,\|\boldsymbol{\mu}_i\|_2}
    \right),
    \qquad
    \overline{\theta}
    =\frac{1}{|\mathcal{P}|}\sum_{i\in\mathcal{P}}\theta_i,
\end{align*}

where $\mathbf{y}_i$ and $\boldsymbol{\mu}_i$ contain all spectral channels at position $i$, and $\mathcal{P}$ is the set of evaluated spatial positions. Smaller angles indicate closer agreement in spectral shape, including the positions and relative intensities of absorption bands that distinguish chemical species. Spectral angle is insensitive to positive multiplicative scaling of either spectrum, allowing spectral fidelity to be assessed separately from overall signal amplitude. It is not a direct measure of peak-position error, since changes in relative band intensities, widths, and baseline also affect the angle.

Segmentation misclassification rate measures whether the reconstruction preserves spatial regions with similar chemistry. Spectra were reduced using principal component analysis and grouped using $K$-means clustering. Because cluster labels are arbitrary, reconstructed labels were matched to ground-truth labels by selecting the permutation that minimizes disagreement:

\begin{align*}
    E_{\mathrm{seg}}
    =\frac{100}{|\mathcal{P}|}
    \min_{\pi\in\mathfrak{S}_K}
    \sum_{i\in\mathcal{P}}
    \mathbf{1}\!\left[z_i\ne\pi(\hat{z}_i)\right],
\end{align*}

where $z_i$ and $\hat{z}_i$ are the ground-truth and reconstructed cluster labels, $\mathfrak{S}_K$ is the set of permutations of the $K$ cluster labels, and $\mathbf{1}[\cdot]$ is the indicator function. The resulting error is the percentage of evaluated positions assigned to a different cluster after label matching. This metric assesses whether spectral differences are preserved sufficiently to distinguish chemically similar domains and reproduce their spatial boundaries. The clusters represent spectral similarity rather than independently verified chemical identities. Lower values indicate better agreement for all three metrics. Better segmentation methods exist for the experimental AFM-IR hyperspectral data presented in this work than the joint PCA-$K$-means, which struggles to identify small regions of unique chemistry. Better segmentation outcomes could likely be achieved with variational autoencoders or Gaussian mixture variational autoencoders combined with more sophisticated clustering methods. PCA-$K$-means is used in this work because of its computational tractability, simplicity, repeatability, and ability to identify overall spectral similarities.

\subsection{Cross-attention Ablation}
\label{subsec::si_attn_ablation}

\begin{figure}
    \centering
    \includegraphics[width=0.9\linewidth]{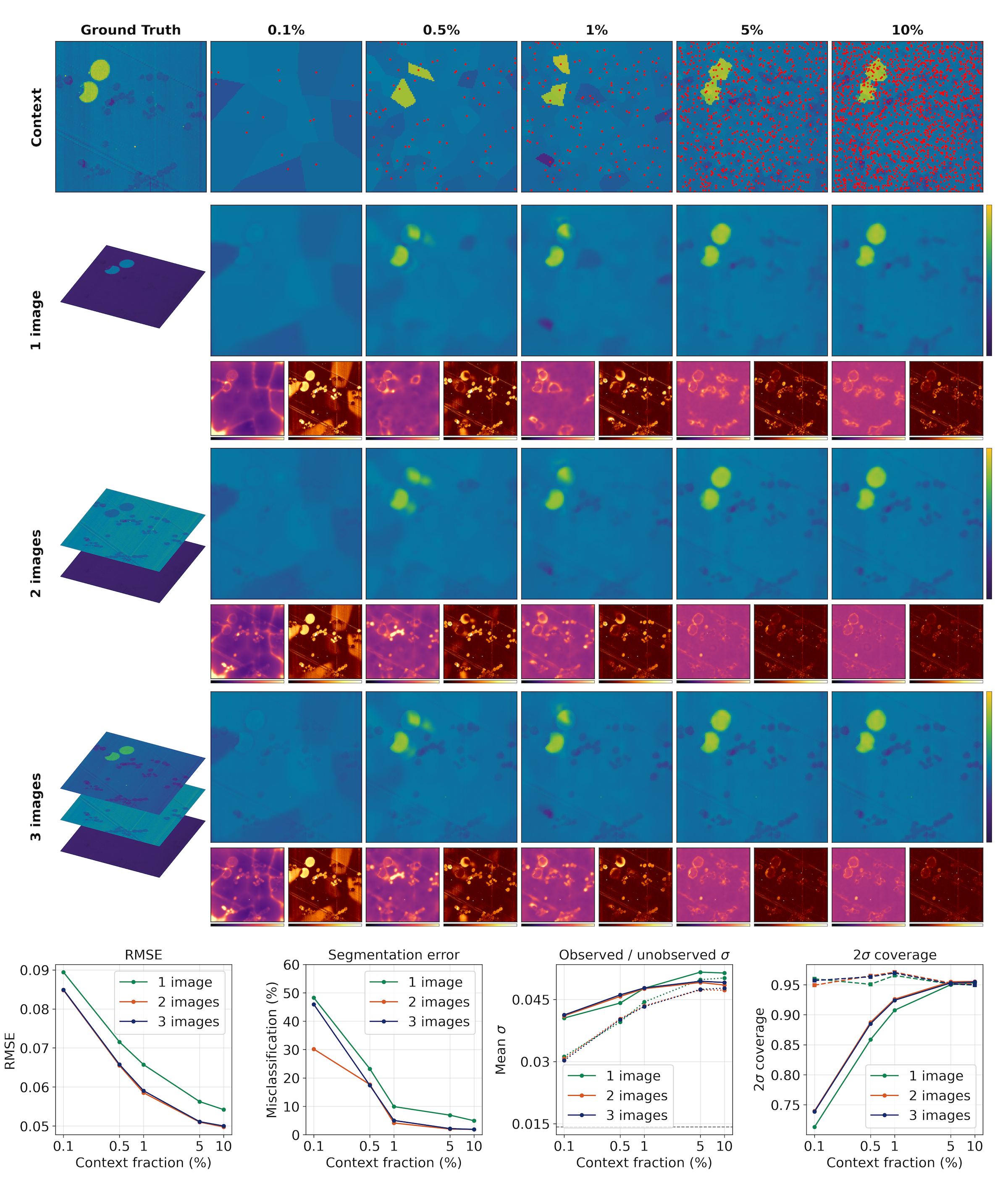}
    \caption{Reconstructions of the multi-particle dataset obtained using the CACNP architecture \textit{without} cross-attention. One to three structure images are used at context fractions from 0.1\% to 10.0\%, as well as corresponding plots of RMSE, segmentation error, $\sigma$, and $2\sigma$ coverage vs context fraction. Reconstructed slices $\mu$ at the \SI{1149}{cm^{-1}} energy are shown, with the pixel-average uncertainty $2\sigma$ and mean error $|y-\mu|$ below. The ground truth \SI{1149}{cm^{-1}} slice is shown in the top left, with nearest neighbor-filled reconstructions at the context points (shown in red) in the top row. The structure images are shown on the left of each reconstruction row. The one-image case uses \SI{1729}{cm^{-1}}, the two-image case uses \SI{1729}{cm^{-1}} and \SI{1049}{cm^{-1}}, and the three-image case additionally uses \SI{1189}{cm^{-1}}. Color bars are shown to the right of the reconstruction plots, and below the $2\sigma$ and $|y-\mu|$ plots, with ranges of (0.0,2.5), (0.0,0.2), and (0.0,0.2), respectively. Dashed/solid lines correspond to observed/unobserved $\sigma$, and raw/calibrated $2\sigma$ coverage.}
    \label{fig:si_attn_ablation}
\end{figure}

Fig.~\ref{fig:si_attn_ablation} shows reconstructions of the multi-particle dataset reconstructed using an architecture the same as the CACNP, but without any cross attention. When compared with the CACNP reconstruction shown in Fig.~\ref{fig:recon_vs_img_multi}, which was performed using the same structure images and context maps, it is clear that the attention-less reconstruction fails to capture the phase boundaries as sharply as the CACNP reconstruction, evidenced by the blurry boundaries and poor registration between the acquired spectra and their correct location within the reconstruction. This poor registration is especially clear at low contexts, where the bright intensity of the PMMA particles at the top of the image blur outside of the boundaries. Reconstruction uncertainties and RMSE are also higher for the attention-less reconstruction than for the CACNP reconstruction in Fig.~\ref{fig:recon_vs_img_multi}. These worse outcomes highlight the added performance provided by the cross-attention between the context spectra and reconstruction throughout the network.

\subsection{Structure Image Impact}
\label{subsec::si_struct_img}

\begin{figure}[htpb!]
    \centering
    \includegraphics[width=\linewidth,height=0.78\textheight,keepaspectratio]{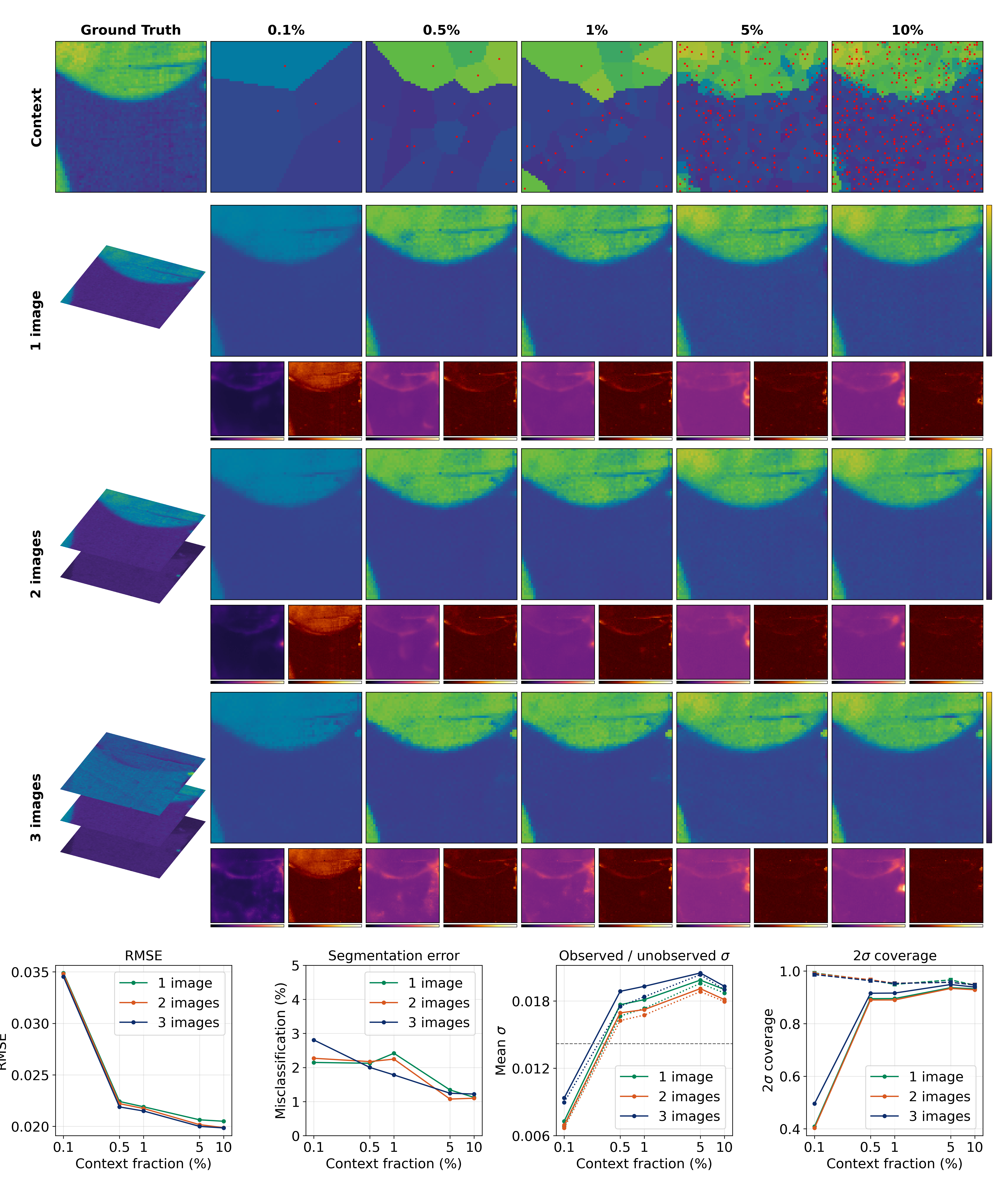}
    \caption{CACNP reconstructions of the single-particle dataset with 1 to 3 structure images with 0.1\% to 10.0\% context fractions. Reconstructed slices $\mu$ at the \SI{1149}{cm^{-1}} energy are shown, with the pixel-average uncertainty $2\sigma$ and mean error $|y-\mu|$ below. The ground truth \SI{1149}{cm^{-1}} slice is shown in the top right, with nearest neighbor-filled reconstructions at the context points (shown in red) in the top row. The structure images (\SI{1505}{cm^{-1}}, \SI{1141}{cm^{-1}}, \SI{1077}{cm^{-1}}) used for each reconstruction are shown on the left of each row of reconstructions. }
    \label{fig:si_recon_vs_img_single}
\end{figure}

The effect of structure images was discussed in \S~\ref{subsec:exp_per}. Fig.~\ref{fig:si_recon_vs_img_single} shows single-particle reconstructions, uncertainty, and error using one to three structure images. The CACNP recovers the PMMA--epoxy boundary and spatial intensity variation within PMMA. Additional images have a smaller effect than in the multi-particle case, consistent with the simpler geometry and two principal phases.

Tables~\ref{tab:si_multi_struct_metrics} and \ref{tab:si_single_struct_metrics} report reconstruction error and uncertainty for the multi- and single-particle datasets. Single-particle RMSE decreases from approximately \SI{0.035}{mV} at 0.1\% context to \SI{0.020}{mV} at 5\%, with little further improvement at 10\%. Misclassification remains low and varies only modestly with the number of structure images. Predicted uncertainty is generally slightly higher with three images, although the dependence is not monotonic.

\begin{table}[htbp!]
\centering
\small
\setlength{\tabcolsep}{3pt}
\begin{tabular}{>{\centering\arraybackslash}p{1.7cm} >{\centering\arraybackslash}p{1.8cm} >{\centering\arraybackslash}p{1.7cm} >{\centering\arraybackslash}p{1.7cm} >{\centering\arraybackslash}p{1.7cm} >{\centering\arraybackslash}p{1.7cm} >{\centering\arraybackslash}p{1.7cm} >{\centering\arraybackslash}p{2.0cm}}
\hline
\makecell[c]{Context \\ Fraction} & \makecell[c]{Number of \\ Images} & \makecell[c]{RMSE} & \makecell[c]{Seg. \\ Error} & \makecell[c]{$\sigma_{\text{obs}}$} & \makecell[c]{$\sigma_{\text{unobs}}$} & \makecell[c]{$2\sigma_{\text{cov}}$ \\ (unobs)} & \makecell[c]{$2\sigma_{\text{cov}}$ \\ (unobs,adj)} \\
\hline \hline
\multirow{3}{*}{0.1\%} & 1 image & 0.0826 & 57.65 & 0.0212 & 0.0335 & 0.7209 & 0.9129 \\
 & 2 images & 0.0761 & 50.40 & 0.0315 & 0.0368 & 0.7993 & 0.9528 \\
 & 3 images & 0.0775 & 52.06 & 0.0275 & 0.0360 & 0.7906 & 0.9460 \\
\hline
\multirow{3}{*}{0.5\%} & 1 image & 0.0574 & 10.21 & 0.0266 & 0.0321 & 0.8632 & 0.9567 \\
 & 2 images & 0.0482 & 10.10 & 0.0278 & 0.0332 & 0.9114 & 0.9668 \\
 & 3 images & 0.0460 & 5.68 & 0.0279 & 0.0327 & 0.9186 & 0.9706 \\
\hline
\multirow{3}{*}{1.0\%} & 1 image & 0.0579 & 13.01 & 0.0348 & 0.0407 & 0.9138 & 0.9566 \\
 & 2 images & 0.0489 & 6.64 & 0.0343 & 0.0401 & 0.9463 & 0.9654 \\
 & 3 images & 0.0461 & 4.87 & 0.0338 & 0.0400 & 0.9500 & 0.9659 \\
\hline
\multirow{3}{*}{5.0\%} & 1 image & 0.0469 & 6.86 & 0.0367 & 0.0375 & 0.9460 & 0.9536 \\
 & 2 images & 0.0384 & 1.80 & 0.0331 & 0.0341 & 0.9541 & 0.9614 \\
 & 3 images & 0.0375 & 1.57 & 0.0333 & 0.0342 & 0.9570 & 0.9621 \\
\hline
\multirow{3}{*}{10.0\%} & 1 image & 0.0441 & 5.46 & 0.0359 & 0.0364 & 0.9478 & 0.9566 \\
 & 2 images & 0.0379 & 1.48 & 0.0317 & 0.0321 & 0.9485 & 0.9575 \\
 & 3 images & 0.0369 & 1.37 & 0.0317 & 0.0321 & 0.9514 & 0.9572 \\
\hline
\end{tabular}
\caption{RMSE, segmentation error, observed $\sigma_{\text{obs}}$ and unobserved $\sigma_{\text{unobs}}$ uncertainty, and raw and calibrated $2\sigma$ coverage for CACNP reconstruction of the multi-particle dataset. Segmentation error is reported as a percentage. Coverage is evaluated on unobserved voxels and reported as a fraction; the nominal Gaussian coverage is 95.4\%.}
\label{tab:si_multi_struct_metrics}
\end{table}

\begin{table}[htbp!]
\centering
\small
\setlength{\tabcolsep}{3pt}
\begin{tabular}{>{\centering\arraybackslash}p{1.7cm} >{\centering\arraybackslash}p{1.8cm} >{\centering\arraybackslash}p{1.7cm} >{\centering\arraybackslash}p{1.7cm} >{\centering\arraybackslash}p{1.7cm} >{\centering\arraybackslash}p{1.7cm} >{\centering\arraybackslash}p{1.7cm} >{\centering\arraybackslash}p{2.0cm}}
\hline
\makecell[c]{Context \\ Fraction} & \makecell[c]{Number of \\ Images} & \makecell[c]{RMSE} & \makecell[c]{Seg. \\ Error} & \makecell[c]{$\sigma_{\text{obs}}$} & \makecell[c]{$\sigma_{\text{unobs}}$} & \makecell[c]{$2\sigma_{\text{cov}}$ \\ (unobs)} & \makecell[c]{$2\sigma_{\text{cov}}$ \\ (unobs,adj)} \\
\hline \hline
\multirow{3}{*}{0.1\%} & 1 image & 0.0349 & 2.15 & 0.0068 & 0.0073 & 0.4089 & 0.9924 \\
& 2 images & 0.0348 & 2.27 & 0.0067 & 0.0069 & 0.4034 & 0.9904 \\
& 3 images & 0.0345 & 2.81 & 0.0090 & 0.0094 & 0.4957 & 0.9864 \\
\hline
\multirow{3}{*}{0.5\%} & 1 image & 0.0224 & 2.12 & 0.0167 & 0.0177 & 0.8946 & 0.9653 \\
& 2 images & 0.0222 & 2.17 & 0.0162 & 0.0169 & 0.8898 & 0.9666 \\
& 3 images & 0.0219 & 2.00 & 0.0175 & 0.0189 & 0.9153 & 0.9631 \\
\hline
\multirow{3}{*}{1.0\%} & 1 image & 0.0219 & 2.42 & 0.0173 & 0.0181 & 0.8953 & 0.9493 \\
& 2 images & 0.0217 & 2.25 & 0.0167 & 0.0172 & 0.8899 & 0.9534 \\
& 3 images & 0.0215 & 1.78 & 0.0184 & 0.0193 & 0.9161 & 0.9527 \\
\hline
\multirow{3}{*}{5.0\%} & 1 image & 0.0206 & 1.34 & 0.0196 & 0.0199 & 0.9364 & 0.9660 \\
& 2 images & 0.0202 & 1.07 & 0.0188 & 0.0191 & 0.9336 & 0.9570 \\
& 3 images & 0.0200 & 1.25 & 0.0203 & 0.0205 & 0.9485 & 0.9571 \\
\hline
\multirow{3}{*}{10.0\%} & 1 image & 0.0205 & 1.12 & 0.0187 & 0.0190 & 0.9321 & 0.9464 \\
& 2 images & 0.0199 & 1.10 & 0.0179 & 0.0181 & 0.9287 & 0.9475 \\
& 3 images & 0.0198 & 1.22 & 0.0190 & 0.0193 & 0.9388 & 0.9467 \\
\hline
\end{tabular}
\caption{RMSE, segmentation error (\%), observed $\sigma_{\text{obs}}$ and unobserved $\sigma_{\text{unobs}}$ uncertainty, and raw and calibrated $2\sigma$ coverage for the CACNP reconstruction of the single-particle dataset. Coverage is evaluated on unobserved voxels and reported as a fraction; the nominal Gaussian coverage is 95.4\%.}
\label{tab:si_single_struct_metrics}
\end{table}

\begin{figure}[htpb!]
    \centering
    \includegraphics[width=\linewidth,height=0.78\textheight,keepaspectratio]{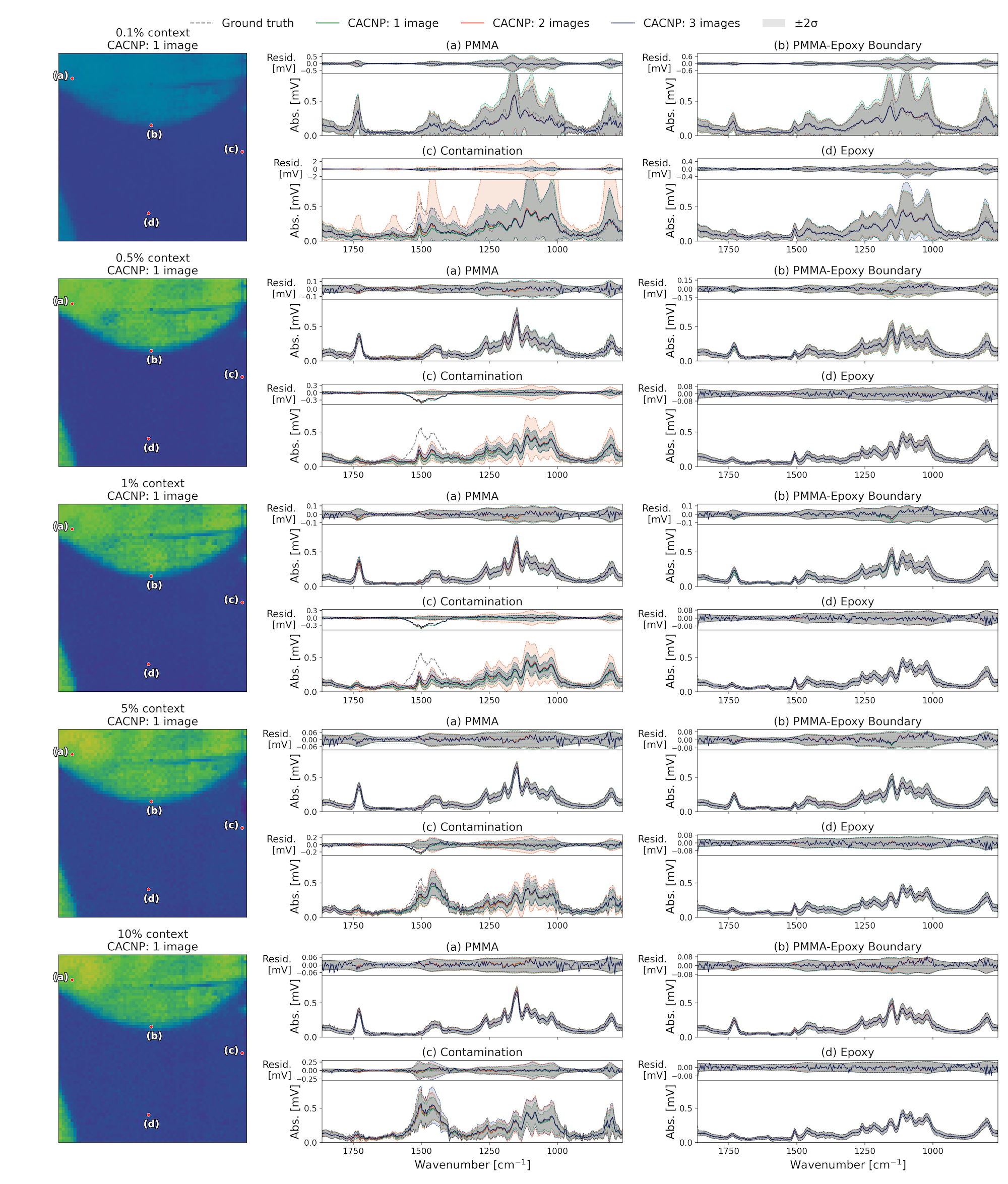}
    \caption{Spectral comparison of CACNP reconstructions of the single-particle dataset using one to three structure images at context fractions of 0.1\%, 0.5\%, 1.0\%, 5.0\%, and 10.0\% (top to bottom). The left column shows reconstructed slices at \SI{1149}{cm^{-1}}, with four locations marked: (a) PMMA, (b) PMMA-Epoxy Boundary, (c) contamination, and (d) epoxy. The corresponding spectral panels compare ground truth (gray dashed lines) with CACNP predictions using increasing numbers of structure images. Colored solid lines indicate the predicted mean $\mu$, while shaded regions and dashed bounds indicate $\mu \pm 2\sigma$ using calibrated standard deviations. Residual strips above each spectrum show the signed error $\mu-y$, with uncertainty bounds of $\pm 2\sigma$ about zero.}
    \label{fig:si_spectra_vs_img_single}
\end{figure}

Fig.~\ref{fig:si_spectra_vs_img_single} compares spectra from PMMA, the PMMA--epoxy boundary, contamination, and epoxy. The contamination feature between 1550 and \SI{1400}{cm^{-1}} is recovered when nearby context spectra contain it, but recovery becomes less reliable when those observations are absent. The reconstructions also retain the stronger \SI{1149}{cm^{-1}} response in parts of the PMMA particle.

\begin{figure}[htpb!]
    \centering
    \includegraphics[width=\linewidth,height=0.78\textheight,keepaspectratio]{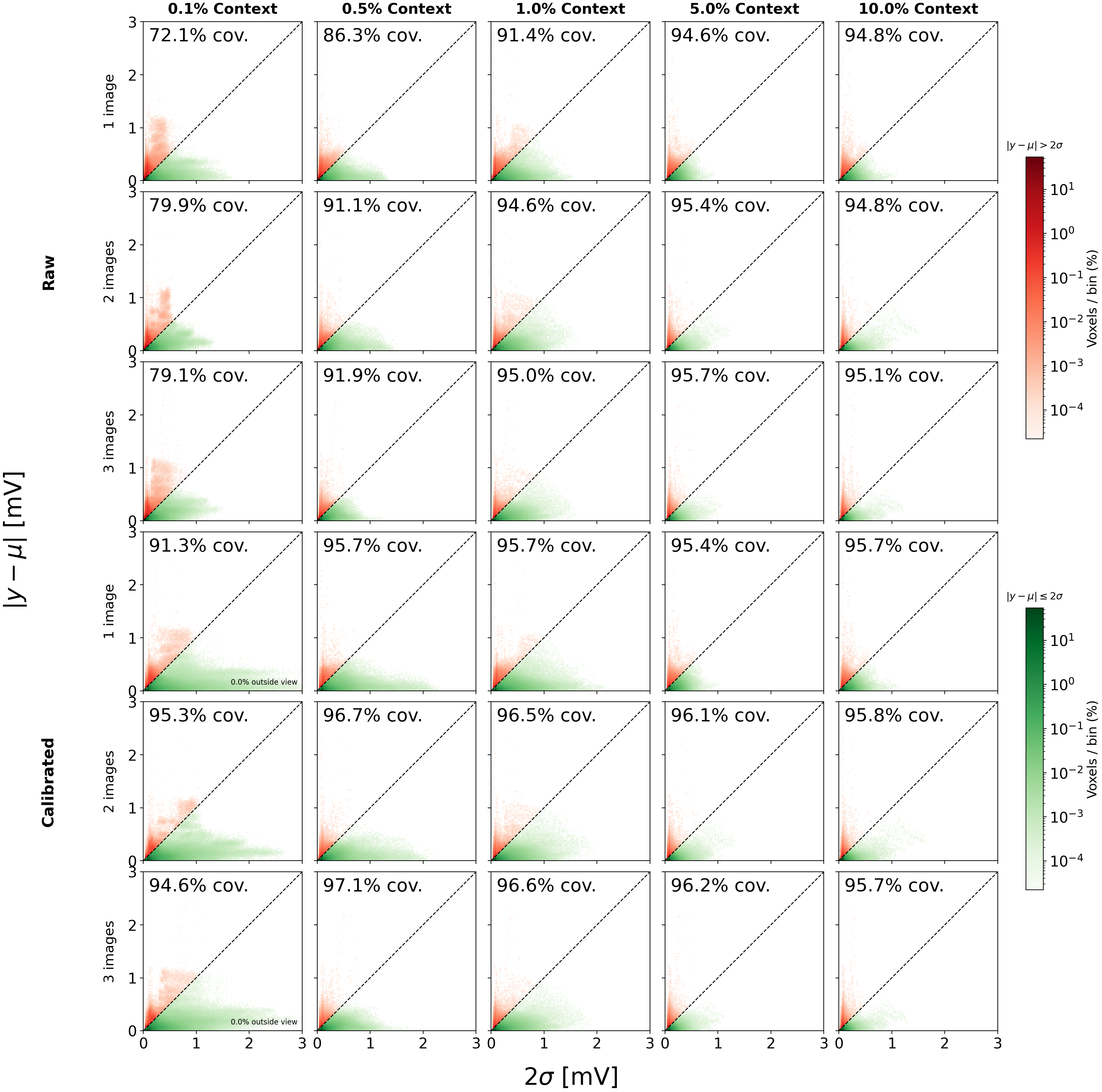}
    \caption{Density plots of $2\sigma$ coverage on unobserved voxels for CACNP-reconstructed multi-particle dataset, using 1 to 3 structure images. Coverage plots are shown for both raw (top 3 rows) and calibrated (bottom 3 rows) uncertainty. The percent of unobserved voxels within $\pm2\sigma$ is displayed in the top left of each plot. Voxels with absolute errors within and outside the $\pm2\sigma$ window are shown in green and red, respectively. Note that the colorscales are logarithmic, and that most of the density is concentrated in the very bottom left of the plot (low uncertainty, low error).}
    \label{fig:si_multi_sigma_cov}
\end{figure}

Fig.~\ref{fig:si_multi_sigma_cov} compares absolute reconstruction error with raw and calibrated $2\sigma$ for unobserved voxels in the multi-particle dataset. Points above the diagonal exceed the uncertainty interval. Calibration reduces the low-context undercoverage, while raw coverage approaches the nominal value as context increases.

\begin{figure}[htpb!]
    \centering
    \includegraphics[width=\linewidth,height=0.78\textheight,keepaspectratio]{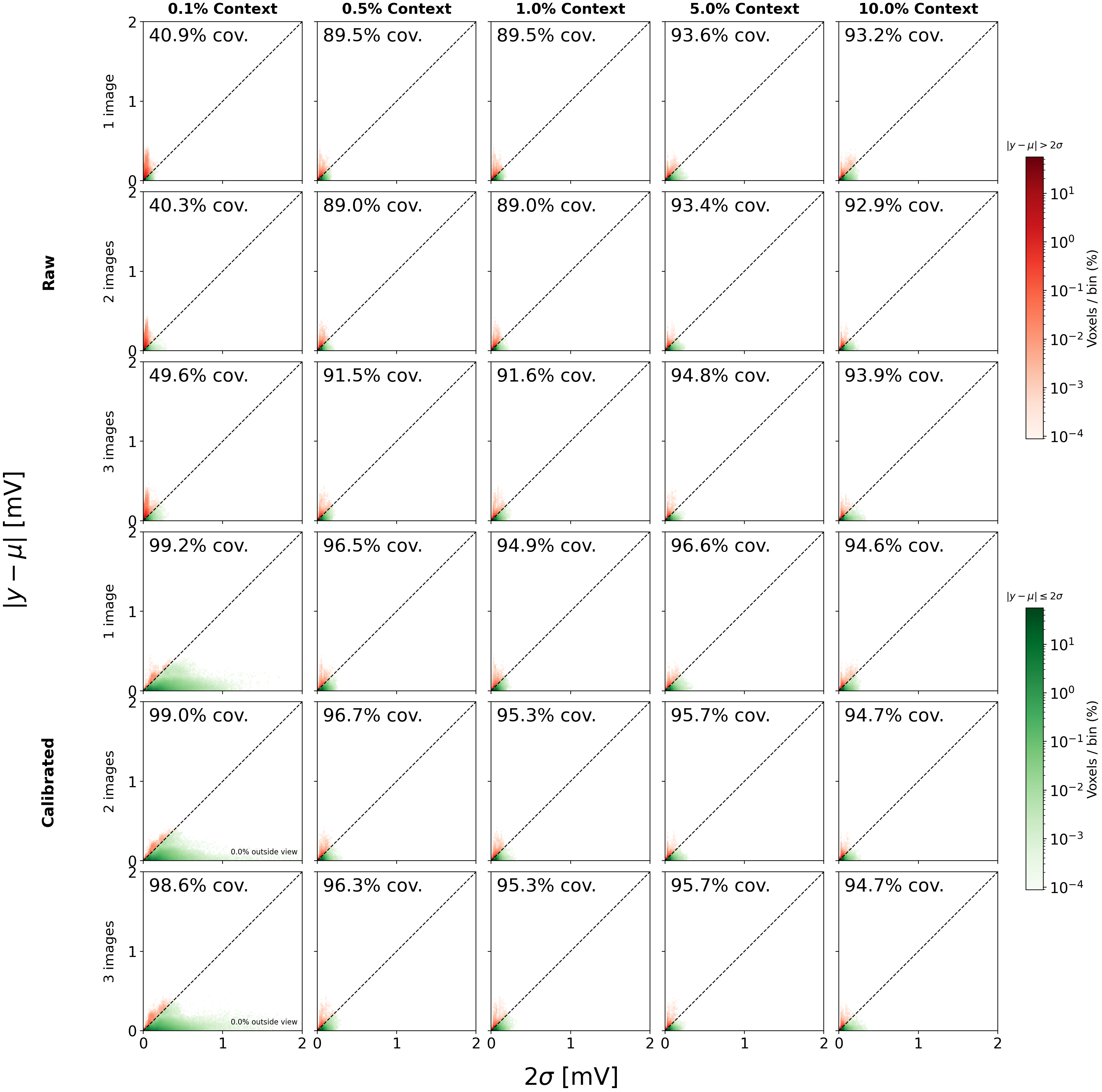}
    \caption{Density plots of $2\sigma$ coverage on unobserved voxels for CACNP-reconstructed single-particle dataset, using 1 to 3 structure images. Coverage plots are shown for both raw (top 3 rows) and calibrated (bottom 3 rows) uncertainty. The percent of unobserved voxels within $\pm2\sigma$ is displayed in the top left of each plot. Voxels with absolute errors within and outside the $\pm2\sigma$ window are shown in green and red, respectively. Note that the colorscales are logarithmic, and that most of the density is concentrated in the very bottom left of the plot (low uncertainty, low error).}
    \label{fig:si_single_sigma_cov}
\end{figure}

Fig.~\ref{fig:si_single_sigma_cov} presents the corresponding single-particle distributions. At 0.1\% context, raw coverage is 40.3--49.6\%, whereas calibrated coverage is 98.6--99.2\%, indicating that calibration overcorrects at this sparse limit. At 1\% context, calibrated coverage is 94.9--95.3\%, close to the nominal 95.4\% (Table~\ref{tab:si_single_struct_metrics}).

\subsection{Other Reconstruction Methods}
\label{subsec::si_other_methods}

Four additional reconstruction methods were compared with the CACNP in \S~\ref{subsec:method_comp}: nearest neighbor filling (NN-filling), coupled non-negative matrix factorization (CNMF) reconstruction \cite{borodinovApplicationPansharpeningAlgorithm2019}, Gaussian process regression, and Gao et al.'s Spatial-Spectral Network (SS-Net) reconstruction \cite{gaoFastNanoIRHyperspectral2024}.

For NN, unmeasured spectra were reconstructed by copying the spectrum at the nearest measured spatial position, identified using the Euclidean distance transform in SciPy. This produces a piecewise-constant reconstruction that assumes local spectral similarity without fitting a spatial or spectral model. The measured structure images were inserted directly into their corresponding spectral channels; their contrast was not used to interpolate the remaining channels.

For GP, the hyperspectral cube was reconstructed using a sparse variational Gaussian process implemented in GPyTorch, with a constant mean, a scaled Mat\'ern-$5/2$ covariance kernel, and a Gaussian observation likelihood. Separate length scales were learned for the two spatial coordinates and the spectral-channel index, each constrained to the interval $[1,25]$ in pixel or channel-index units, imposing a stationary correlation model across the cube. The model was conditioned on the measured context spectra and structure images, with up to 2000 learnable inducing points, and optimized for 1000 Adam steps using batches of up to 20\,000 observed voxels.

For CNMF, we adapted the coupled non-negative matrix factorization approach of Borodinov et al. \cite{borodinovApplicationPansharpeningAlgorithm2019}. We specified a rank-three factorization, following their use of a prescribed component count and reflecting the expectation of up to three constituents--PS, PMMA, and epoxy--in our experimental datasets. Non-negative spectral factors and spatial coefficients were first estimated from the measured context spectra. The coefficients were then interpolated onto the full spatial grid and refined through a second non-negative factorization of the supplied structure images, initialized using the corresponding spectral-factor values. The final cube combined the original spectral factors with the refined spatial coefficients. No explicit spatial-smoothness penalty or sum-to-one constraints were imposed.

\begin{figure}[htbp!]
    \centering
    \includegraphics[width=1.0\linewidth]{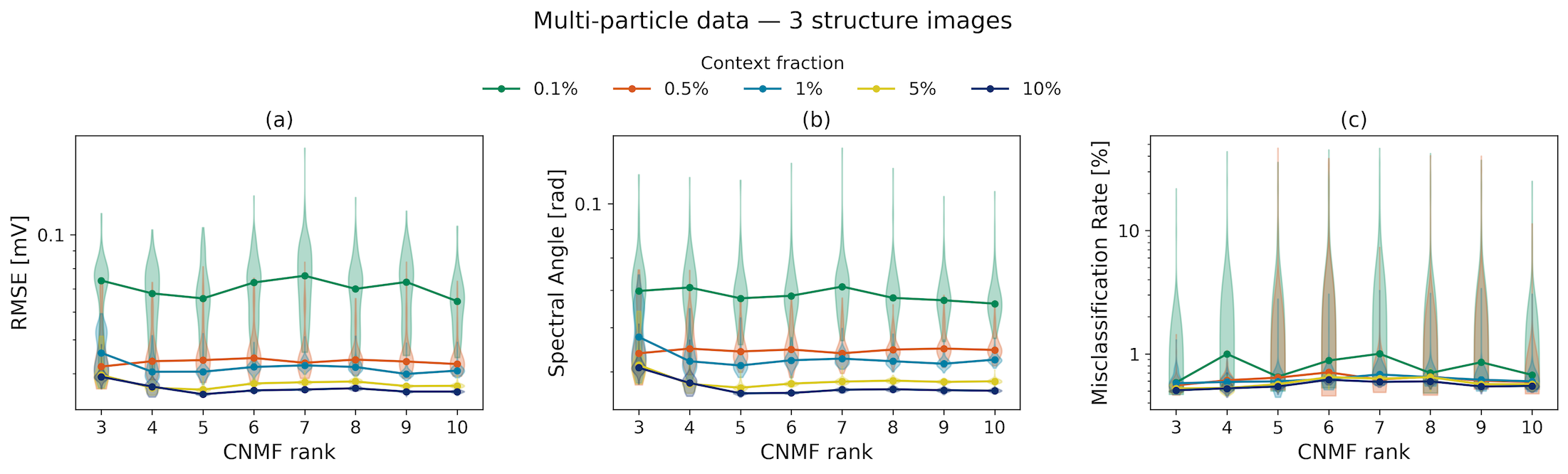}
    \caption{Effect of CNMF rank on reconstruction of the multi-particle dataset using three structure images at \SI{1729}{cm^{-1}}, \SI{1189}{cm^{-1}}, and \SI{1049}{cm^{-1}}. Panels show (a) RMSE, (b) spectral angle, and (c) misclassification rate for ranks 3--10. Colors indicate context fractions of 0.1\%, 0.5\%, 1\%, 5\%, and 10\%. Violin plots show the distributions across shared random context draws, with connected markers indicating median errors. Comparisons use 100 draws per fraction, except at 10\% context, where 89 draws completed at every rank are included. All error axes use logarithmic scales.}
    \label{fig:cnmf_rnak_sweep}
\end{figure}

To assess sensitivity to the prescribed component count, CNMF reconstructions of the multi-particle dataset were repeated with ranks from 3 to 10 using the same three structure images and shared context draws. Fig.~\ref{fig:cnmf_rnak_sweep} compares the resulting RMSE, spectral angle, and misclassification rate. Quantitative differences between ranks are relatively modest and depend on the context fraction and error metric: increasing rank improves spectral reconstruction in several comparisons, particularly between ranks three and five at higher context, but further increases do not produce consistent improvements, and segmentation does not improve systematically. We therefore retained rank three as a practical choice based on prior knowledge of the three principal constituents, PS, PMMA, and epoxy, rather than selecting a rank retrospectively using the fully measured cube. This follows the chemistry-informed rationale of Borodinov et al.~\cite{borodinovApplicationPansharpeningAlgorithm2019}, who related the component count to the number of distinguishable spectral signatures and used two components for their experimental PMMA--PS--epoxy sample because PS and epoxy were nearly indistinguishable over their measured spectral range. Rank three is consequently a physically motivated baseline for our datasets, rather than a claim of universally optimal reconstruction performance.

For SS-Net, we adapted the structure-to-spectrum convolutional network described by Gao et al \cite{gaoFastNanoIRHyperspectral2024}. A separate model was trained for each context realization and structure-image combination, using patches of the supplied structure images as inputs and measured context spectra as supervision. Five convolutional layers mapped these patches to local hyperspectral predictions. Training minimized spectral mean-squared error at observed positions and an image-consistency term based on structural similarity at the supplied image channels. Ten percent of context spectra were withheld for validation, and the model with the lowest validation loss was retained over a maximum of 256 epochs. Overlapping spatial neighborhoods were then evaluated across the image to reconstruct the complete hyperspectral cube.

\subsection{Comparison with Other Reconstruction Methods}
\label{subsec::si_method_comp}

\begin{figure}[htbp!]
    \centering
    \includegraphics[width=\linewidth,height=0.78\textheight,keepaspectratio]{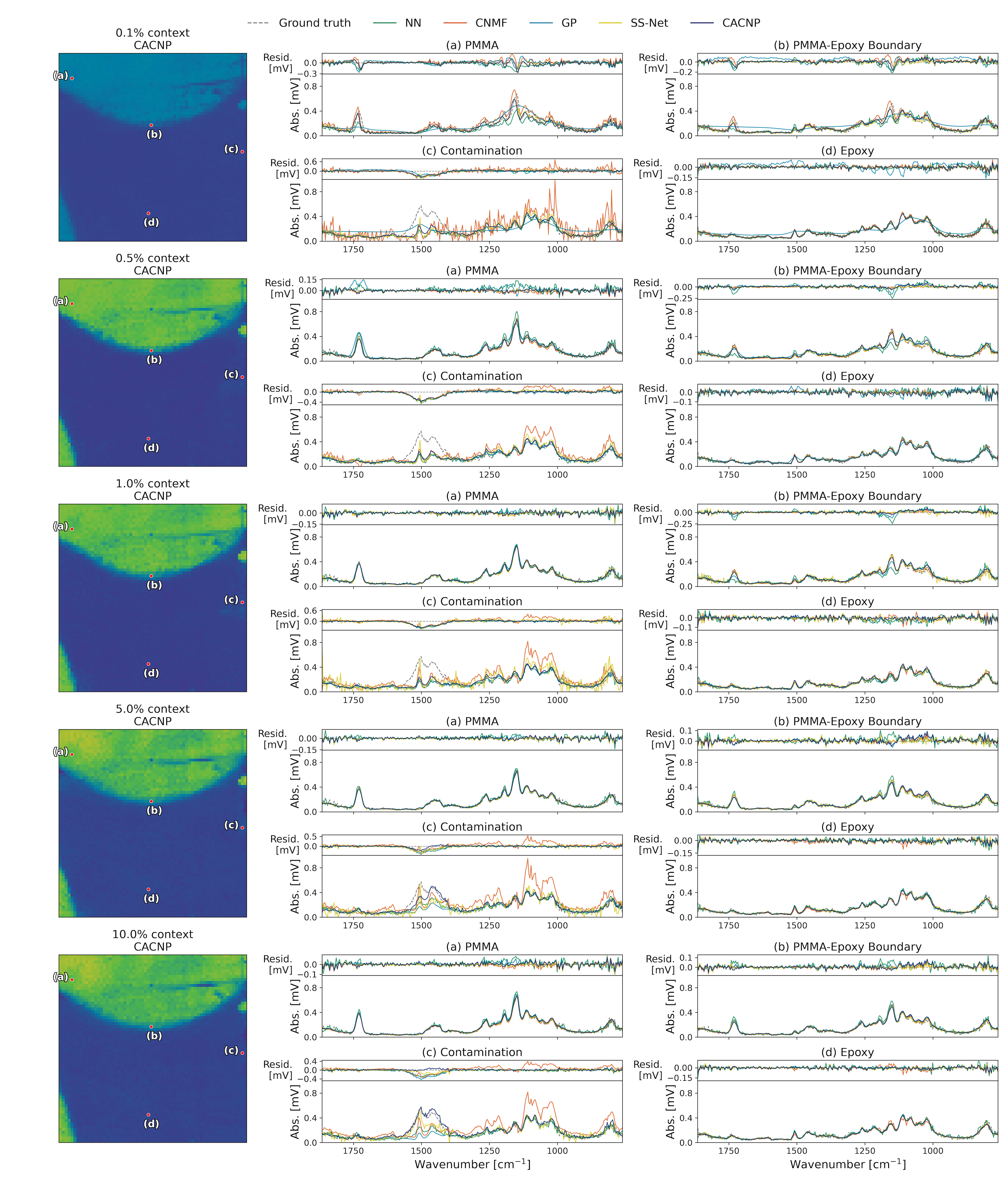}
    \caption{Spectral comparison of NN, CNMF, GP, SS-Net, and CACNP reconstructions of the single-particle dataset at context fractions of 0.1\% to 10.0\%. Each row shows the CACNP reconstruction at \SI{1149}{cm^{-1}} on the left, with four labeled sampling locations. The corresponding ground-truth spectra (dashed gray) and reconstructed spectra are shown on the right. Residuals $\mu-y$ are plotted above each spectral panel.}
    \label{fig:si_meth_spec_single}
\end{figure}

The reconstruction-method comparisons were introduced in \S~\ref{subsec:method_comp}. Fig.~\ref{fig:si_meth_spec_single} compares NN, CNMF, GP, SS-Net, and CACNP spectra at four locations in the single-particle dataset across the five context fractions. The residual strips show the signed discrepancies from ground truth, including those around the localized contamination feature.

\begin{figure}[htbp!]
    \centering
    \includegraphics[width=0.8\linewidth]{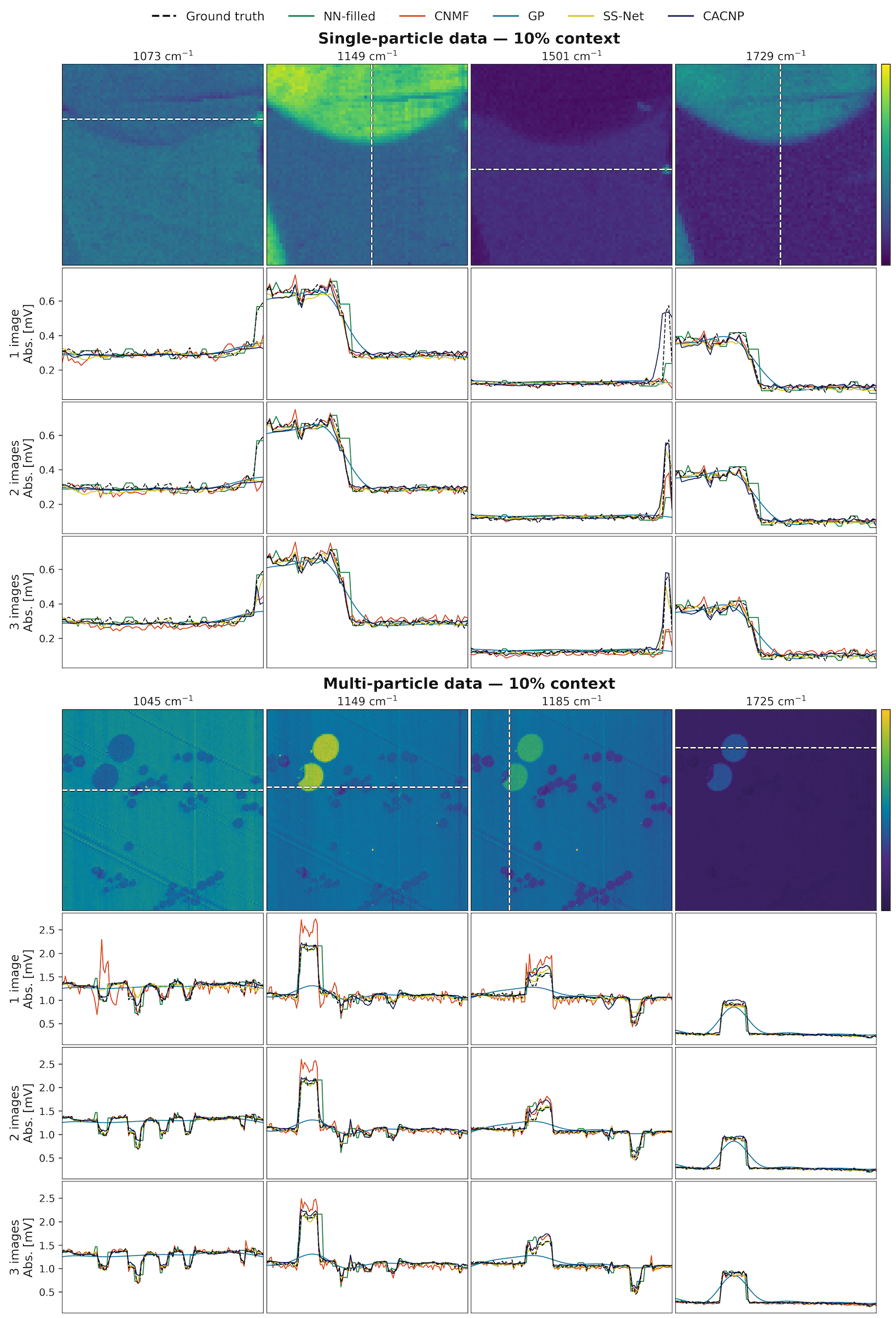}
    \caption{Line profiles of the single- (top) and multi- (bottom) particle datasets reconstructed using NN-filling, coupled non-negative matrix factorization (CNMF), Gaussian process (GP), SS-Net, and the CACNP at 10\% context. The ground truth images at specified energies are shown on top, and the locations of the plotted line profiles is shown in a dashed white line. Line profiles for reconstructions using 1, 2, and 3 structure images are shown.}
    \label{fig:si_line_recon}
\end{figure}

\begin{figure}[htbp!]
    \centering
    \includegraphics[width=1.0\linewidth]{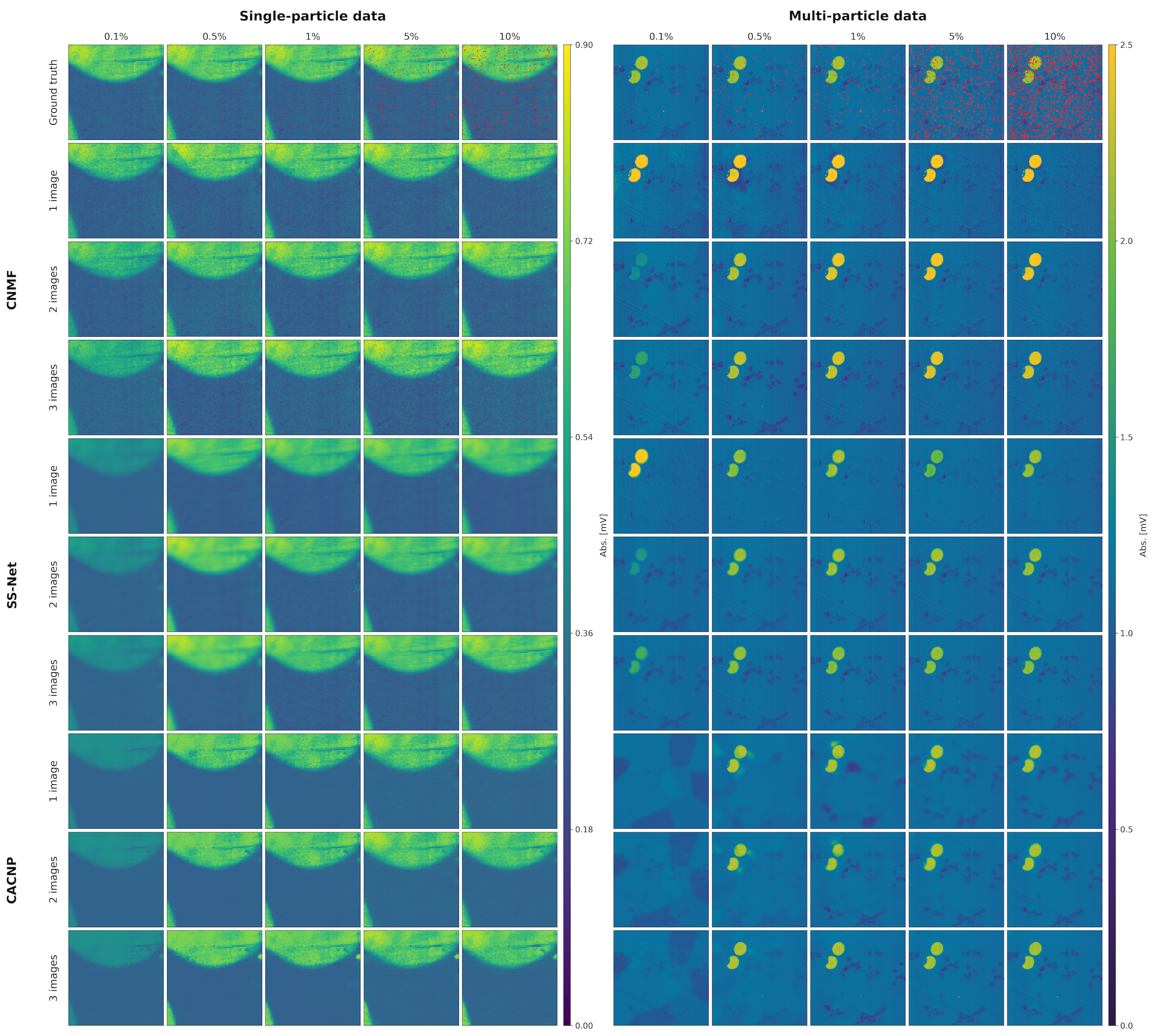}
    \caption{Comparisons of \SI{1149}{cm^{-1}} slices of CNMF, SS-Net, and CACNP reconstructions of the single- (left) and multi- (right) particle dataset using 1, 2, and 3 structure images. Ground truth images are shown at the top with locations of context spectra shown as red dots. The structure images used for the single- and multi-particle datasets correspond to the \SI{1505}{cm^{-1}}, \SI{1141}{cm^{-1}}, \SI{1077}{cm^{-1}}, and \SI{1729}{cm^{-1}}, \SI{1189}{cm^{-1}}, \SI{1049}{cm^{-1}}, energies, respectively.}
    \label{fig:cnmf_ssnet_cacnp_img_comp}
\end{figure}

\begin{figure}[htbp!]
    \centering
    \includegraphics[width=\linewidth,height=0.78\textheight,keepaspectratio]{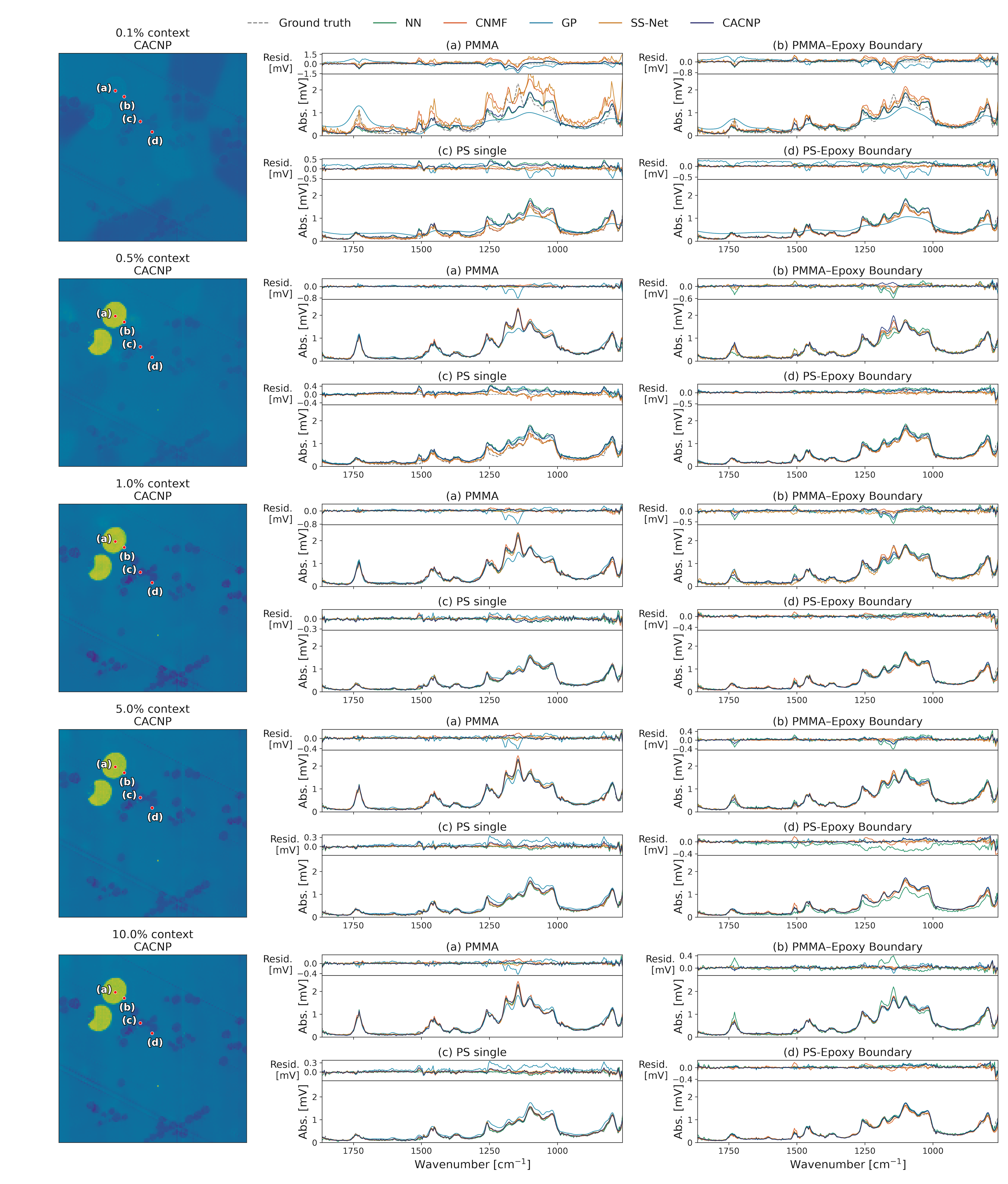}
    \caption{Spectral comparison of NN, CNMF, GP, SS-Net, and CACNP reconstructions of the multi-particle dataset at context fractions of 0.1\% to 10.0\%. Each row shows the CACNP reconstruction at \SI{1149}{cm^{-1}} on the left, with four labeled sampling locations. The corresponding ground-truth spectra (dashed gray) and reconstructed spectra are shown on the right. Residuals $\mu-y$ are plotted above each spectral panel.}
    \label{fig:si_meth_spec_big}
\end{figure}

Fig.~\ref{fig:si_meth_spec_big} shows the corresponding multi-particle comparison at PMMA, a PMMA--epoxy boundary, an isolated PS particle, and a PS--epoxy boundary. CACNP and CNMF have similar overall spectral fidelity, while CACNP better captures some outlier features and NN and GP show larger discrepancies.

\subsection{Estimation of Reconstruction Times}
\label{subsec::si_timing}

\S~\ref{subsec:method_comp} compared the reconstruction performance of NN, CNMF, GP, and SS-Net with the CACNP, including their comparative reconstruction times. Fig.~\ref{fig:err_vs_time} shows the estimated reconstruction times for different context fractions of the single- and multi-particle dataset. We expand on how the times were estimated here.

The cumulative acquisition and reconstruction time was estimated assuming that the hyperspectral cube is reconstructed after each newly acquired spectrum:
\begin{equation}
T_m(N) = K t_{\mathrm{img}} + N t_{\mathrm{spec}}
+ \sum_{n=1}^{N} \widehat{t}_m(n),
\end{equation}
where $T_m(N)$ is the total estimated time for method $m$ after acquiring $N$ context spectra, $K$ is the number of structure images, $t_{\mathrm{img}}$ is the acquisition time per structure image, $t_{\mathrm{spec}}$ is the acquisition time per spectrum, and $\widehat{t}_m(n)$ is the estimated computation time for an update using $n$ context spectra. Three structure images were assumed to be acquired once at the beginning of the experiment. At an assumed imaging rate of one line per second, each image requires 64~s for the single-particle dataset and 128~s for the multi-particle dataset. Each spectrum was assumed to require 4~s, as was the case for acquiring the single- and multi-particle datasets.

Computation times were averaged across repeated timing trials at each measured context size and linearly interpolated to intermediate integer context counts. For CNMF, GP, and SS-Net, each update includes fitting and reconstruction; for CACNP, it includes reconstruction and uncertainty calibration, with offline model training excluded. Acquisition and computation were assumed to occur sequentially, with instantaneous transfer of newly acquired experimental data to an NVIDIA A100 GPU; data-transfer overhead within the timed reconstruction routines remains included. These times are intended as estimates rather than predictions of complete experimental duration. Additional factors, including tip positioning, instrument settling, scan overhead, communication latency, and drift correction, could increase the actual acquisition time. Reconstruction times, along with corresponding reconstruction errors are shown for the single- and multi-particle dataset in Table~\ref{tab:si_time_errors_small} and Table~\ref{tab:si_time_errors_big}, respectively. 

\begin{table}[htbp!]
\centering
\small
\setlength{\tabcolsep}{4pt}
\caption{Single-particle reconstruction errors and estimated acquisition plus reconstruction times using three structure images. Errors are median (sample standard deviation) over 100 context draws. Times reproduce the cumulative estimates in Fig.~\ref{fig:err_vs_time}. Acquisition assumes 4 s per spectrum and 64 s per structure image, with reconstruction after each acquired spectrum and no overlap. CACNP uncertainty calibration is included; offline training and additional experimental overhead are excluded. All methods were timed on an NVIDIA A100 GPU.}
\label{tab:si_time_errors_small}
\begin{tabular}{cccccc}
\toprule
Method & Context (\%) & Time (h:min:s.ms) & RMSE (mV) & Spectral angle (rad) & Misclassification (\%) \\
\midrule
\multirow[c]{5}{*}{CNMF} & 0.1 & 00:03:28.679 & 0.0358 (0.0096) & 0.1737 (0.0426) & 0.73 (5.45) \\
 & 0.5 & 00:04:36.019 & 0.0254 (0.0014) & 0.1297 (0.0035) & 0.49 (0.35) \\
 & 1 & 00:06:04.406 & 0.0239 (0.0009) & 0.1212 (0.0022) & 0.37 (0.26) \\
 & 5 & 00:17:31.173 & 0.0214 (0.0005) & 0.1111 (0.0011) & 0.27 (0.07) \\
 & 10 & 00:31:51.513 & 0.0209 (0.0004) & 0.1093 (0.0010) & 0.24 (3.32) \\
\midrule
\multirow[c]{5}{*}{GP} & 0.1 & 00:06:44.087 & 0.0902 (0.0051) & 0.3999 (0.0142) & 23.71 (3.38) \\
 & 0.5 & 00:22:27.478 & 0.0321 (0.0031) & 0.1601 (0.0132) & 17.74 (5.87) \\
 & 1 & 00:45:55.561 & 0.0260 (0.0012) & 0.1304 (0.0046) & 10.75 (2.62) \\
 & 5 & 03:52:26.229 & 0.0226 (0.0002) & 0.1152 (0.0007) & 9.69 (0.45) \\
 & 10 & 07:45:08.998 & 0.0222 (0.0001) & 0.1132 (0.0004) & 9.62 (0.33) \\
\midrule
\multirow[c]{5}{*}{SS-Net} & 0.1 & 00:04:07.098 & 0.0269 (0.0142) & 0.1433 (0.0389) & 0.95 (1.75) \\
 & 0.5 & 00:18:57.827 & 0.0234 (0.0009) & 0.1246 (0.0050) & 0.61 (0.19) \\
 & 1 & 00:52:51.838 & 0.0243 (0.0017) & 0.1293 (0.0094) & 0.59 (0.18) \\
 & 5 & 05:32:58.498 & 0.0216 (0.0008) & 0.1154 (0.0041) & 0.48 (0.11) \\
 & 10 & 11:25:13.913 & 0.0203 (0.0003) & 0.1092 (0.0017) & 0.44 (0.09) \\
\midrule
\multirow[c]{5}{*}{CACNP} & 0.1 & 00:03:28.219 & 0.0326 (0.0084) & 0.1582 (0.0344) & 8.63 (12.11) \\
 & 0.5 & 00:04:33.560 & 0.0225 (0.0014) & 0.1150 (0.0045) & 1.27 (1.65) \\
 & 1 & 00:05:59.328 & 0.0215 (0.0006) & 0.1108 (0.0020) & 1.07 (0.62) \\
 & 5 & 00:17:09.209 & 0.0203 (0.0001) & 0.1073 (0.0003) & 1.07 (4.92) \\
 & 10 & 00:31:06.792 & 0.0200 (0.0001) & 0.1067 (0.0003) & 1.03 (4.57) \\
\bottomrule
\end{tabular}
\end{table}

\begin{table}[htbp!]
\centering
\small
\setlength{\tabcolsep}{4pt}
\caption{Multi-particle reconstruction errors and estimated acquisition plus reconstruction times using three structure images. Errors are median (sample standard deviation) over 100 context draws. Times reproduce the cumulative estimates in Fig.~\ref{fig:err_vs_time}. Acquisition assumes 4 s per spectrum and 128 s per structure image, with reconstruction after each acquired spectrum and no overlap. CACNP uncertainty calibration is included; offline training and additional experimental overhead are excluded. All methods were timed on a NVIDIA A100 GPU.}
\label{tab:si_time_errors_big}
\begin{tabular}{cccccc}
\toprule
Method & Context (\%) & Time (h:min:s.ms) & RMSE (mV) & Spectral angle (rad) & Misclassification (\%) \\
\midrule
\multirow[c]{5}{*}{CNMF} & 0.1 & 00:07:30.014 & 0.0739 (0.0168) & 0.0698 (0.0113) & 0.66 (5.50) \\
 & 0.5 & 00:12:04.389 & 0.0419 (0.0112) & 0.0540 (0.0078) & 0.49 (1.15) \\
 & 1 & 00:17:46.883 & 0.0458 (0.0063) & 0.0577 (0.0069) & 0.53 (0.66) \\
 & 5 & 01:03:50.258 & 0.0396 (0.0047) & 0.0514 (0.0052) & 0.47 (0.29) \\
 & 10 & 02:01:52.415 & 0.0391 (0.0014) & 0.0508 (0.0015) & 0.46 (0.39) \\
\midrule
\multirow[c]{5}{*}{GP} & 0.1 & 00:26:12.542 & 0.1351 (0.0237) & 0.1426 (0.0265) & 46.39 (7.52) \\
 & 0.5 & 01:47:46.415 & 0.0766 (0.0048) & 0.0791 (0.0043) & 28.67 (8.20) \\
 & 1 & 03:29:03.197 & 0.0673 (0.0017) & 0.0710 (0.0018) & 26.09 (8.29) \\
 & 5 & 16:58:15.975 & 0.0615 (0.0004) & 0.0660 (0.0004) & 26.84 (4.10) \\
 & 10 & 33:51:05.393 & 0.0607 (0.0002) & 0.0654 (0.0003) & 27.54 (2.95) \\
\midrule
\multirow[c]{5}{*}{SS-Net} & 0.1 & 00:18:29.362 & 0.0640 (0.0131) & 0.0584 (0.0031) & 2.53 (0.51) \\
 & 0.5 & 02:15:19.818 & 0.0413 (0.0069) & 0.0499 (0.0023) & 1.34 (0.50) \\
 & 1 & 04:45:14.899 & 0.0379 (0.0024) & 0.0473 (0.0012) & 1.11 (0.46) \\
 & 5 & 25:12:19.850 & 0.0348 (0.0030) & 0.0439 (0.0016) & 0.78 (0.46) \\
 & 10 & 51:20:12.332 & 0.0338 (0.0031) & 0.0431 (0.0017) & 0.65 (0.44) \\
\midrule
\multirow[c]{5}{*}{CACNP} & 0.1 & 00:07:30.703 & 0.0688 (0.0130) & 0.0617 (0.0080) & 28.81 (12.48) \\
 & 0.5 & 00:12:07.004 & 0.0483 (0.0064) & 0.0506 (0.0027) & 6.27 (11.66) \\
 & 1 & 00:17:49.980 & 0.0421 (0.0042) & 0.0480 (0.0015) & 2.95 (1.64) \\
 & 5 & 01:03:30.930 & 0.0375 (0.0041) & 0.0461 (0.0014) & 1.47 (0.86) \\
 & 10 & 02:00:55.730 & 0.0369 (0.0055) & 0.0458 (0.0018) & 1.34 (0.91) \\
\bottomrule
\end{tabular}
\end{table}

\subsection{Acquisition Stopping Conditions}
\label{subsec::si_stopping}

Stopping was evaluated using the calibrated mean predicted standard deviation over all spectral channels at currently unobserved spatial positions, denoted $\bar{\sigma}_n$ after acquiring $n$ spectra. At the end of each window of ten new acquisitions, the uncertainty variation was calculated as
\begin{equation}
R_n = \max_{j=n-10,\ldots,n}\bar{\sigma}_j
      - \min_{j=n-10,\ldots,n}\bar{\sigma}_j.
\end{equation}
Here, $R_n$ measures the full uncertainty range within the window, capturing both increases and decreases rather than only the net change between its endpoints. A stopping condition was satisfied when $R_n < \delta$ for three consecutive windows, using tolerances $\delta = 0.002$, $0.003$, $0.004$, $0.005$, and $0.010$~mV. Windows contained non-overlapping acquisition increments and shared their boundary state. Any window that failed the criterion reset the consecutive-window count; the earliest possible stopping point was therefore after 30 additional spectra. This criterion constrains variation within each window, rather than the total drift across all three windows.

\begin{figure}
    \centering
    \includegraphics[width=1.0\linewidth]{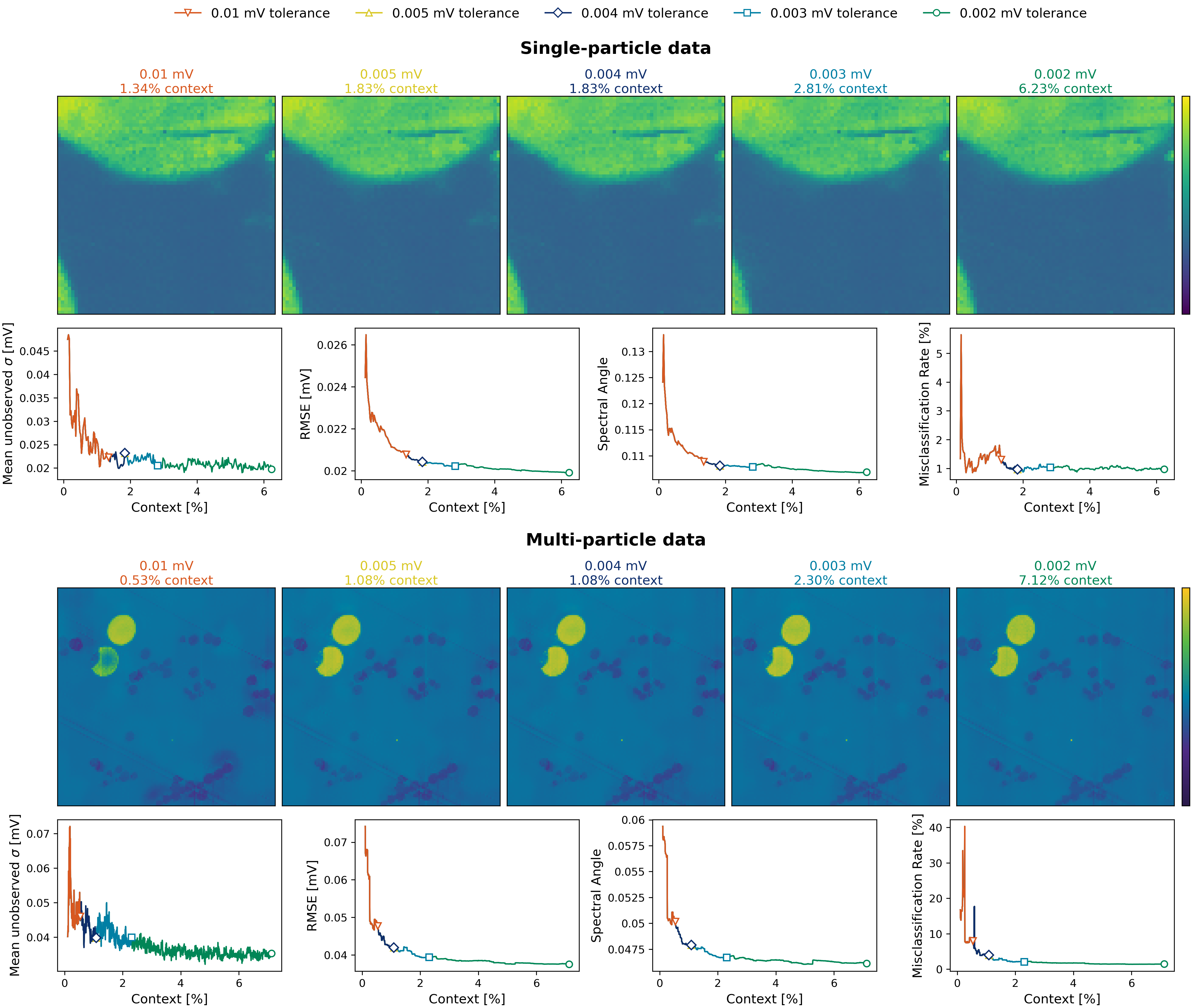}
    \caption{Stopping criteria results for the single- (top) and multi- (bottom) particle datasets.  \SI{1149}{cm^{-1}} slices of the reconstructions are shown. Acquisitions were run until the mean unobserved $\sigma$ has not changed by a given threshold (0.01, 0.005, 0.004, 0.003, \SI{0.002}{mV}) for three consecutive, non-overlapping windows of ten newly acquired spectra.}
    \label{fig:si_stopping}
\end{figure}

Ten random-sampling trajectories were evaluated per dataset using different acquisition seeds and a fixed initial hexagonal grid containing five and seventeen spectra for the single- and multi-particle datasets, respectively. One spectrum was added per update, and uncertainty calibration was repeated using a fixed-seed 70/30 split of the available context. Each trajectory recorded the first satisfaction of every tolerance, continuing until all criteria were satisfied or the total spectrum budget of 800 or 3200 was reached. Thus, comparisons between tolerances within a run used the same acquisition history. Fig.~\ref{fig:si_stopping} presents one representative trajectory per dataset: reconstructed slices at \SI{1149}{cm^{-1}} are shown at each stopping point, together with the corresponding mean uncertainty, RMSE, spectral angle, and misclassification histories. Image labels report the stopping tolerance and acquired context fraction, and matching markers identify the stopping points on the curves.

\subsection{Acquisition Strategy Impacts}
\label{subsec::si_acq_strat}

\begin{figure}[htbp!]
    \centering
    \includegraphics[width=\linewidth,height=0.78\textheight,keepaspectratio]{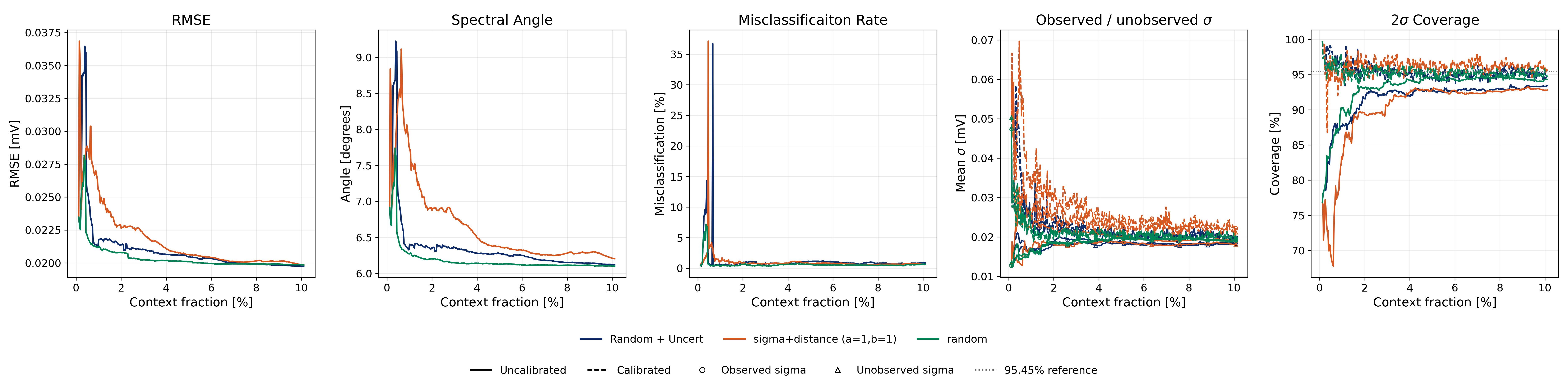}
    \caption{RMSE, spectral angle, misclassification rate, observed and unobserved $\sigma$, and $2\sigma$ coverage for the CACNP-reconstructed single-particle dataset using random, combined random-uncertainty, and distance-uncertainty acquisition functions.}
    \label{fig:si_strat_comp_single}
\end{figure}

The acquisition strategies and their effects were discussed in \S\S~\ref{subsec::acq} and \ref{subsec:acq_strat}. Fig.~\ref{fig:si_strat_comp_single} compares reconstruction error, uncertainty, and coverage during acquisition of the single-particle dataset. Distance-uncertainty selection produces larger RMSE and spectral angle than random or random-uncertainty selection, despite resolving localized features in the main-text renderings.

\begin{figure}[htbp!]
    \centering
    \includegraphics[width=\linewidth,height=0.78\textheight,keepaspectratio]{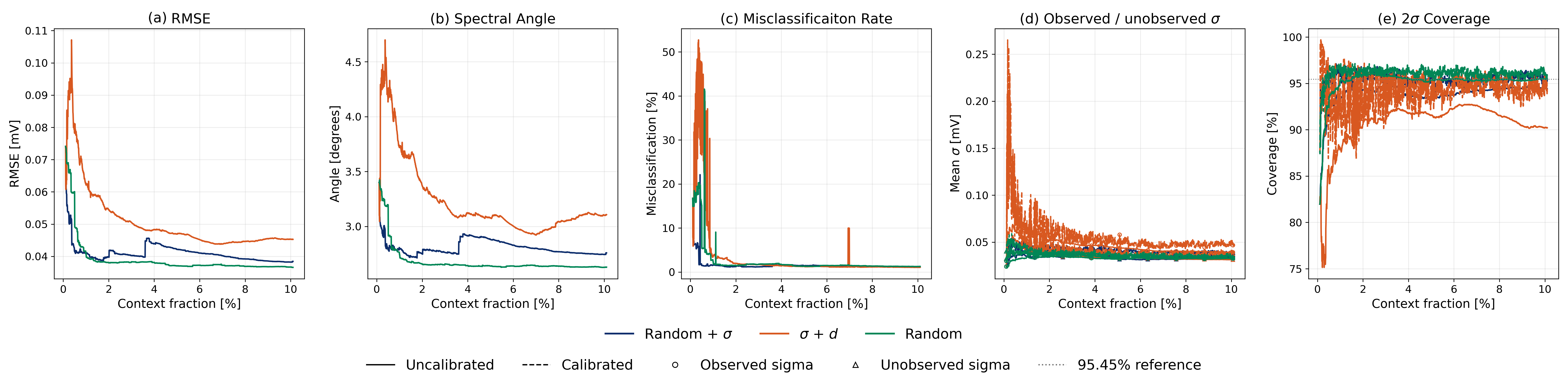}
    \caption{RMSE, spectral angle, misclassification rate, observed and unobserved $\sigma$, and $2\sigma$ coverage for the CACNP-reconstructed multi-particle dataset using random, combined random-uncertainty, and distance-uncertainty acquisition functions.}
    \label{fig:si_strat_comp_multi}
\end{figure}

Fig.~\ref{fig:si_strat_comp_multi} presents the same metrics for the multi-particle dataset. Under distance-uncertainty acquisition, uncertainty rises rapidly before declining, raw coverage is less accurate, and calibrated coverage fluctuates more strongly. These results illustrate the limitations of interpreting uncertainty as an acquisition or stopping criterion when the context distribution differs from training.

\begin{figure}[htpb!]
    \centering
    \includegraphics[width=\linewidth,height=0.78\textheight,keepaspectratio]{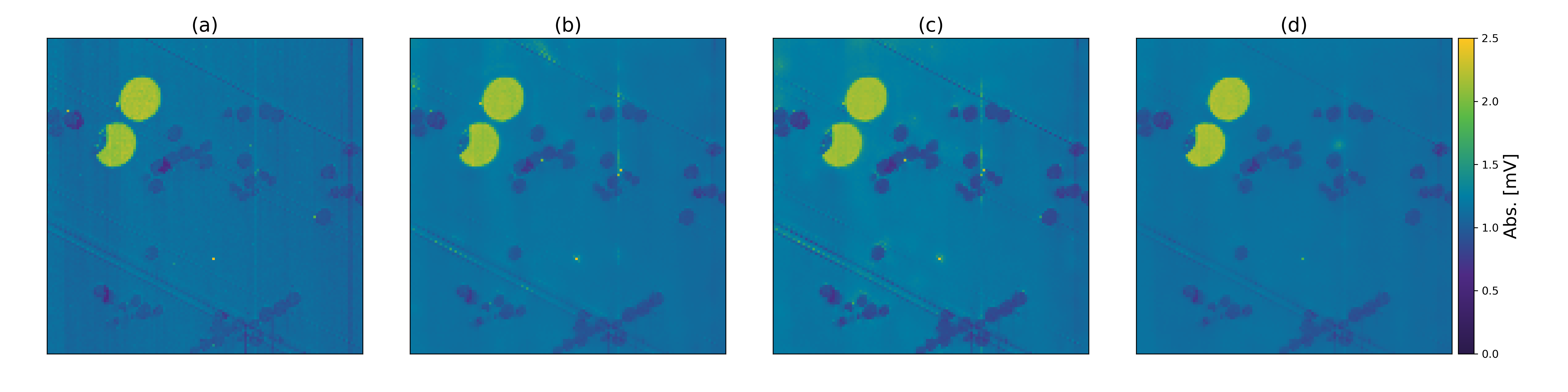}
    \caption{CACNP reconstructions using different strategies. (a) shows the ground truth (GT), (b) shows the uncertainty-random acquisition, (c) shows the uncertainty-distance acquisition, and (d) shows the random acquisition. Each slice is of the \SI{1149}{cm^{-1}} energy, and each reconstruction has 10\% context.}
    \label{fig:si_acq_img_comp}
\end{figure}

Fig.~\ref{fig:si_acq_img_comp} compares multi-particle reconstructions at 10\% context. The uncertainty-driven strategies propagate outlier intensities into surrounding regions, whereas the random reconstruction does not show these artifacts. This spatial comparison complements the aggregate metrics in Fig.~\ref{fig:si_strat_comp_multi}.

\bibliographystyle{IEEEtran}

\bibliography{references.bib} 

\end{document}